\documentclass{statsoc}
\usepackage[utf8]{inputenc}
\usepackage[english]{babel}
\usepackage[a4paper]{geometry}
\usepackage{bm}
\usepackage{amsmath}
\usepackage{amssymb}
\usepackage[authoryear]{natbib}
\usepackage{relsize}
\usepackage[table]{xcolor}
\usepackage{multirow}
\usepackage{mathptmx}
\usepackage{float}
\usepackage{etoolbox}
\usepackage{url}
\usepackage[table]{xcolor}
\usepackage{array}
\usepackage{tikz}
\usetikzlibrary{trees}
\DeclareGraphicsExtensions{.pdf,.png,.jpg,.eps} 
\usepackage[ruled]{algorithm2e}
 \definecolor{algobackgroup}{gray}{0.05}

\makeatletter
\patchcmd{\@makecaption}
{\parbox}
{\advance\@tempdima-\fontdimen2} 
{}{}
\makeatother  

\usepackage{graphicx}
\usepackage[caption=false]{subfig}
\usepackage{enumerate}
\usepackage{comment}
\usepackage{float}

\newcommand\Pair[4]{%
  \arrayrulecolor{cyan!60!black!40}%
  \arrayrulewidth=1pt
  \renewcommand\extrarowheight{1.5pt}%
  \begin{tabular}{|p{2cm}|>{\centering\arraybackslash}p{10pt}|}
  \hline
  \rowcolor{cyan!60!black!10}\textcolor{red!60!black}{#1} & \textcolor{red!60!black}{#2} \\
  \hline 
  \rowcolor{cyan!60!black!10}\textcolor{red!60!black}{#3} & \textcolor{red!60!black}{#4} \\
  \hline
  \end{tabular}%
}

\title[]{A Bayesian Quest for Finding a Unified Model for Predicting Volleyball Games}
\author[Egidi and Ntzoufras]{Leonardo Egidi}
\address{Dipartimento di Scienze Economiche, Aziendali, Matematiche e Statistiche `Bruno de Finetti',
	Universit\`{a} degli Studi di Trieste,
	Trieste,
	Italy.}
\email{legidi@units.it}
\author[Egidi and Ntzoufras]{Ioannis Ntzoufras}
\address{AUEB Sports Analytics Group, Computational and Bayesian Statistics Lab, Department of Statistics,
	Athens University of Economics and Business,
	Athens, Greece.}
\email{ntzoufras@aueb.gr}

\begin{document}

\maketitle

\begin{abstract}

Volleyball is a team sport with unique and specific characteristics. 
We introduce a new two level-hierarchical Bayesian model which accounts for theses volleyball specific characteristics. 
In the first  level, we model the set outcome with a simple logistic regression model. 
Conditionally on the winner of the set, in the second level,  we use a truncated negative binomial distribution for the points earned by the loosing team. An additional Poisson distributed inflation component is introduced to  model the extra points played in the case that the two teams have point difference less than two points. The number of points of the winner within each set is deterministically specified by the winner of the set and the points of the inflation component. 
The team specific abilities and the home effect are used as covariates on all layers of the model (set, point, and extra inflated points). 
The implementation of the proposed model on  the Italian Superlega 2017/2018 data
shows an exceptional reproducibility of  the final league table and a satisfactory predictive ability.

\end{abstract}

\section{Introduction}
\label{sect:intro}

Sports analytics and modelling has long tradition among the statistical community with initial works published back to 1950s and 1960s. 
For example, seminal works have been initiated in the bibliography in the most popular  sports like 
baseball \citep{Mosteller_1952, albert1992bayesian}, 
association football--soccer \citep{Reep_Benjamin_1967}, 
American football  \citep{Mosteller_1970,Harville_1977}, 
and basketball \citep{Stefani_1980,Schwertman_etal_1991}.
World wide web and recent technologies have given access to many scientists to interesting sport related data which are now widely and freely available (see for example in \url{www.football-data.co.uk} for association football and \url{http://www.tennis-data.co.uk/} for tennis). 
Moreover,  new interesting problems have been raised due to the big data that can be derived by inplay sensor and camera driven technologies; see for example in \cite{Metulini_etal_2017} 
and \cite{Facchinetti_etal_2019} for applications in basketball.

Sports analytics is currently a fashionable and attractive topic of research with a growing community of both academics and professionals. 
Regarding team sports prediction modelling, two are the main outcomes of response: 
(a) the win/draw/loss outcome modelled by logistic or multinomial regression models and other similar approaches \citep{carpita2019exploring}, and (b) the goals or points scored by each team, which are modelled by sport specific models depending on the nature of the game. 
In this work, we focus on the second approach, where the response is richer in terms of information used for the estimation of team abilities and allows to obtain a model with better prediction accuracy. 
The biggest group of team sports is the one where the score is measured with a discrete number of goals (of equal value) such as association football, water polo, handball, and hockey (among others). 
In such cases, Poisson log-linear models and their extensions are the most popular choices of models. 
The historic timeline of models developed for such team sports includes the use of simple double Poisson models in the works of \cite{Maher_1982}, \cite{Lee_1997}, the extension of \cite{Dixon_Coles_1997} with the adjustment for $0-0$ and $1-1$ draws, 
the diagonal bivariate Poisson model of \cite{karlis2003analysis}, the Poisson difference model \citep{Karlis_Ntzoufras_2009},  
and the dynamic models of \cite{Rue_Salvesen_2000}, \cite{Owen_2011} and \cite{Koopman_Lit_2015}; see \cite{Tsokos_etal_2019} and references therein for further details and an up-to-date review.

Unlike what happens for other major team sports, modelling volleyball match outcomes has not been thoroughly addressed by  statisticians and data scientists. 
Early attempts to model volleyball data date back to \cite{Lee2004}, who analysed the effect of a team deciding serve or to receive the service at the fifth set.  
Concerning prediction models in volleyball, several authors considered implementing Markov chain models to estimate the winning probabilities of a set and a game; see \cite{Barnett_etal_2008} for a first attempt and 
\cite{ferrante2014winning} for a more recent and complete treatment of the problem. 
\cite{Miskin_etal_2010} used a Bayesian multinomial model based on Markovian transition matrix to model each point and the effect of volleball skills. A similar approach was used by \cite{Drikos_etal_2019} to analyse top-level international teams at several age categories. 
\cite{Sepulveda2017} used a Markov chain model as a useful tool to analyse players’ probability of attack  in terms of team rotation. 
A simpler alternative was based on logistic regression models for the probability of winning 
a set \citep{Marcelino2009,Fellingham_etal_2013}  or a point \citep{Miskin_etal_2010}. 
Recently, \cite{Gabrio2020} proposed a Bayesian hierarchical model for the prediction
of the rankings of volleyball national teams, which also allows
to estimate the results of each match in the league.  
In the point level, \cite{Gabrio2020} used a double Poisson model component. 
\cite{Sonnabend_2020} published an empirical study on the characteristics of the beach volleyball, including details about the distribution of points. 
He used a normal regression model to study the effect of several game characteristics (game heteregoneity, referees, home effect, tournament phase, the fact of winning/loosing the previous sets, gender and age) on the difference of points in a set. 
Finally, concerning the research direction of player performance evaluation, \cite{Mendes_etal_2008} used Bayesian multi-level models to analyse sports activities of elite adult Brazilian players while \cite{Hass2018} implemented a plus/minus approach in order to obtain volleyball players' evaluation metrics.

Unlike volleyball, in most sports (as in basketball and football, for example) there is a single performance outcome, namely the number of points or goals, which is measured cumulatively from the beginning to the end of the game. 
In these situations, a model with the total goals or points as a response is required. 
On the other hand, in volleyball, the winner is announced in two stages/levels of outcomes:  sets and  points within each set. Hence, the winner is the team that reaches first the three sets. 
For this reason, the second level outcome, i.e. the total number of sets, is a random variable which ranges from a minimum of  three to a maximum of five. 
Each set is won by the team that reaches first a pre-specified number of points which is equal to 25 for the first four sets and equal 15  to the final tie-break set.\footnote{This point system was adopted in 1998, during the Men and Women’s World Championships held in Japan (source: \url{https://ncva.com/info/general-info/history-of-volleyball/}).}
Nevertheless, the number of points required by the team winning the set further varies depending on whether there is a margin of two points. 
Hence, volleyball outcomes consist of a natural hierarchy of sets and points within sets, with both  measurements to be random variables. 

In this work, we follow the approach of modelling both outcomes of volleyball: sets and points. 
By this way, the response data are richer in terms of information which enables us to estimate team abilities more accurately and increase the prediction accuracy of our model.

In our perspective, the task of modelling volleyball match results should follow a top-down strategy, from the sets to the single points. Thus, defining the probability of winning a set is the first step; 
building up a generative discrete model for the points realized in each set is the second step. Although following this kind of hierarchy is not mandatory, we maintain it into all our fitted models. 
Hence, we propose a set-by-set statistical model for the  points of the loosing team, conditionally to the set result. 
Another aspect to consider is the strength difference among the teams: weaker teams are of course not favoured when competing against stronger teams, and a parametric assumption about teams' skills is needed. 
In the Bayesian approach, teams' abilities are 
easily incorporated into the model by the use of weakly-informative prior distributions \citep{gelman2008weakly}: similarly to what happens for football models \citep{karlis2003analysis}, the abilities may regard both attack and defense skills, and, moreover, be considered as dynamic over the season \citep{Owen_2011}.  

The rest of the paper is organized as follows. The main features of the game are presented in Section \ref{sec:feat}. In Section \ref{sec:models} we introduce the basic negative binomial model for volleyball outcomes. Model extensions are thoroughly presented in Section \ref{sec:extension}, whereas model estimation, goodness of fit diagnostics and out-of-sample prediction measures are detailed in Section \ref{sec:est}. MCMC replications for the selected negative binomial model are used in Section \ref{sec:pp} to assess its plausibility in comparison with the observed results and to reconstruct the final rank of the league. The paper concludes after a detailed discussion.

\section{The Features of the Game} 
\label{sec:feat}

Volleyball is different than other team sports of invasion (like football and basketball) since the two teams are separated and there is no contact between the players of the two competing teams. 
It belongs to a category of net and ball sports (volleyball, footvolley, headis or sepak takraw, tennis, badminton, pickleball, table-tennis) and therefore it has some unique characteristics that cannot be modelled by using the approaches adopted in other sports  such as the Poisson regression models commonly used in football. 

Here we summarize these characteristics and, in the latter, we address theses issues one-by-one. 
\begin{enumerate}
	\item The first and most important characteristic is that the main outcome of the game is split into two levels: the sets and the points inside each set. 
	Roughly speaking, a set is played until one of the two teams wins first 25 points. This team is the winner of the set. The game is played until a team wins 3 sets. Hence we have two levels of outcomes (sets and points) which are interconnected and should modelled simultaneously. 
	
	\item Moreover, the sets in a volleyball game range from three to five and hence, 
	it is reasonable to test for the assumption of repeated measures of the points which are correlated across different sets.  
	The existence of repeated measurements of points needs to be addressed stochastically and tested within our modelling approach. 
	\color{black}

	\item The points of the winning team are (almost) fixed by the design and the rules of the game. So, given that we know who won the set, the only outcome variability is reflected by the points of the team that lost the specific set. 
	
	\item An additional rule, that creates further complication, is that the winning team should have at least two points margin of difference to win a set. So conceptually if two teams are close in terms of abilities, they could play for infinite time and points until the required difference of two points is achieved. 
	
	\item Finally, the fifth set of the game is terminated at 15 points (and not at 25 points) and it is called tie-break. The two points margin of difference is also required for the tie-break.

\end{enumerate} 	

In this work, we deal with each of the unique characteristics of the volleyball by adding a corresponding component to the model formulation. The resulted model is  a unified approach for the volleyball data and it is unique in the literature. 
To be more specific, we model the two response outcomes (sets and points) hierarchically, using a binomial model for each set and, conditionally on the winner of the set, we use a negative binomial distribution for the points of the loosing team assuming $r=25$ or $r=15$ successes for normal sets and tie-breaks, respectively  (features (a), (c) and (e)). We further truncate this distribution to deal with the two points margin of difference required in each set (feature (d)), and we model the excess of points due to ties (sets with less than two points difference) using a zero inflated Poisson distribution (feature (d)). 
Furthermore, we consider normal random effects to account for the correlation between sets of the same game (feature (b)). 
Finally, we take into consideration the connection between sets and points by considering general team abilities in contrast to point or set specific team abilities (feature (a)). 

As the reader might initially think, our approach is counter-intuitive and apparently in contrast to what a usual sports model may consider. But this counter-intuitive logic is the main innovation of the model we propose. 
By using this approach, our aim is to exploit the fact that we (almost) know the points of the winning team. So if we consider modelling the win/loss of each set in the first level of the model then,  conditionally on the set winner, we can specify a sensible distribution for the points of the loosing team (while the points of the winning team are specified deterministically). 
On the other hand, considering the usual approach, i.e. modelling directly the number of points using a bivariate distribution is more cumbersome and challenging due to the restrictions imposed by the game regulations.
\color{black} 

In Section \ref{sec:models} which follows, we formalise the basic  structure and assumptions of our proposed model while further considerations and extensions of the model are provided in Section \ref{sec:extension}.

\section{The Basic Model for Volleyball}
\label{sec:models}

\subsection{Truncated negative binomial model}
\label{sec:negbin}

Let $Y^A_{s}$ and $Y^B_{s}$ be the random variables of the points in set $s=1,2,\dots, S$ of  two competing teams $A$ and $B$ playing at home and away stadium, respectively. Furthermore, $W_s$  is a binary indicator denoting the win or loss of the home team. 
To begin with, assume for the moment that each set finishes at fixed number of points (25 or 15 depending on the type of set), 
then the points of the winning team are fixed and not random. 
Hence interest lies on the random variable $Y_s$ which denotes the number of points for the team loosing the $s$-th set. 
Concerning the observed realization of the points gained by the loosing team, this will be obtained by
$$
y_s = w_s  y^B_{s} +(1-w_s)   y^A_{s}. 
$$

So in our dataset, we will eventually model the data for two responses: the binary $W_s$ and the count variable $Y_s$. 
Our model is built hierarchically. For the outcome of each set, we use a simple logistic regression model given by 
\begin{eqnarray}
	W_s &\sim& \mathsf{Bernoulli}( \omega_s ), \label{eq_pset0} \\
\mbox{logit}(\omega_s) &= & H^{set}+\alpha_{A(s)}-\alpha_{B(s)}, \label{eq_pset} 
\end{eqnarray}  
where $\alpha_T$ is a parameter capturing the ability of team $T$ to win a set (set abilities henceworth), 
$A(s)$ and $B(s)$ are the home and away team indices, respectively, competing each other at set $s$. 
Now conditionally on the winner of the set, we then model the points of the loosing team for each set using a negative binomial model (ignoring at the moment that the game may continue if the margin of points' difference is less than two points). 
Hence, the model formulation will be now given by 
\begin{equation} 
Y_s | W_s \sim \mathsf{NegBin}(r_s, p_s){\mathcal I}( Y_s \le r_s-2 ),
\label{eq:model}
\end{equation} 
which is the right truncated negative binomial distribution 
with parameters $r_s$ and $p_s$. $\mathcal{I}(A)$ denotes the event indicator, equal one if the event $A$ is true and zero otherwise.
The first parameter, $r_s$, is the number of successes (points here) required to finish the set and it is equal to 25 for sets 1--4, and equal to 15 for the last (fifth) set. 
Mathematically this can be written as 

\begin{equation}
r_s = 25 - 10 \times \mathcal{I} \left( R_s=5 \right), 
\label{total_points}
\end{equation}
where $R_s$ is the sequential set number for the specific game $G(s)$.
Parameter $p_s$ is the probability of realizing a point for the team winning set $s$. 
Equivalently, $q_s=1-p_s$ denotes the probability of realizing a point for the team loosing set $s$. The right truncation has been fixed at $r_s-2$ (23 or 13) points since this is the highest number of points that can be achieved by the loosing team (under the assumption of no ties).
Moreover, the point success probability will be modelled as 
\begin{eqnarray}
\eta_s &=&  \mu+ (1-W_s)H^{points}+(\beta_{A(s)}-\beta_{B(s)})(1-2W_s), 
\label{pointeta} \\ 
p_s&=& \frac{1}{1+e^{\eta_s}}, 
\label{point_prob}
\end{eqnarray} 
where $\eta_s$ is the (fixed effects) linear predictor for the  points of the loosing team given by \eqref{pointeta}.
The constant $\mu$ is a common baseline parameter, $H^{point}$ is the point home advantage for the host  team, $\beta_{A(s)}, \beta_{B(s)}$ are the point abilities for teams $A(s)$  and $B(s)$, respectively. 
Consider the first equation: the larger is the difference between the abilities of team $A$ and 
team $B$, $\xi_s = \beta_{A(s)}-\beta_{B(s)}$, the higher is the expected number of points team $A$ will win when loosing a set. 
Equivalently, in this case, the lower will be the number of points of team $B$ when loosing a set. Hence the multiplier $(1-W_s)$ in Eq.~\eqref{pointeta} controls the presence of the home effect, while the multiplier $(1-2W_s)$ controls the sign of the difference in the abilities of the two teams (depending on which team is playing at home).

Before we proceed, let us focus for a moment on the untruncated negative binomial, for which the average number of points for team $A$ (evaluated if $W_s=0$) and team $B$ (evaluated if $W_s=1$) in the $s$-th set are, respectively:
\begin{align}
\begin{split}
E[ Y^A_{s}| W_s=0]  &=r_s \exp \left \{ \mu + H^{points} + \beta_{A(s)}-\beta_{B(s)} \right \}  = r_s \times M \cdot \xi_s \cdot e^{H^{points}}\\
E[ Y^B_{s}| W_s=1]  &=r_s \exp \left \{\mu -\beta_{A(s)}+\beta_{B(s)} \right \} =  r_s \times \frac{M}{ \xi_s }, \\ 
\mbox{where~} M &= e^{\mu} \mbox{~and~}  \xi_s =  \exp \left( \beta_{A(s)}-\beta_{B(s)} \right)~.
\end{split}
\label{eq_averages}
\end{align}

\begin{figure}
\centering
\includegraphics[scale=0.7]{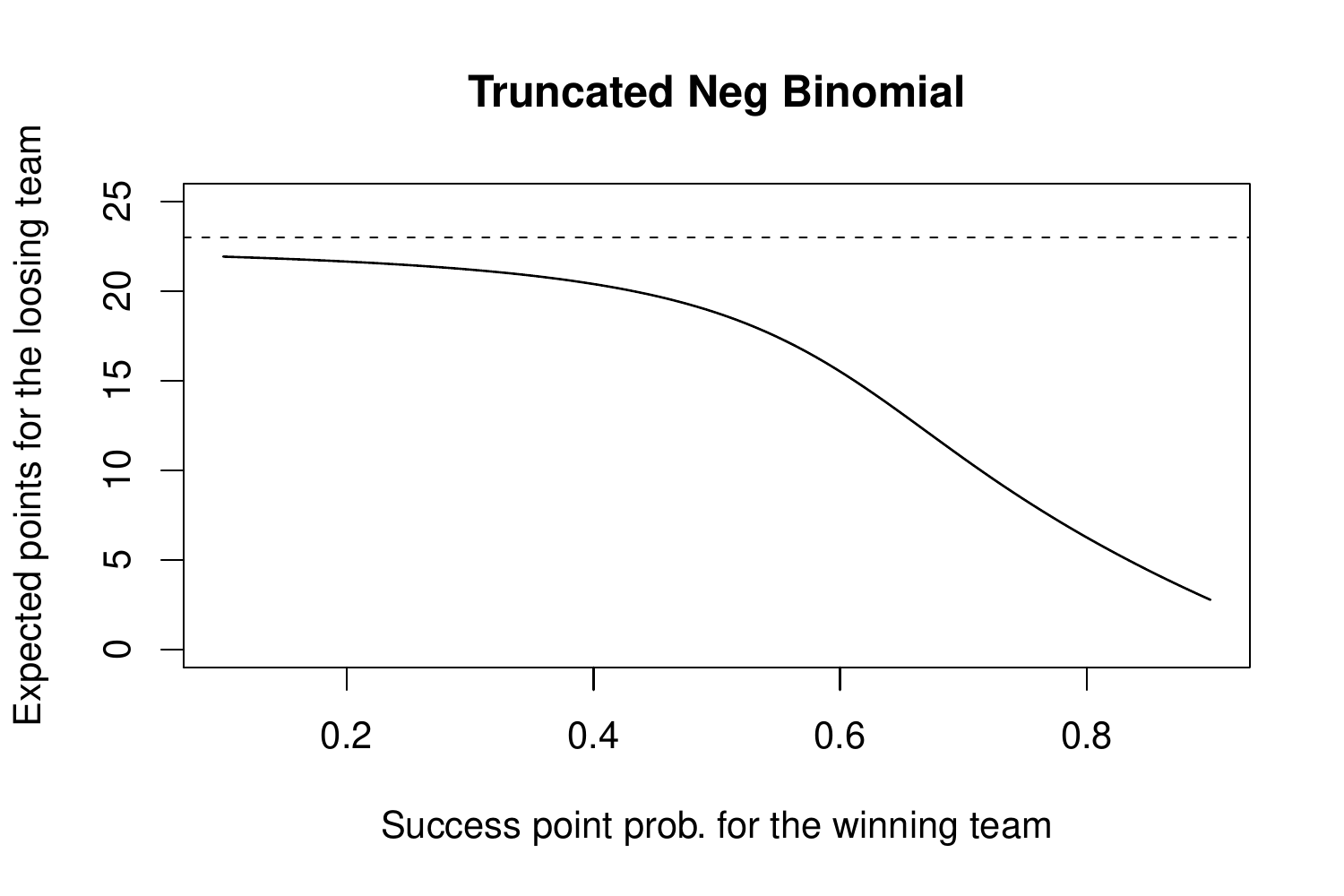}
\caption{ Expected number of points collected by the team loosing the set $s$ against the success point probability $p_s$ for the winning team, truncated negative binomial with upper truncation at $r_s-2$. As the point probability for the team winning the set increases, the expected number of points for the team loosing the set decreases.}
\label{fig1} 
\end{figure}

However, in this initial model formulation the loosing-set team can reach at most $r_s-2$ points (in case of no extra points), then we need to reconsider the expected number of points of the loosing team (i.e. Eq. \eqref{eq_averages}) in the light of the upper truncation. 
\cite{shonkwiler2016variance} reports the mathematical expression for the truncated negative binomial distribution which in our case becomes equal to: 
\begin{align}
\begin{split}
E[ Y^A_{s}| Y^A_{s} \leq r_s-2, W_s=0]
&\ =E[ Y^A_{s}| W_s=0] - \frac{c^*_s}{p_s}\\
E[ Y^B_{s}| Y^B_{s}\leq r_s-2, W_s=1]
&\ = E[ Y^B_{s}| W_s=1] - \frac{c^*_s}{p_s},\\ 
c^*_s  =\frac{(r_s-1)f_{NB}(r_s-1)}{F_{NB}(r_s-2; r_s, p_s)},\ & c^*_s>0, 
\end{split}
\label{eq_averages_trunc}
\end{align}
where $f_{NB}$, $F_{NB}(x; r, p)$ are the probability mass function and the cumulative function, respectively, of negative binomial with parameters $r$ and $p$.
The interpretation is identical to the untruncated case: the higher is the point ability of a team, the higher will be the number of points when loosing a set. However, the untruncated mean is subtracted by the positive factor $c^*_s/p_s$, which forces the mean of the points of the loosing team to be lower or equal than $r_s-2$. For illustration purposes, Figure \ref{fig1} displays the expected number of points collected by the team loosing the set $s$ against the success point probability $p_s$:  as the point probability for the team winning the set increases, the expected number of points for the team loosing the set decreases. In Section \ref{sec:zip} we will extend the model to allow for extra points after $r_s$ due to the required margin of two points difference.

The random variables of the points of each team, under the assumed model, can be now written as 
$$
Y^A_s = W_s r_s + (1-W_s) Y_s \mbox{~~and~~}
Y^B_s = W_s Y_s + (1-W_s) r_s~. 
$$
while the expected points of each set are given by
\begin{eqnarray}
E(Y^A_s) &=& \omega_s r_s + (1-\omega_s) r_s \xi_s M e^{H^{points}} - (1-\omega_s)  \left(1+\xi_s M e^{H^{points}}\right)c_s^*,  \nonumber\\ 
E(Y^B_s) &=& \omega_s r_s \frac{M}{\xi_s} + (1-\omega_s) r_s -
\omega_s  \left(1+\frac{M}{\xi_s}\right) c_s^*,  
\label{expected_values_TRNB} 
\end{eqnarray}
where $c^*_s$ is given in \eqref{eq_averages_trunc} while $\xi_s, \, M$ are defined in \eqref{eq_averages}. 
\color{black}

The Bayesian model is completed  by assigning some weakly informative priors \citep{gelman2008weakly} to the  set and point abilities, for each team $T=1,\ldots, N_T$:
\begin{align}
\begin{split}
\alpha_T^*, \beta_T^*  &\sim \mathcal{N}(0,2^2), \\
 \mu, H^{point}, H^{set} & \sim  \mathcal{N}(0,10^6),
\end{split}
\label{eq_priors}
\end{align}
where $N_T$ is the total number of teams in the league. 
In order to achieve identifiability, set and point abilities need to be constrained; in such a framework we impose sum-to-zero (STZ) constraints for both $\alpha$ and $\beta$ by centering the free parameters $\alpha_T^*$ and $\beta_T^*$ using the equations:
\begin{eqnarray*} 
\alpha_T &=&\alpha_T^* - \overline{\alpha}^* \\
\beta_T &=&\beta_T^* - \overline{\beta}^* ,
 \end{eqnarray*} 
 for $T=1,\ldots, N_T$, where $\overline{\alpha}^*$ and $\overline{\beta}^*$ are the means of the unconstrained abilities given by 
 $\overline{\alpha}^* = \frac{1}{N_T} \sum _{T=1}^{N_T}\alpha_T^*$ and 
 $\overline{\beta}^* = \frac{1}{N_T} \sum _{T=1}^{N_T}\beta_T^*$, respectively.  
Note that the constrained abilities $\alpha_T$ and $\beta_T$ are finally used in the model which automatically satisfies the sum-to-zero constraint and this centering is applied in every iteration of the MCMC algorithm. 
In terms of interpretation, the STZ parametrization implies that an average team will have ability parameter close to zero.

%
%
%

\color{black}

\subsection{Using random effects to capture within game correlation} 
\label{sec:random_effects} 

We further introduce game additive random effects to capture the induced correlation between the set repetition and the fact that we have 3--5 measurements of the points of the loosing team. 
Hence, the point probability in each set given by \eqref{point_prob} is slightly changed to 
\begin{eqnarray}
p_s &=& \frac{1}{1 + e^{\eta_s+ \varepsilon_{G(s)}}}, \nonumber  
\end{eqnarray} 
where $\eta_s$ is the (fixed effects) linear predictor 
as defined in \eqref{pointeta} 
and $\varepsilon_{G(s)}$ are the game random effects which are used to capture any potential correlation across the measurements of the points within each game. 

\begin{figure}
\centering
\includegraphics[scale=0.44]{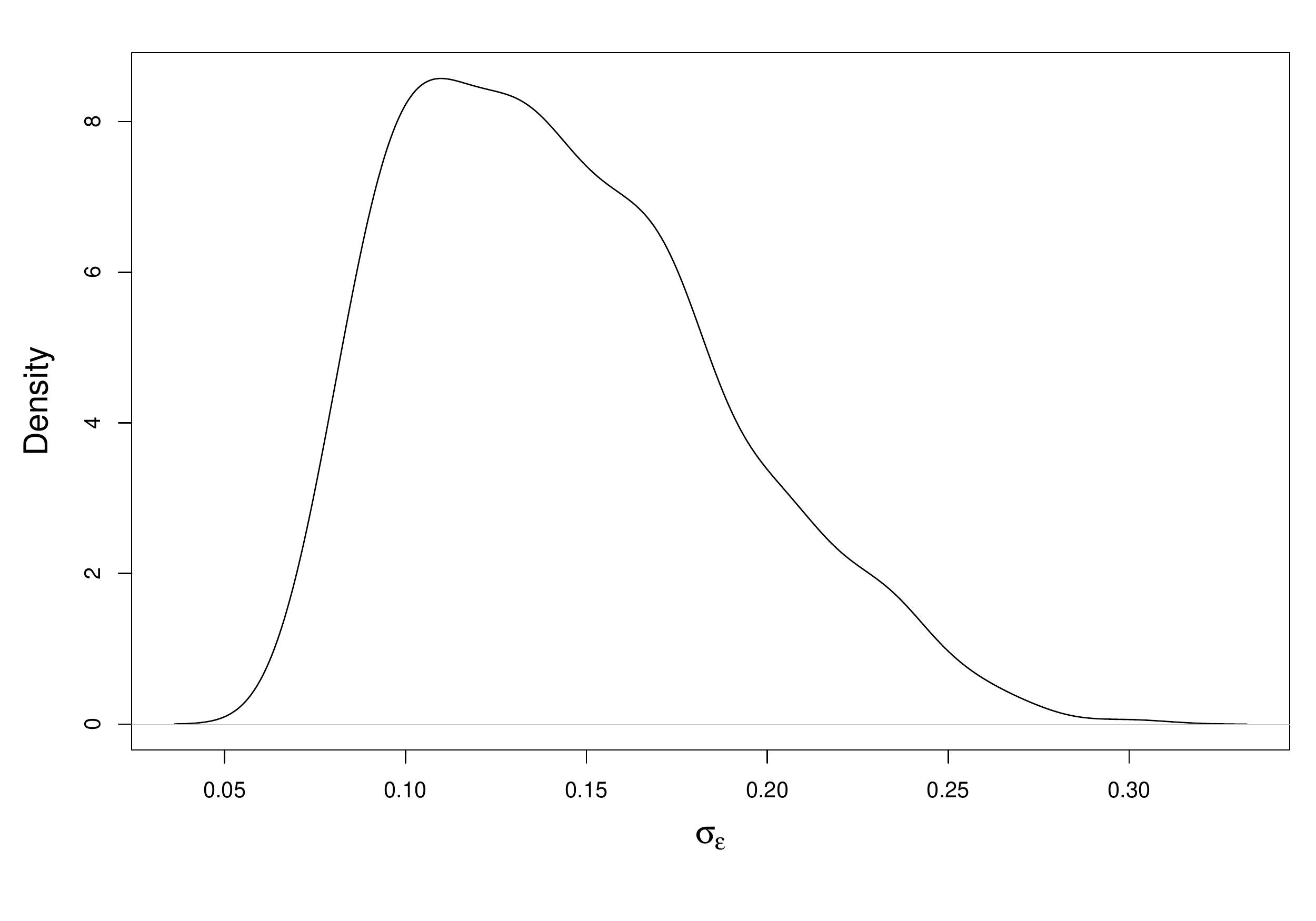}
\caption{Estimated posterior marginal distribution of the standard deviation $\sigma_\varepsilon$ of the random effects $\epsilon_{G(s)}$  for the Italian SuperLega 2017/2018 data.}
\label{fig2}
\end{figure}

To complete the model formulation, we include a hierarchical step to assume exchangability of the game random effects by 
$$
\varepsilon_{g} \sim \mathcal{N}( 0, \sigma_\varepsilon^2 ),
$$
and a hyper-prior for the variance of the random effects 
$$
\sigma_\varepsilon^2 \sim \mathsf{InvGamma}( a_\varepsilon, b_\varepsilon ),
$$
with fixed hyperparameters $a_{\epsilon}, b_{\epsilon}$.
Small posterior values of $\sigma_\varepsilon$ indicate that there is no need for such game effects, while large value indicates the need of unconnected (fixed) game effects (and possibly bad fit of the model without any game effects). Figure \ref{fig2} displays the posterior marginal distribution for $\sigma_\varepsilon$ for the Italian SuperLega 2017/2018 data: there is little evidence of any set effect here, as will be further investigated in Section \ref{sec:basicdic}.

\subsection{Zero inflated Poisson (ZIP) for the extra points}
\label{sec:zip}

To allow for the extra points arising due to the 24-deuce (or 14-deuce), the model proposed in Section \ref{sec:negbin} is extended by specifying a zero-inflated Poisson (ZIP) latent variable for the extra points collected by the loosing-set team. The number of extra points is zero if the loosing-set team does not reach 24 points, and 
greater than zero otherwise. 
So the model for the random variable of the points collected by the loosing team is now defined as:
\begin{align}
\begin{split}
Z_s &= Y_s + O_s  \\ 
Y_s &\sim  \mathsf{NegBin}(r_s, p_s){\mathcal I}( Y_s \le r_s-2 ) \\ 
O_s &\sim \mathsf{ZIPoisson}( \pi_{s}, \lambda ). 
\end{split}
\end{align}
The zero inflated Poisson (ZIP) distribution for the number of extra points $O_s$ collected by the team loosing the $s$-th set is then defined as:
\begin{equation}
f_{ZID}(o_s)= \pi_{s}{\mathcal I}(o_s=0)+(1-\pi_{s})f_P(o_s; \lambda),
\label{eq:zip}
\end{equation}
where $\pi_s$ describes the proportion of  zeros and $f_P(x; \lambda)$ is the probability mass function of a  Poisson distribution with rate parameter $\lambda$ evaluated at $x$.  
In this section, we assume constant inflation probability for all games, but in Section \ref{cov_zip} we explore the possibility of expressing $\pi_s$ as a function of the team abilities. 

Following \eqref{expected_values_TRNB}, under this model the random variables of the points are now given by
$$
Z_s^A=Y_s^A+O_s=W_s r_s + (1-W_s) Y_s+O_s \mbox{~and~} Z_s^B=Y_s^B+O_s=W_s Y_s + (1-W_s) r_s+O_s
$$
for the home and the away team, respectively. 
Now the expected points are adjusted for the extra points, hence they are given by 
$$
E(Z^A_s) = E(Y^A_s) + (1-\pi_s) \lambda \mbox{~and~} 
E(Z^B_s) = E(Y^B_s) + (1-\pi_s) \lambda, 
$$ 
where $E(Y^A_s)$ and $E(Y^B_s)$ are given in \eqref{expected_values_TRNB} and represent the expected number of points under the truncated model of Section \ref{sec:negbin} which does not considers any extra points for each set.  
\color{black}

\subsection{Model Comparisons for the Basic Model Formulation Using DIC}
\label{sec:basicdic}

Table \ref{tab:01} reports the DIC \citep{spiegelhalter2002bayesian} values and the effective number of parameters on the Italian SuperLega 2017/2018 for a simple Poisson model and the basic models presented in Sections \ref{sec:negbin}--\ref{sec:zip}, computed by running 3000 iterations obtained by 3 parallel chains of 1000 iterations of Gibbs sampling via the {\tt R} package  {\tt rjags} \citep{rjags}. In the Poisson model, the rates have a log-linear specification depending on the point abilities. Models 1 and 2 use unrestricted data (with no explicit modelling of the ties) and both report higher DIC than the truncated negative binomial model with extra points (model 3). 
As far as we can conclude from the DIC, using random effects to capture within game correlation (model 4) improves the fit only slightly (DIC=4537.2 vs. 4537.7); see also the posterior marginal distribution of $\sigma^2_\varepsilon$ in Figure \ref{fig2} and the considerations in Section \ref{sec:random_effects}. 
So we recommend to use the truncated negative binomial model allowing for extra points (model 3 in Table \ref{tab:01}; see Section \ref{sec:zip} for details) since it has similar predictive accuracy (in terms of DIC) to the corresponding random effects model (model 4 in Table \ref{tab:01}), while the computational burden and its model model complexity is considerably lower.

\begin{table}
\caption{Details of the fitted models with different distributional assumptions for the points of the loosing team for the Italian SuperLega 2017/2018 season.\label{tab:01}}

\begin{tabular}{|p{3cm} p{2.5cm} p{3.8cm} c c@{~~}|}
  \hline
\emph{Point Distribution of the loosing team$^*$} & \emph{Equations}& \emph{Additional model details at the point level} & \emph{\# eff. param.} & \emph{DIC}\\ 
\hline
1. Poisson           & log-linear model$^\dag$ & No upper limit                   & 29  & 4557.2 \\
2. Tr. Neg. binomial & 3--6 & Upper limit, no extra points                     & 29   & 4674.2 \\
3. ZIP Tr. Neg. bin. & 4--6 \& 10--11 & Upper limit \& extra points         & 133 & 4537.7   \\
4. ZIP Tr. Neg. bin. & 4--5, 10--11 \& Section 3.2     & Model 3 \& game random effects& 151 &  4537.2  \\
\hline  
\multicolumn{5}{p{14cm}}{\footnotesize\it MCMC sampling, 3000 iterations, {\tt rjags} package} \\ 
\multicolumn{5}{p{14cm}}{\footnotesize\it $^*$In all models we use:  (a) a logistic regression model for the sets (Eq. 1--2);} \\ 
\multicolumn{5}{p{14cm}}{\footnotesize\it \hspace{2em} (b) Overall disconnected team abilities $\alpha_T$ and $\beta_T$ for the set and the point level.} \\ 
\multicolumn{5}{p{14cm}}{\footnotesize\it $^\dag Y_s|W_s \sim Poisson(\lambda_s)$ with $\log (\lambda_s/r_s)=\eta_s$ where $r_s$ and $\eta_s$ are given by \eqref{total_points} and \eqref{point_prob}, respectively.} \\ 
\end{tabular}
\end{table}

\section{Model extensions concerning team abilities}
\label{sec:extension}

\subsection{Attacking and defensive abilities}
\label{sec:attdef}
A common practice is many team sports (such as football, basketball and hockey) is to separately model the attacking and the defensive team abilities. This is also relevant for coaches and sports scientists
because modern sports are highly specialized: estimates of attack and defence abilities
give an indication of athlete/team performance. 
Following this practice also in our proposed model for volleyball,  we can assume the following decomposition of the point abilities of team $T, \ T=1,\ldots,N_T$:
\begin{align}
\begin{split}
\beta_{T}=&\ \beta^{\text{att}}_{T} + \beta^{\text{def}}_{T},\\ 
\end{split}
\end{align}
where the global point abilities $\beta$ are defined as the sum between the attack and the defence abilities at point level for each team. It is worth noting that assuming different attacking and defensive abilities at the set level would make the logit model~\eqref{eq_pset} not identifiable.

\color{black}
\subsection{Connecting the abilities}
\label{sec:connecting}

In equations~\eqref{eq_pset} and~\eqref{pointeta}, set and point abilities  influence separately the set and  point probabilities, respectively: conditionally on winning/loosing a set, point abilities are then estimated from the probability to realize a point. However, we could combine them by defining a global ability measure. 
Here we consider a model where the abilities of winning a point also influence the probability of winning a set by a different scaling factor (controlled by the parameter $\theta$). 
Hence the probability of winning a set is now given by 
\begin{align}
\begin{split}
\mbox{logit}(\omega_s) = &
H^{set}+ v_1(\alpha_{A(s)}-\alpha_{B(s)}) + v_2\theta (  \beta_{A(s)}-\beta_{B(s)}   ),
\end{split}
\label{eq_p3}
\end{align} 
where $v_1, v_2$ are indicator variables, and $\theta$ summarizes the effect of the point abilities on winning a set. If $v_1=1, v_2=0$ we obtain the basic model of Section \ref{sec:negbin} with set probability as defined by Eq. \eqref{eq_pset}; if $v_1=0, v_2=1$ we assume connected point and set abilities where the set ability parameters are simply proportional to point abilities; whereas if $v_1 = v_2=1$ we assume connected point and set abilities and extra set specific abilities. 
For illustration purposes only, just view everything from the perspective of team $A$. 
Let us now consider the model with connected abilities and extra set abilities ($\nu_1=\nu_2=1$). 
If two teams are almost equally strong in terms of points, then the point abilities 
difference $\beta_{A(s)}-\beta_{B(s)}$ will be very small, 
and the  set probability will be solely driven by the extra set abilities. 
Conversely, when two teams are expected to be quite far in terms of point performance, then the set winning probability will be mainly affected by the point performance.

In this generalised version of the model (case $v_1 = v_2=1$), the set abilities will capture diversions of teams in the set efficiency in comparison to the point efficiency. 
For most of the teams, intuitively we do not expect an excess of set abilities and the probability of winning set will be mainly driven by a unified (set and point) ability. 
But a limited number of teams is expected to be more or less efficient on the set level than on the point level. Therefore, we have used posterior intervals and DIC to identify which teams behave in a different way in terms of sets and therefore an extra parameter is needed to handle for these differences. 

In Table \ref{tab:02} the DIC values and the effective number of parameters for each model are reported with respect to the Italian SuperLega 2017/2018. 
According to this analysis the ZIP truncated negative binomial model  with connected abilities and extra set abilities only for   Verona and Padova and constant zero inflated probability is the best fitted model.

\color{black}
\begin{center}
\begin{table}
\caption{Details of the fitted Logistic--ZIP Truncated Negative binomial models for the Italian SuperLega 2017/2018 season.\label{tab:02}}

\begin{tabular}{|p{1cm}p{2cm}p{6cm}cc|}
  \hline
\emph{Model$^*$} & \emph{Connected team abilities} & \emph{Additional  model features} & \emph{\# eff. param.} & \emph{DIC}\\ 
\hline
3.   & No & --- & 133 & 4537.7   \\
\hline
5.  & No & Separate attacking and defensive abilities at the point level & 147  &  4541.1 \\
6.  & Yes & Connected abilities only ($v_1=0, v_2=1$)                    & 122  & 4524.3  \\
7.  & Yes & $+$ extra set abilities ($v_1=v_2=1$)                        & 133  & 4536.3  \\
8.  & Yes & $+$ extra set abilities for Verona                           & 123  & 4522.2 \\
9.  & Yes & $+$ extra set abilities for Verona and Padova                & 124 & 4521.1  \\
10. & No & $ \beta_{t}$ dynamic (point level)                            & 174  & 4569.5   \\
11. & No & $\alpha_{t}$ dynamic (set level)                              & 132  & 4530.1 \\
\hline  
\multicolumn{5}{p{13cm}}{\footnotesize\it MCMC sampling, 3000 iterations, {\tt rjags} package} \\ 
\multicolumn{5}{p{13cm}}{\footnotesize\it $^*$In all models we use:  (a) a logistic regression model for the sets (Eq. 1--2);} \\ 
\multicolumn{5}{p{13cm}}{\footnotesize\it (b) The Zero-inflated Poisson for extra points and the Truncated Negative binomial model was used for the points (ZIP Tr. Neg. bin.).} \\ 
\multicolumn{5}{p{13cm}}{\footnotesize\it (c) Constant probability and Poisson rate for the ZIP component of the extra points is assumed.} \\ 
\end{tabular}
\end{table}
\end{center}

\subsection{Dynamic abilities}

\begin{figure}[b!]
	\centering
	\includegraphics[scale=0.65]{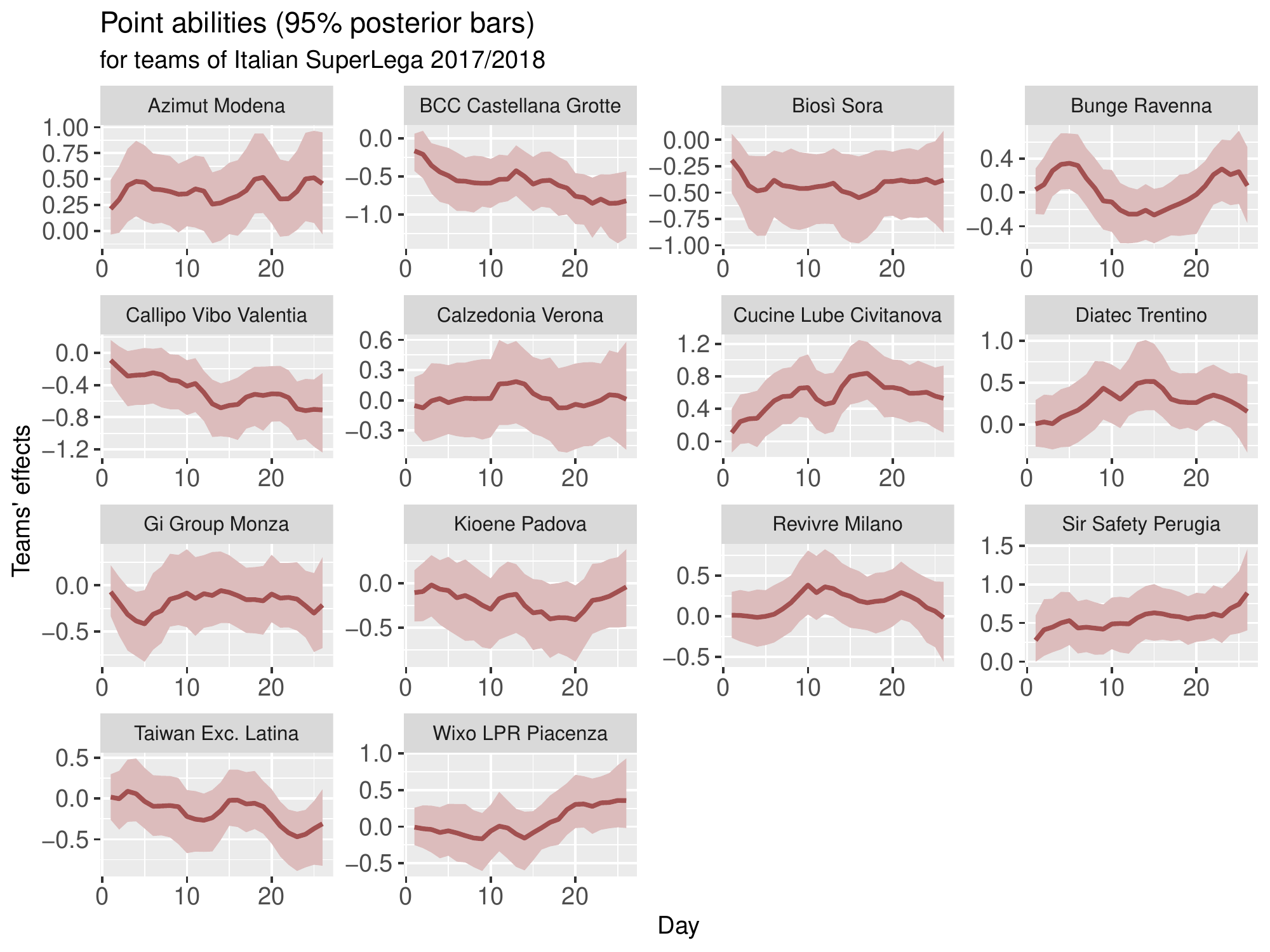}
	\caption{Posterior medians and 95\% density intervals for the dynamic point abilities parameters $\bm{\beta}$ for the  Italian SuperLega 2017/2018.}
	\label{fig3}
\end{figure}

\begin{figure}
\centering
\includegraphics[scale=0.3]{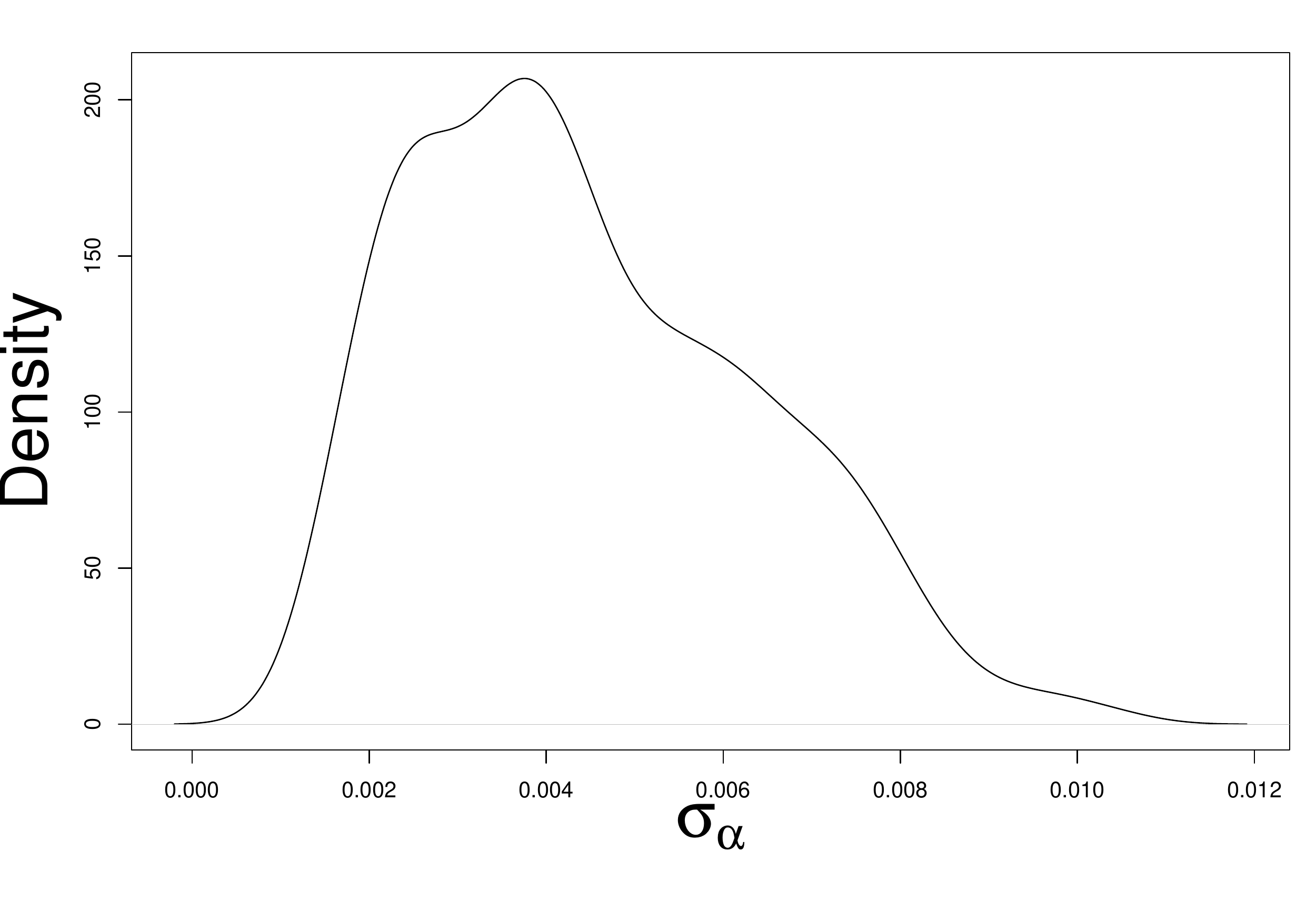}~
\includegraphics[scale=0.3]{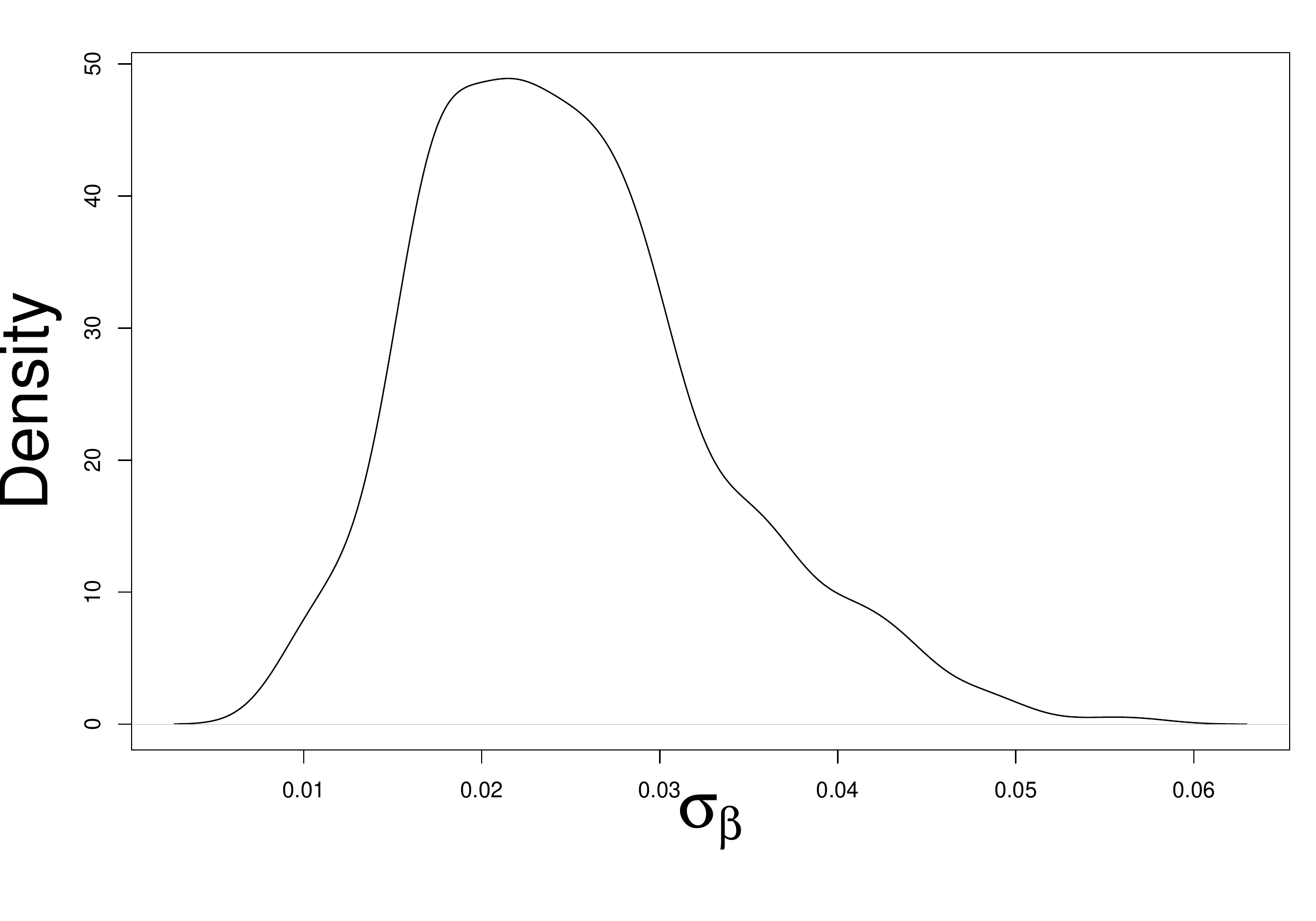}
\caption{Posterior marginal distribution for the standard deviations of the dynamic set and point abilities for the  Italian SuperLega 2017/2018;   for sets: $\sigma_\alpha$ (left plot) and for points: $\sigma_{\beta}$  (right plot).}
\label{fig4}
\end{figure}

The performance of each team is likely to change within a season. Hence, temporal trends may be helpful for modelling the ability of each team within a season. A dynamic structural assumption for the ability parameters is a step forward. A natural choice is an auto-regressive model for the set and point abilities. For each team $T=1,\ldots, N_T$ and game $G=2,\ldots,N_G$ we specify:
\begin{align}
\begin{split}
\alpha_{T,G}  \sim & \mathcal{N}(\alpha_{T, G-1}, \sigma^2_{\alpha})\\
\beta_{T,G}  \sim & \mathcal{N}(\beta_{T, G-1}, \sigma^2_{\beta}),
\end{split}
\end{align}
whereas for the first match we assume:
\begin{align}
\begin{split}
\alpha_{T,1}  \sim & \mathcal{N}(0, \sigma^2_{\alpha})\\
\beta_{T,1}  \sim & \mathcal{N}(0, \sigma^2_{\beta}).
\end{split}
\label{dynamic_ab}
\end{align}
Analogously as in Section \ref{sec:negbin}, sum-to-zero constraints are required for each match-day to achieve identifiability. The variance parameters  $\sigma^2_{\beta}, \sigma^2_{\beta}$ are assigned the following hyper-priors:
\begin{equation}
\sigma^{2}_{\alpha}, \sigma_\beta^2 \sim \mathsf{InvGamma}(0.001, 0.001 ). 
\label{sigma_beta}
\end{equation}
As it is evident from Table \ref{tab:02}, the assumption of dynamic ability parameters does not improve the fit of the model for the Italian SuperLega data we consider here. However, modelling dynamic patterns may be very useful in other leagues when considering distinct subsets of a league (such as regular season and play off). Figure \ref{fig3} displays posterior 95\% intervals for the dynamic point abilities for the Italian SuperLega 2017/18 data, whereas the corresponding marginal posterior distributions for the standard deviations $\sigma_{\alpha}, \sigma_\beta$ are plotted in Figure \ref{fig4}: the time variability is negligible in the Italian data we analyse in this paper. 
	Although the within-team variability of the point abilities may look high, 
	we believe that it is reasonable, since it corresponds only to a small portion of the total variability of the response measurement of this model's component (which is the logit of the proportion of points earned by the loosing team after removing the extra points played due to ties). 
	To confirm this, we have calculated the proportion of points won by the loosing team (after removing extra points played due to ties) which, on average, was found to be equal to $0.795 \pm 0.12$. 
	The corresponding logits of these proportions were found equal to $1.5$ on average with 
	$0.742$ standard deviation. 
	According to our posterior results, the posterior standard deviations of the point parameters were found to be approximately $0.10$--$0.15$ which corresponds only to 12--20\% of the total variability of the response measurement (as we stated previously).

\subsection{Modelling extra points as a function of team abilities}
\label{cov_zip} 

In this section we explore whether the probability of observing extra points due to ties (i.e. the inflation component probability) can be written as a function of the set and/or point abilities. 
In principle, we would expect a negative association between the team ability differences and the probability of playing extra points in each set. 
Therefore, the closer the two teams are, the higher is the probability of being tied in each set and as a result to play for extra points. 
For the probability $\pi_s$ we considered various versions of the linear predictor which can be summarized by:
\begin{align}
\begin{split}
\mbox{logit}(\pi_{s})= &  m +  \delta \, \varPhi\mbox{\footnotesize $\big(\alpha_{A(s)}- \alpha_{B(s)}\big)$} +  \gamma \, \varPhi\mbox{\footnotesize $\big(\beta_{A(s)}- \beta_{B(s)}\big)$}\\
&m \sim   \mathcal{N}(0, 1); ~\delta, \gamma  \sim \mathcal{N}(0, 10^6); ~\lambda \sim \mathsf{LN}(0,1),
\end{split}
\label{eq:Zip_priors}
\end{align}
where $\mathsf{LN}(\mu, \sigma^2)$ denotes the log-normal distribution with parameters $\mu$ and $\sigma^2$, $m$ is a constant parameter, $\delta, \gamma$ are the coefficients associated to the set and point abilities differences, $\varPhi(\cdot)$ is a specific function for the set/point abilities differences, depending on the model that we consider: 
\begin{itemize}
	\item constant probability by using the null function: $\varPhi(x) = 0$ , 
	\item linear effect of ability differences by using the linear function: $\varPhi(x) = c_1x$, 
	\item linear effect of absolute ability differences by using the linear absolute function: $\varPhi(x) = c_1|x|$, 
	\item quadratic effect of ability differences by using the quadratic function: $\varPhi(x) = c_1 x + c_2x^2$,
	\item quadratic effect of absolute ability differences by using the quadratic function of absolute values of $x$ : $\varPhi(x) = c_1 |x| + c_2x^2$,
\end{itemize} 	
where $c_0, c_1$ and $c_2$ are further parameters for estimation.
In Table \ref{tab:02bis} the DIC values and the effective number of parameters for each model are reported with respect to the Italian SuperLega 2017/2018. No model improves the fit we obtain when using the constant Poisson rate for the ZIP component of the extra points (model 9).

\begin{center}
\begin{table}
\caption{Details of the fitted Logistic--ZIP Truncated Negative binomial models with different structure on the probability of extra points for the Italian SuperLega 2017/2018 season.\label{tab:02bis}}

\begin{tabular}{|lp{5cm}lcc|}
  \hline
\emph{Model$^*$} &  \emph{Structural assumption about the probability of extra points} & $\varPhi(x)$&\emph{\# eff. param.} & \emph{DIC}\\ 
\hline
\color{blue}
9.  &  Constant probability      & $\varPhi(x)=0$                 & 124 & 4521.1  \\
\color{black}
12. & Linear effects             & $\varPhi(x)=c_1x$         & 125 & 4523.0 \\
13. & Linear absolute effects    & $\varPhi(x)=c_1|x|$       & 125 & 4525.8 \\
14. & Quadratic effects          & $\varPhi(x)=c_1x+c_2x^2$  & 126 & 4524.6  \\
15. & Quadratic absolute effects & $\varPhi(x)=c_1|x|+c_2x^2$& 126 & 4526.6 \\
\hline  
\multicolumn{4}{p{13cm}}{\footnotesize\it MCMC sampling, 3000 iterations, {\tt rjags} package} \\ 
\multicolumn{4}{p{13cm}}{\footnotesize\it $^*$In all models we use:  (a) a logistic regression model for the sets (Eq. 1--2);} \\ 
\multicolumn{4}{p{13cm}}{\footnotesize\it (b) The Zero-inflated Poisson for extra points and the Truncated Negative binomial model for points (ZIP Tr. Neg. bin.).} \\ 
\multicolumn{4}{p{13cm}}{\footnotesize\it (c) Connected team abilities and extra set abilities for Verona and Padova.} \\ 
\end{tabular}
\end{table}
\end{center}

\color{black} 
 \section{Analysis and Results of the Italian SuperLega 2017/2018}
\label{sec:est}

\subsection{Data and computational details}

Data come from the regular season of the Italian SuperLega 2017/2018 and consist of a seasonal sample of 680 set observations, for a total number of 182 matches and 14 involved teams.\footnote{Source: Webpage of the Italian SuperLega \url{https://www.legavolley.it/category/superlega/}} 
Posterior estimates are obtained with the {\tt rjags} {\tt R} package (MCMC sampling from the posterior distribution using the Gibbs sampling), for a total of 3000 iterations obtained by 3 parallel chans of 1000 iterations and a burn-in period of 100. 
Following the suggestions of \cite{gelman2013bayesian}, 
we monitored the convergence of our MCMC algorithms by checking the effective sample size of each chain parameter and by implementing the Gelman-Rubin statistic \citep{gelman1992inference}  which resulted to be lower than the usual threshold 1.1 for all the parameters (details given in Table 5).

In 39 matches out of 182 (21.4\%) the final winner was determined in the tie break (i.e. in the fifth set), whereas in 101 out of 680 (14.8\%) sets it was required to play for extra points in order to declare the winner of the corresponding set.

\subsection{Interpretation of the selected model} 
\label{sec:final}

Here we focus on the analysis of the Italian SupeLega 2017/18 data using the  model selected in Sections \ref{sec:basicdic} and \ref{sec:connecting} (model 9 in Table \ref{tab:02}), that is the model with connected abilities for all teams and extra set abilities only for Verona and Padova. 
The complete model formulation, including likelihood specification, priors and identifiability constraints, is summarized in Table \ref{tab:04}. Posterior estimates for the set home advantage $H^{set}$, the point home advantage $H^{points}$, the  intercept $\mu$ and the ZIP parameters $\lambda, m$ are reported in Table \ref{tab:03}: there is a clear indication of home advantage which seems to be smaller for the set level (posterior median of 0.16, 95\% posterior interval marginally containing zero), and higher at the point level (posterior median of 0.20, 95\% posterior interval not containing the zero). 
In terms of percentage change, this means that in a game between two teams of equal strength we expect that the home team will have $17\%$ (posterior 95\% interval: (0\%, 42\%)) and $22\%$ (posterior 95\% interval: (8\%, 40\%)) higher odds of winning a set and a point, respectively. 

The scaling factor $\theta$ (posterior mean of 4.60) shows a very strong positive association between the point abilities and the probability to win the set, as assumed in Eq.~\eqref{eq_p3}. No evidence was found for the parameters $ \gamma$ and $\delta$ in Eq.~\eqref{eq:Zip_priors}, describing the influence of set and point abilities differences, respectively, on the probability of observing extra points; however, we feel these parameters could be beneficial for other datasets or other leagues.

Number of effective sample sizes ($n\_eff$) and Gelman-Rubin statistics ($\hat{R}$) appear to be quite satisfactory for each parameter.

\begin{table} 
	\caption{Posterior summaries for the set home $H^{set}$, the point home $H^{points}$, the connecting abilities scaling factor $\theta$, the  intercept $\mu$, and ZIP parameters  $\lambda, \ m$ for the Italian SuperLega 2017/18 using the ZIP truncated negative binomial model with connected abilities and extra set abilities for Verona and Padova (model 9 in Table \ref{tab:02}). Also reported: the effective sample size ($n\_eff$) and the Gelman-Rubin statistics ($\hat{R}$).  \label{tab:03}}
	\begin{tabular}{|lcccccccc|}
		\hline
		Description& Parameter & Mean & Median & sd & 2.5\% & 97.5\% & $n\_eff$ & $\hat{R}$\\ 
		\hline
		Set home advantage&$H^{set} $& 0.16  & 0.15 & 0.09 &-0.01&  0.33 & 2839 & 1 \\ 
		Point home advantage & $H^{points}$ & 0.20 &  0.20& 0.06&  0.08&  0.34 & 1664 & 1 \\
		Connecting abilities & $\theta$ & 4.60 &  4.52 &0.80 & 3.36 & 6.30 & 2310 & 1\\ 
		Intercept&  $\mu$ & 0.36  & 0.36 & 0.05 &  0.27 & 0.46& 2213 & 1  \\ 
		ZIP Poisson rate&  $\lambda$ & 3.97 & 3.97 & 1.07 & 3.45 & 4.52& 2115 & 1 \\ 
		Tie probability intercept & $m$ & 2.12 &  2.13& 0.13&  1.87&  2.39 & 2460 &1 \\ 
		\hline
	\end{tabular}
\end{table}

\begin{small}
\begin{table}
\caption{Final model formulation for model 9 of Table \ref{tab:02}: likelihood, priors and identifiability constraints (24 parameters in total); STZ: Sum-to-zero constraints. \label{tab:04}}
\colorbox{gray!25}{\parbox{0.9\textwidth}{
\begin{eqnarray*}
\begin{split}
\underline{\bm{Likelihood}} & \\
  \mbox{\textcolor{blue}{Total set points}} &\ \ \ \ Z_s = Y_s + O_s  \\ 
\mbox{\textcolor{blue}{Loosing team points}} &\ \ \ \ Y_s | W_s  \sim \mathsf{NegBin}(r_s, p_s){\mathcal I}( Y_s < r_s-2 ) \\
\mbox{\textcolor{blue}{Extra points}} &\ \  \ \ O_s  \sim  \mathsf{ZIPoisson}( \pi_{s}, \lambda )\\
\mbox{\textcolor{blue}{Home win indicator}} &\ \  \ \ W_s \sim \mathsf{Bernoulli}( \omega_s ) \\
\mbox{\textcolor{blue}{Logit of set win}} &\ \ \ \ \mbox{logit}(\omega_s) = 
H^{set}+ v_1(\alpha_{A(s)}-\alpha_{B(s)}) + v_2\theta (  \beta_{A(s)}-\beta_{B(s)}   ) \\
\mbox{\textcolor{blue}{ZIP: Log-odds of extra points}} &\ \ \ \ \mbox{logit}(\pi_{s})=   m + \delta \, \varPhi(\alpha_{A(s)}- \alpha_{B(s)})+ \gamma \, \varPhi(\beta_{A(s)}- \beta_{B(s)} )\\
\mbox{\textcolor{blue}{Linear predictor for Points}} &\ \ \ \ \eta_s =  \mu+ (1-W_s)H^{points}+(\beta_{A(s)}-\beta_{B(s)})(1-2W_s) \\ 
\mbox{\textcolor{blue}{Win. team point prob.}}&\ \ \ \ \ p_s = \frac{1}{1+e^{\eta_s}} \\
\mbox{\textcolor{blue}{Required success points}}&\ \ \ \ \ r_s =  25 - 10 \times \mathcal{I} \left( R_s=5 \right)\\[1.0em]
\underline{\bm{Constraints}} & \\[1em]
\mbox{\textcolor{blue}{Extra Set Abilities}} &\ \ \ \ \alpha_T =\alpha_T^*  
	\mbox{~with~} \alpha_T^*  \equiv 0, \ T\ne 10,12 \\
\mbox{\textcolor{blue}{(Only for specific teams)}$^\ddag$} & \\ 
\mbox{\textcolor{blue}{STZ for Point Abilities}} &\ \ \ \ 
\beta_T =\beta_T^* - \overline{\beta}^*; \ \  \overline{\beta}^* = \frac{1}{N_T} \sum _{T=1}^{N_T}\beta_T^*\\
\mbox{\textcolor{blue}{Connecting Abilities Setup }} &\ \ \ \ v_1 = v_2 =1\\[1em]
\mbox{\textcolor{blue}{ZIP general probability function}} &\ \ \ \ \varPhi(x) = c_0x + c_1 |x| +c_2 x^2\\[1em]
\mbox{\textcolor{blue}{ZIP finally selected model}} &\ \ \ \ c_0=c_1 = c_2 =0\\[1em]
\underline{\bm{Priors}} & \\[-0.5em]
\mbox{\textcolor{blue}{Set Abilities (Padova \& Verona)}}
&\ \ \ \ \alpha^*_{10}, \alpha^*_{12}  \ \sim  \mathcal{N}(0,2^2) \\
\mbox{\textcolor{blue}{Point Abilities (Unconstrained)}}
&\ \ \ \ \beta_{T}^* \sim  \mathcal{N}(0, 2^2)\\
\mbox{\textcolor{blue}{Constant for Points}}
&\ \ \ \ \mu  \sim   \mathcal{N}(0,10^6)\\
\mbox{\textcolor{blue}{Point Ability Coef. for Sets}}
&\ \ \ \  \theta  \sim   \mathcal{N}(0,10^6)\\
\mbox{\textcolor{blue}{Home effects}}
&\ \ \ \ H^{point}, H^{set} \sim   \mathcal{N}(0,10^6)\\
\mbox{\textcolor{blue}{Constant for Extra Points}}
&\ \ \ \  m \sim  \mathcal{N}(0,1)\\
\mbox{\textcolor{blue}{Poisson rate for ZIP}}
&\ \ \ \  \lambda \sim  \mathsf{LN}(0,1)\\
\end{split}
\end{eqnarray*} }} \\[2em] 
\parbox{0.9\textwidth}{\footnotesize\it $^\ddag$This parametrization is used in the final model, where extra abilities are considered only for two specific teams; In the general formulation with all set abilities then the STZ parametrization is recommended where  $\alpha_T =\alpha_T^* - \overline{\alpha}^*$, $\overline{\alpha}^* = \frac{1}{N_T} \sum _{T=1}^{N_T}\alpha_T^*$ and $N_T$ is the number of teams under consideration. 
	Prior when all set abilities are used with STZ: $\alpha_T^* \sim \mathcal{N}(0,2^2)$.}  
\end{table}
\end{small}

\begin{figure}[t!]
	\centering
	\includegraphics[scale=0.6]{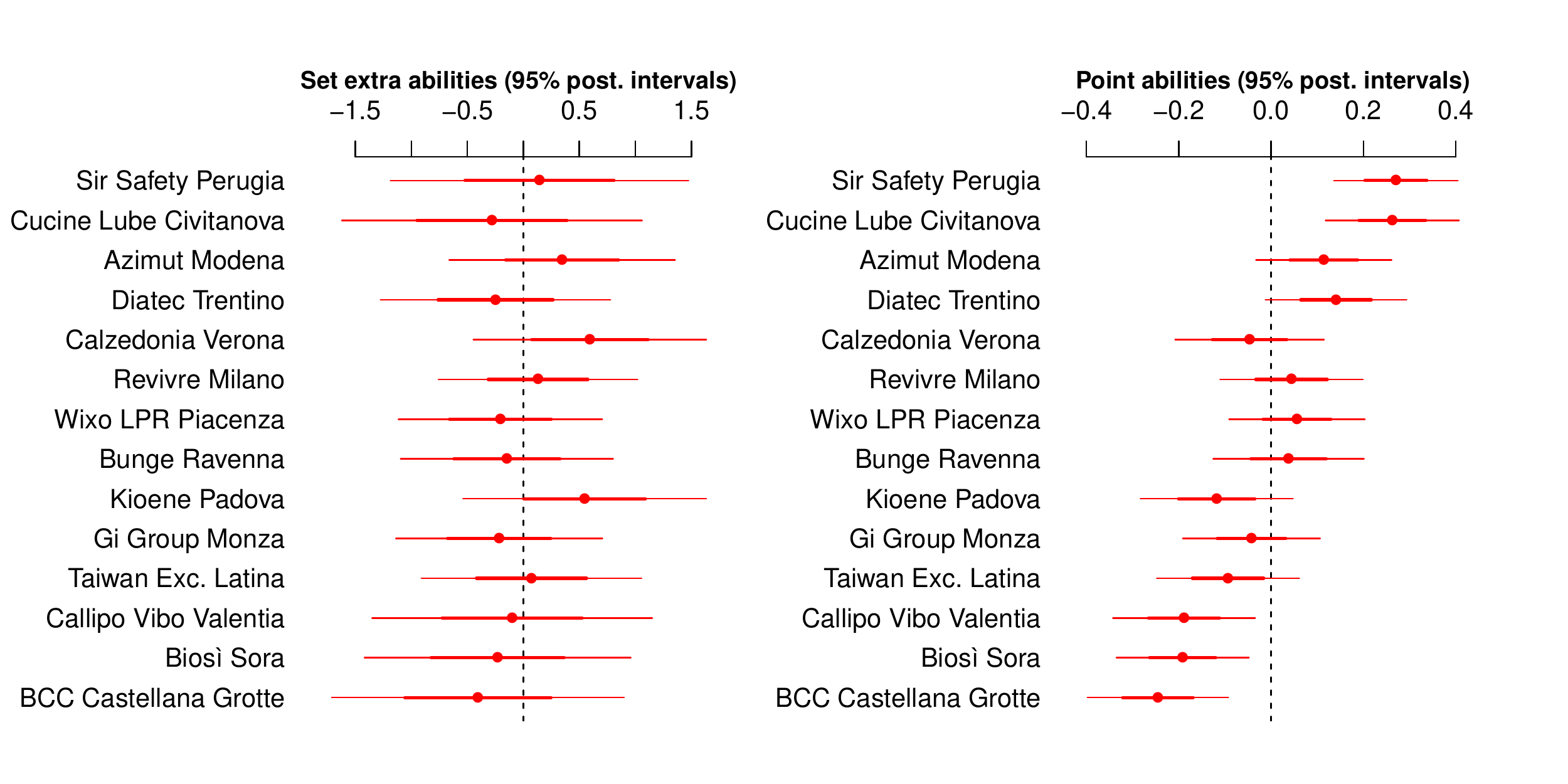}\\
	\caption{ 95\% posterior intervals for set and point team abilities for the Italian SuperLega 2017/18 using the ZIP truncated negative binomial model with connected abilities (model 7 in Table \ref{tab:02}); intervals are ordered by the actual final ranking of each team.}
	\label{fig5}
\end{figure}

\begin{figure}[b!]
	\centering
	\includegraphics[scale=0.6]{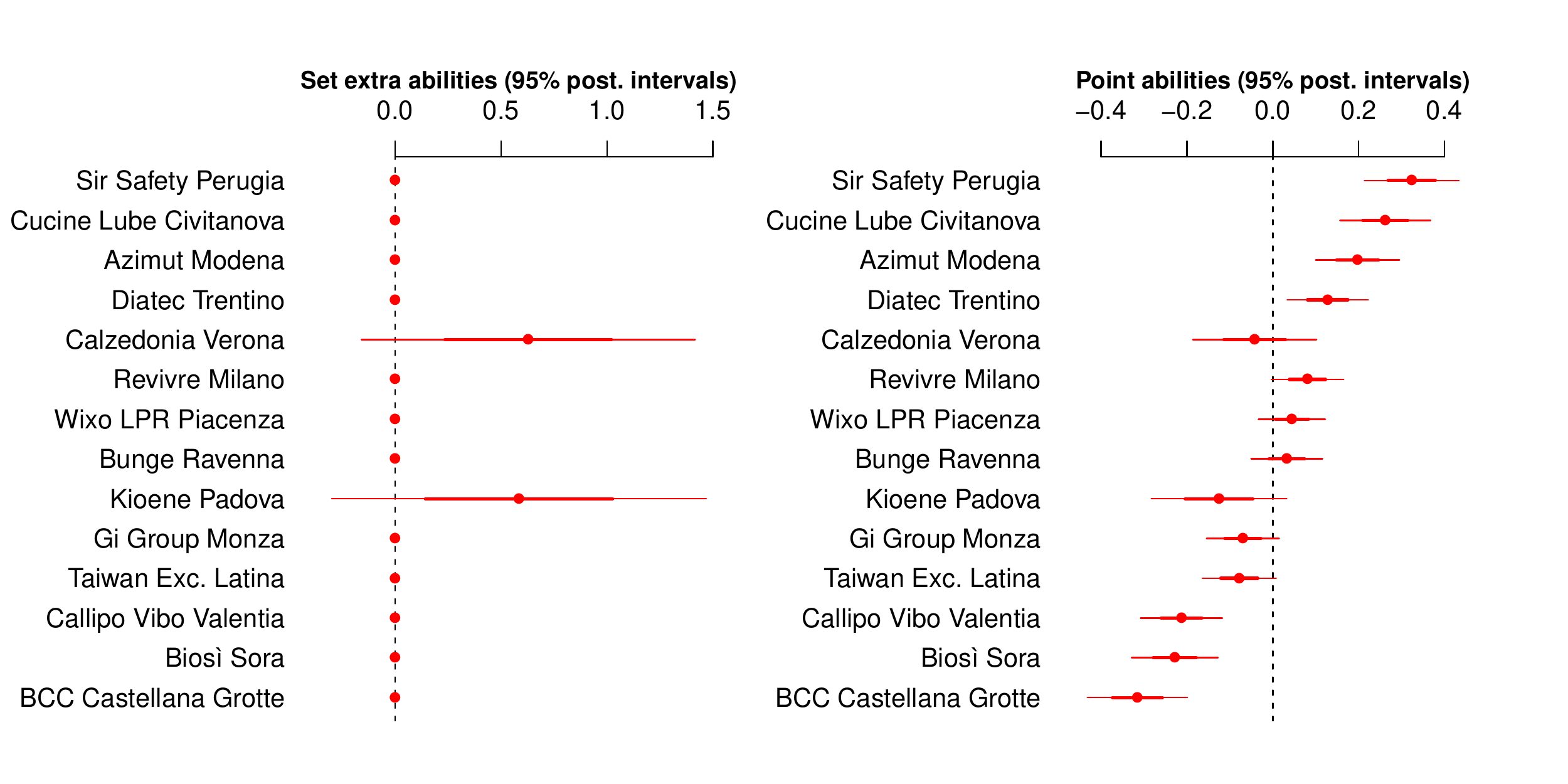}\\
	\caption{95\% posterior intervals for set and point team abilities for the the Italian SuperLega 2017/2018 using the ZIP truncated negative binomial model with connected abilities and extra set abilities for Verona and Padova (model 9 in Table \ref{tab:02}); ordered by the actual final ranking of each team.}
	\label{fig6}
\end{figure}

\begin{figure}
	\centering
	\includegraphics[scale=0.6]{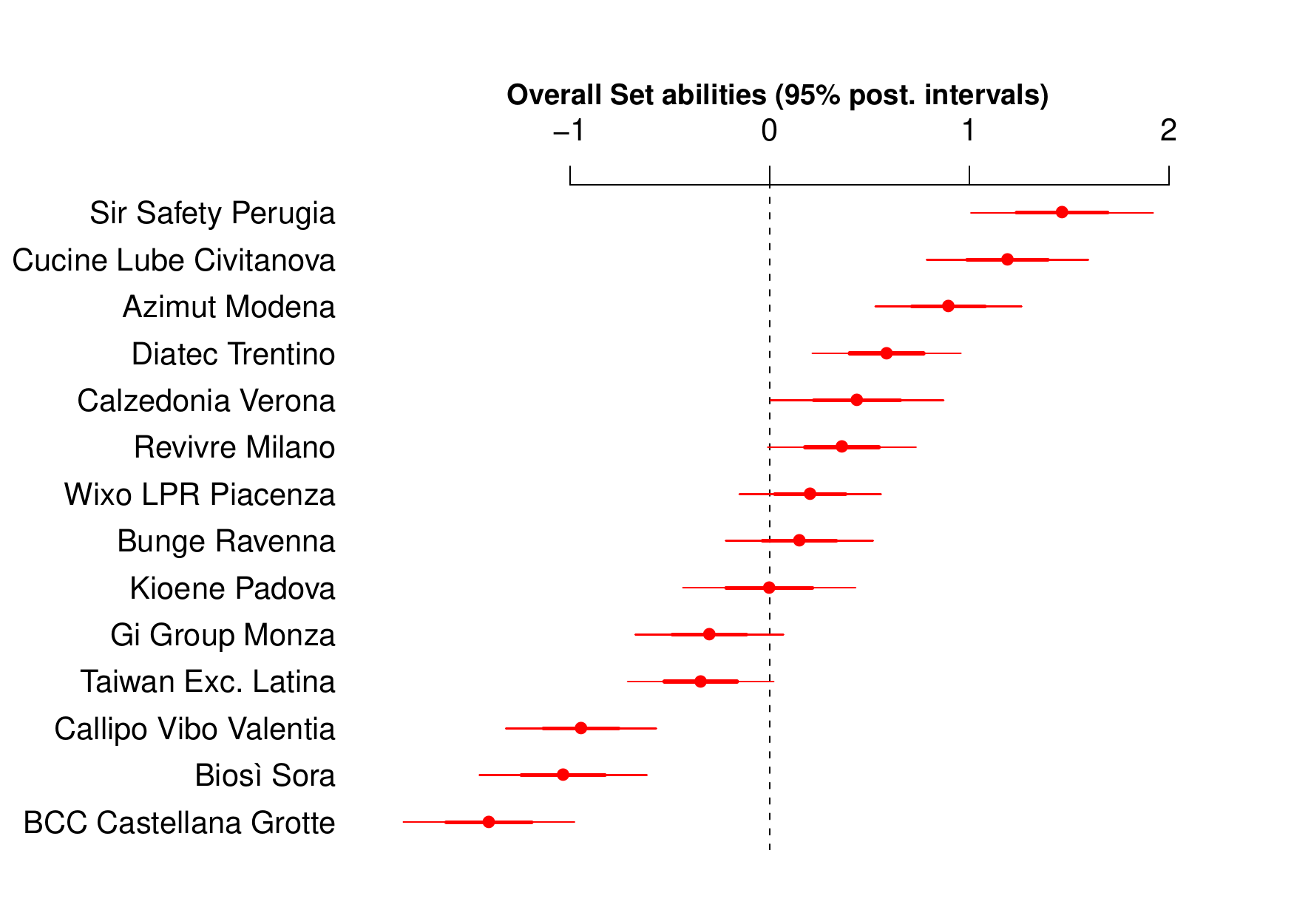}\\
	\caption{
		95\% posterior intervals for overall  set team abilities $\alpha'_T=\alpha_T+\theta \beta_T$ for the Italian SupeLega 2017/18 using the  ZIP truncated negative binomial model with connected abilities and extra set abilities only for Verona and Padova and constant ZIP probability for extra points (model 9 in Table \ref{tab:02}); Intervals are ordered by the actual final rank of each team.}
	\label{fig7}
\end{figure}

The 95\% posterior intervals for set and point team abilities are displayed in Figures \ref{fig5} and \ref{fig6} for the model with connected abilities and extra set abilities for all teams (model 7 in Table \ref{tab:02}) and the corresponding model with extra set abilities only for Verona and Padova (model 9 in Table \ref{tab:02}), respectively. 
From Figure \ref{fig5} (right plot) we can observe that the point abilities of Verona and Padova are slightly misaligned compared to the actual rank. Moreover, from the left plot we notice that all the 95\% posterior intervals of the extra set abilities contain the value of zero. However, for Verona and Padova we obtained a marginal effect in terms of set extra abilities. For this reason, we moved to the model with connected abilities and  set extra abilities only for these two teams (Figure \ref{fig6}), by forcing all the remaining extra set abilities to be restricted to zero. In such a way, we reduced the model complexity by 12 parameters while we improved our final model in terms of predictive accuracy (see Section \ref{sec:connecting}).
Finally, for this model we can specify the actual effects (abilities) of each team on the winning probability of a set as: 
	$$
	\alpha_T{'} = \alpha_T + \theta \beta_T.
	$$
These overall set abilities are depicted in Figure~\ref{fig7} where we can clearly see that the overall abilities depict the actual observed rankings since the extra set abilities for Padova and Verona correct for any inconsistencies between the set and point level concerning the efficiency of the teams. \\

\subsection{League reconstruction and predictive measures of fit}
\label{sec:pp}

To assess the in-sample predictive accuracy of our final model, we reconstruct the league in terms of final points and rank positions from the predictive distribution of the model. To do so, for each iteration of the MCMC sampling, we draw values from the model's sampling likelihood (see Table \ref{tab:04}) for the given set of parameter values generated at each each iteration resulting in a new sample of match results obtained from posterior predictive distribution of the model. 
It is worth mentioning that predicting future matches in volleyball is not as easy as in other sports. First, we need to simulate the actual number of sets for each game using Equations~\ref{eq_pset0} and~\ref{eq_pset}; we terminate the sets' simulation when one of the two teams wins three sets first.
Then, we  calculate the number of points the teams collected after winning a game at each reconstructed league of each iteration. 
For each MCMC iteration, each match is simulated in terms of both points and sets from their posterior predictive distribution. This means that a new replicated dataset ${y}^{rep}$ is sampled from:
\begin{equation}\label{eq_pp:distr}
p({y}^{rep}|y) = \int_{\Theta}p ({y}^{rep}|\theta) p(\theta|y) d\theta,
\end{equation}
where $p(\theta|y)$ is the posterior distribution for the parameter $\theta$, and $p({y}^{rep}|\theta)$ is the sampling distribution for the hypothetical values. In many 
applications, the posterior predictive distribution in Eq.~\ref{eq_pp:distr} is not 
available in a closed form, and therefore we sample from it by using MCMC methods. Algorithm 1 presents the entire procedure for the stochastic league reconstruction from the posterior predictive distribution: the entire league is simulated and the final points and rankings are obtained for further analysis. 

\begin{table}
	\caption{Final reconstructed league for the Italian SuperLega 2017/18 using the ZIP truncated negative binomial model with connected abilities and extra set abilities for Verona and Padova (model 9 in Table \ref{tab:02}).
		\label{tab:05}}
	\begin{tabular}{|clcc|}
		\hline
		\emph{Predicted} &       & \multicolumn{2}{c}{Points}\\  
		\cline{3-4} 
		\emph{Rank$^*$}      & Teams & \emph{Expected (Actual)} &  \emph{95\% CI}$^\dag$ \\ 
		\hline
		~1  (--)&  Sir Safety Perugia     & 66 (70) & (57--73)  \\ 
		~2  (--)&  Cucine Lube Civitanova & 61 (64) & (51--69)  \\ 
		~3  (--)&  Azimut Modena          & 56 (60) & (45--66)  \\ 
		~4  (--)&  Diatec Trentino        & 52 (51) & (40--63)  \\ 
		~5  (--)&  Calzedonia Verona      & 49 (50) & (35--60)  \\ 
		~6  (--)&  Revivre Milano         & 45 (44) & (32--57)  \\ 
		~7  (--)&  Wixo LPR Piacenza      & 42 (42) & (29--54)  \\ 
		~8  (--)&  Bunge Ravenna          & 40 (41) & (27--52)  \\ 
		~9  (--)&  Kioene Padova          & 35 (35) & (22--48)  \\ 
		10 (--)&   Gi Group Monza        & 28 (28) & (17--42)  \\ 
		11 (--)&   Taiwan Exc. Latina    & 28 (25) & (18--40)  \\ 
		\;\:12 (+1)&   Biosì Sora            & 17 (13) & (8--26)  \\ 
		\;13 (--1)&   Callipo Vibo Valentia & 16 (13) & (9--27)  \\ 
		14 (--)&   BCC Castellana Grotte & 13 (10) & (5--21)  \\
		\hline
		\multicolumn{4}{p{10cm}}{\footnotesize\it $^*$ Rank based on the expected predicted points; in brackets: the change in predictive ranking in comparison to the actual ranking} \\ 
		\multicolumn{4}{p{10cm}}{\footnotesize\it $^\dag$ 95\% credible interval based on the 2.5\% and 97.5\% percentiles of the posterior predictive distribution.} \\ 
	\end{tabular}
\end{table}

Table \ref{tab:05} reports the expected final points and the 95\% predictive intervals estimated from the MCMC algorithm along with the observed points and the actual team rankings: the points reported in the table are obtained by computing the median and the 95\% predictive intervals of the posterior predictive distribution of the final points, for each team. 
The agreement between the actual and the expected number of points is remarkable since the difference is at most equal to 4 points, and the simulated positions mirror perfectly the final observed rank, with the exception of switch in the expected positions between Sora and Vibo Valentia.  
Generally speaking, the model's in-sample predictions mirror almost perfectly the observed results in terms of expected points and final rank positions.

\begin{figure}
\centering
\includegraphics[scale=0.3]{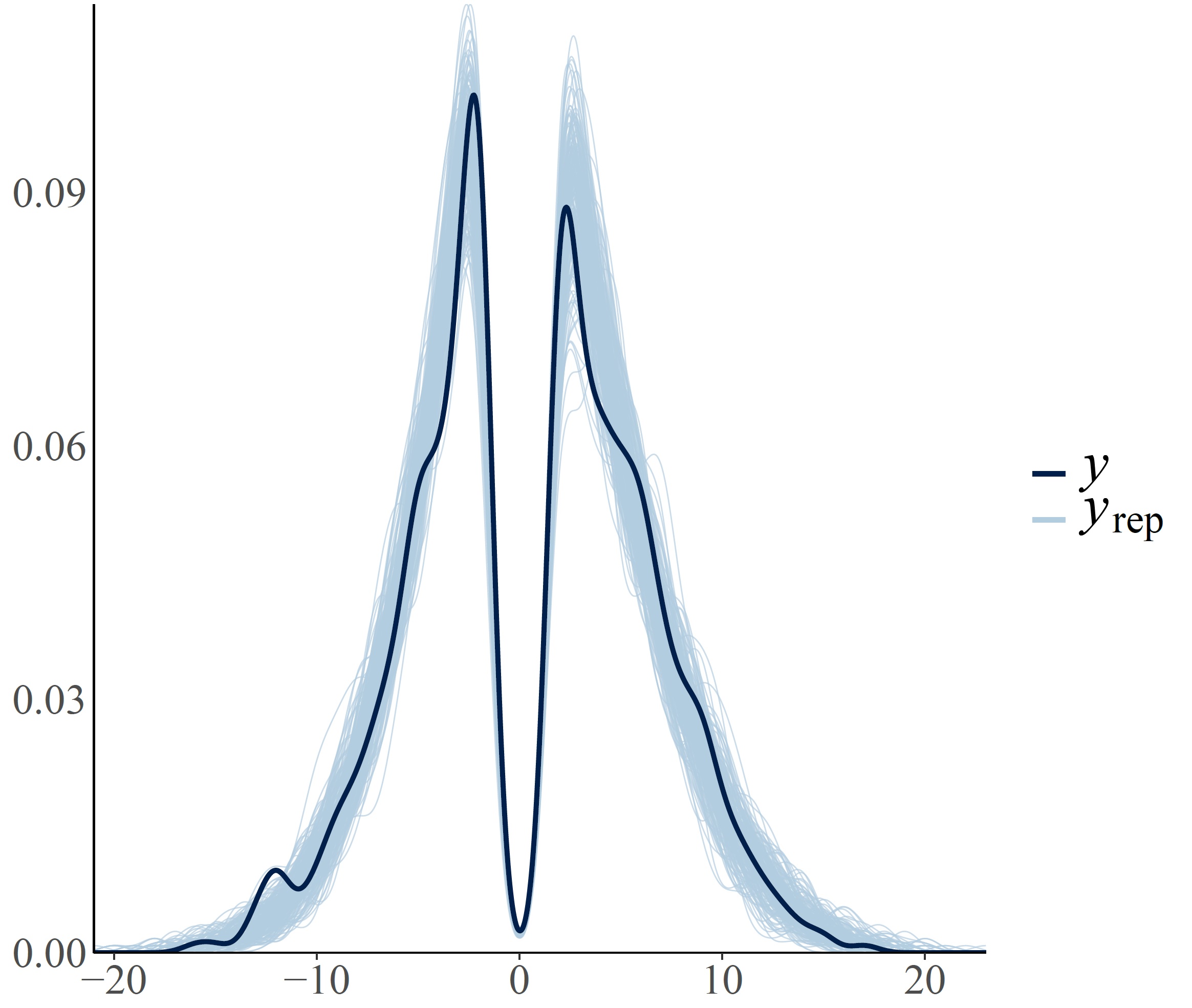}
\caption{Distribution of the observed set points differences $d_s= Y^A_{s}- Y^B_{s}$ (dark blue) plotted against the posterior predictive distribution of $\tilde{d}^{(t)}_{s}={y}^{A\ rep(t)}_{s}-{y}^{B\ rep(t)}_{s}$ (light blue) for the Italian SuperLega 2017/18 using the ZIP truncated negative binomial model with connected abilities and extra set abilities for Verona and Padova (model 9 in Table \ref{tab:02}).}
\label{fig8}
\end{figure}

Beyond that, it is straightforward to obtain a measure of model goodness of fit at the point level.
For each set $s,\  s =1,\ldots,S$, we denote by $d_s$ the set points difference $ Y^A_{s}- Y^B_{s}$, and with $\tilde{d}^{(t)}_{s}$  the corresponding points difference arising from the $t$-th MCMC replications, ${y}^{A\ rep(t)}_{s}-{y}^{B\ rep(t)}_{s}$. Once we replicate new existing values from our model, it is of interest to assess how far they are if compared with the actual data we observed. Figure \ref{fig8} displays the posterior predictive distribution of each $\tilde{d}^{(t)}_{s}$ (light blue) plotted against the true observed distribution for $d_s$: there is a quite good agreement between the replicated distributions and the observed distribution, and this is another corroboration of the goodness of fit of our final model (the plot is obtained through the {\tt bayesplot} package \citep{bayesplot}, which  always provides a continuous approximation for discrete dstributions).

\begin{algorithm}
	\colorbox{gray!25}{\parbox{0.95\textwidth}{
			\SetKwInOut{Input}{Input}
			\SetKwInOut{Output}{Output}
			\Input{$(\pi_G^{(t)}, p_G^{(t)}, \lambda^{(t)}, \omega_{G}^{(t)}, HT_G, AT_G)$: MCMC values of the parameters of the model for each game $G$ and MCMC iteration $t$; $HT_G, AT_G$: denote the home and the away teams in each game $G$}
			\Output{${\bf L}^{(t)}=\big(P^{(t)}_T, SW^{(t)}_T, SL^{(t)}_T, GPW^{(t)}_T, GPL^{(t)}_T \big)$: 
				$T_{mcmc}$ leagues with number of league points, total sets won and lost and total number of game points won and lost, respectively, for each team $T$ and MCMC iteration $t$. }
			\For{$t = 1$ \KwTo $T_{mcmc}$}{ 
				\# Initialize the league output for iteration $t$ \\  ${\bf L}^{(t)}={\bf 0}$; \\
				\For{$G = 1$ \KwTo $N_G$}{ 
					\# Initialize set and points for each game \\
					$S_H=S_W=P_H=P_W=0;$\\
					\While{$\max\{S_{H},S_W\}<3$ \# number of sets won by each team lower than three}{
						$R=S_H+S_W$; \# Calculate total number of sets until now \\ 
						$r <- 15 + 10 \times {\mathcal I}(R=5)$; \# Calculate maximum number of points\\ 
						$W \sim {\sf Bernoulli}(\omega_G^{(t)})$; \# Generate the winner of the set\\
						$O \sim {\sf ZIPoisson}(\pi_G^{(t)}, \lambda^{(t)})$;\# Generate the extra points \\
						$Y \sim {\sf NegBin}(r, p_G^{(t)}){\mathcal I}(Y<r-2)$ \# Generate the points of the loosing team \\[0.2em] 
						$S_H=S_H+W$; 
						$S_A=S_A+(1-W)$; \# Update the winning sets of the teams\\[0.2em]
						$P_W=r+O$; 
						$P_L=Y+O$; \# Points of the winning and loosing team\\
						$P_H=P_H+W\times P_W+(1-W)\times P_L$;\# Update the total points of the home team\\ [0.2em]  
						$P_A=P_A+(1-W)\times P_W+ W\times P_L$; \# Update the total points of the away team\\[0.2em]  
					}
					\# Updating the league parameters for the home team \\ 
					$P_{HT_G}^{(t)} = P_{HT_G}^{(t)} + 3 \times {\mathcal I}(S_H-S_A>1) +  {\mathcal I}(S_H-S_A=1)$; 
					\# League points\\
					$SW_{HT_G}^{(t)} = SW_{HT_G}^{(t)}  + S_H$; \# Sets won  
					$SL_{HT_G}^{(t)} = SW_{HT_G}^{(t)}  + S_A$; \# Sets lost  \\
					$PW_{HT_G}^{(t)} = PW_{HT_G}^{(t)}  + P_H$; \# Game points won \\
					$PL_{HT_G}^{(t)} = PL_{HT_G}^{(t)}  + P_A$; \# Game points lost \\[0.5em] 
					\# Updating the league parameters for the away team \\ 
					$P_{AT_G}^{(t)} = P_{AT_G}^{(t)} +3 \times {\mathcal I}(S_A-S_H>1) +  {\mathcal I}(S_A-S_H=1)$; 
					\# League points \\
					$SW_{AT_G}^{(t)} = SW_{AT_G}^{(t)}  + S_A$; \# Sets won \\
					$SL_{AT_G}^{(t)} = SW_{AT_G}^{(t)}  + S_H$; \# Sets lost \\
					$PW_{AT_G}^{(t)} = PW_{AT_G}^{(t)}  + P_A$; \# Game points won \\
					$PL_{AT_G}^{(t)} = PL_{AT_G}^{(t)}  + P_H$; \# Game points lost \\
				}	
			}
			{
				return league results ${\bf L}^{(t)}$ for $t=1,\dots, T_{mcmc}$\;
			}
			{\scriptsize Indexes: $t=,1, \dots T_{mcmc}$; $T_{mcmc}$: number of MCMC iterations; \\ 
				$G=1,\dots, N_G$; $N_G$: number of games; \\ 
				$T=1,\dots, N_T$; $N_T$: number of teams in the league.}
			\caption{Volleyball stochastic league re-construction algorithm}
			\label{algo1} 
		}
	}
\end{algorithm}

\subsection{Out-of-sample prediction}
\label{sec:out}

Our final task is to assess the out-of-sample predictive ability of our proposed model. As usual, we expect a lower predictive accuracy than the one obtained for in-sample measures. 
Nevertheless, it is crucial to ensure that our proposed model has satisfactory predictive performance. 

The procedure is similar to the stochastic league regeneration described in Section \ref{sec:pp} and Algorithm \ref{algo1}. The main difference here is that a specific number of games is now known and fixed (i.e. data) and only the remaining games are generated from the predictive distribution, while in Section \ref{sec:pp} the data of the whole season were available and the data of a new full league were re-generated assuming that we have exactly the same characteristics and team abilities as the one observed. 
Here we procceed with two scenarios: (a) the mid-season prediction scenario and (b) the play off prediction scenario. 
In the first case we assume that we are at the middle of the season where half of game results are available and we try to predict the final league standings. 
In the second scenario, we use the full league data to predict the final results in the play off phase. 

With the latter case, a further complication arises which is due to the formation of the \emph{play off} phase. In this after-season tournament, the best eight teams are competing from the quarter of finals: the team that wins three matches first goes to the next step. Thus, each game say between team $A$ and $B$ consists of a random number of repeated measurements, ranging from three to five, whereas the set point system is the same as the one described in the previous sections. 


\subsubsection{Mid-season prediction}

In this section we predict the second half of the season using the data of the first half of the season as a training set. 
This time point is important psychologically for the sports fans.
For example, in national football (soccer) leagues, there is the informal title of the ``Winter Champion'' which is mainly promoted by sports media and newspapers.  
In terms of data, at this time point, a considerable amount of games is available and all teams have played against all their opponents once. 
Hence, reliable estimates of the model parameters and team abilities can be obtained leading to accurate enough predictions about the final league ranking.

\begin{figure}
\includegraphics[scale=0.28]{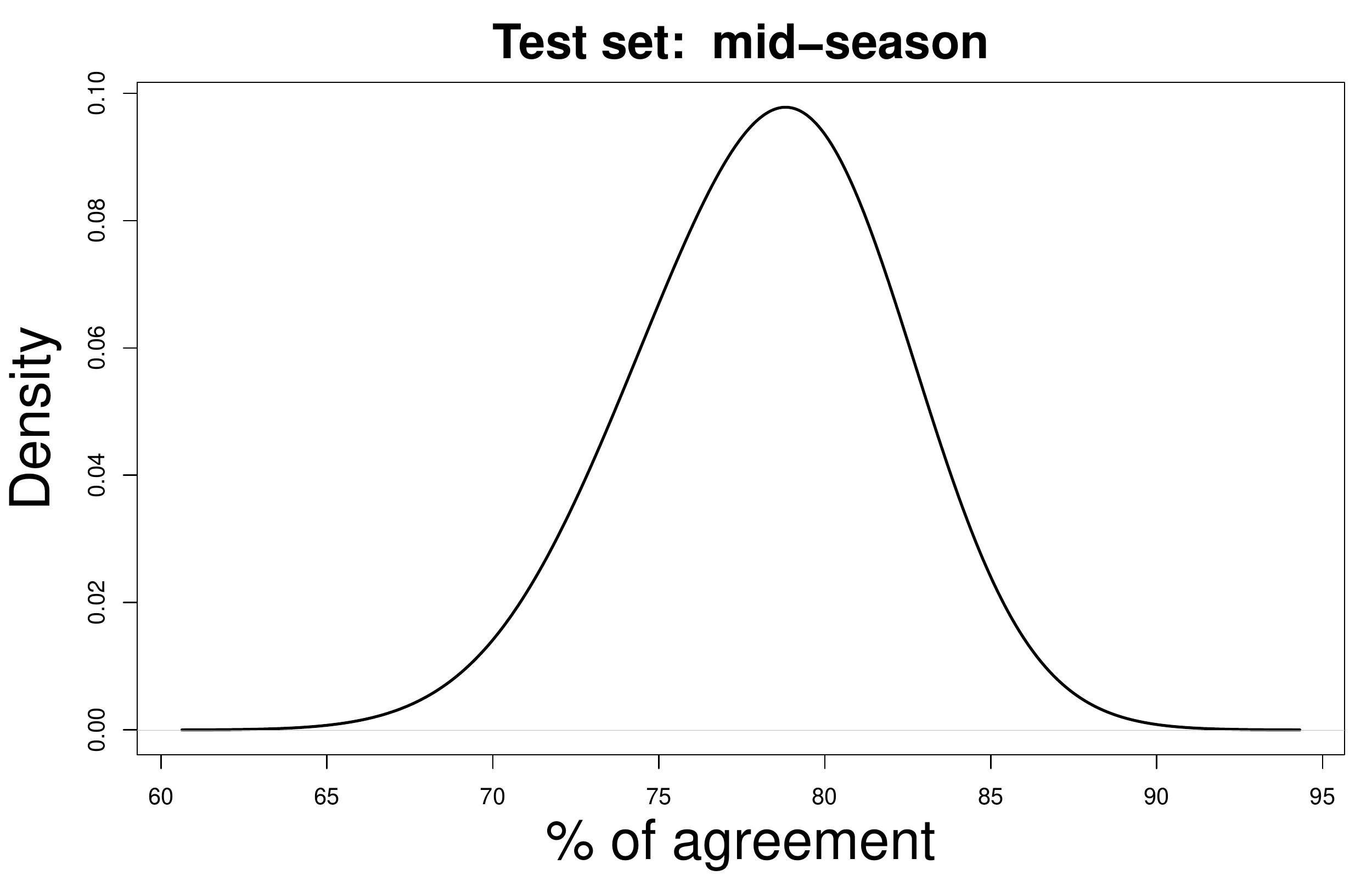}~
\includegraphics[scale=0.28]{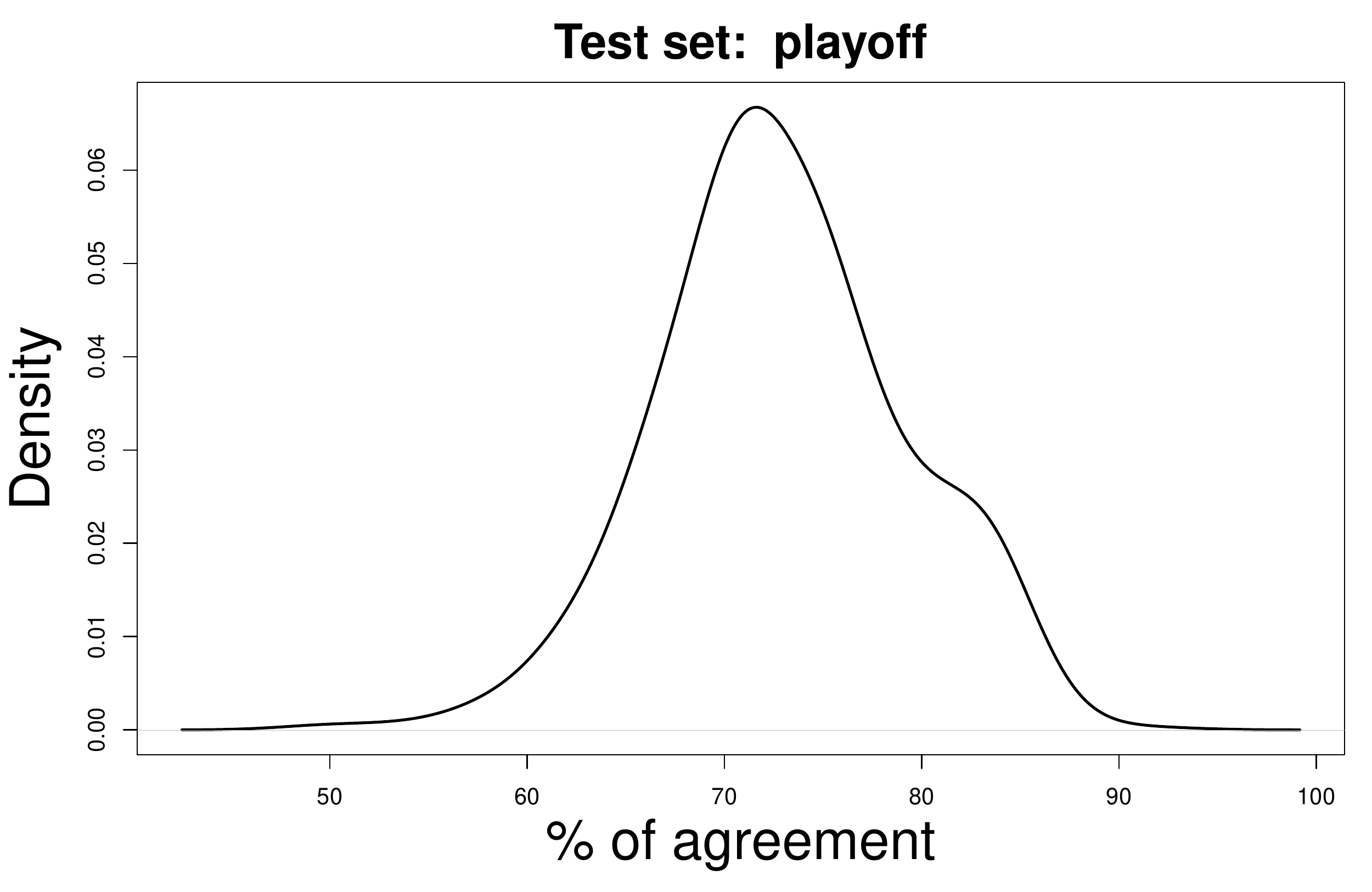}
\caption{Out-of-sample predictions: posterior distribution of percentage of correctly predicted games for the mid-season and the play off phase for the Italian SuperLega 2017/18 using the ZIP truncated negative binomial model with connected abilities and extra set abilities for Verona and Padova (model 9 in Table \ref{tab:02}).}
\label{fig9}
\end{figure}

\begin{figure}
\centering
\includegraphics[scale=0.7]{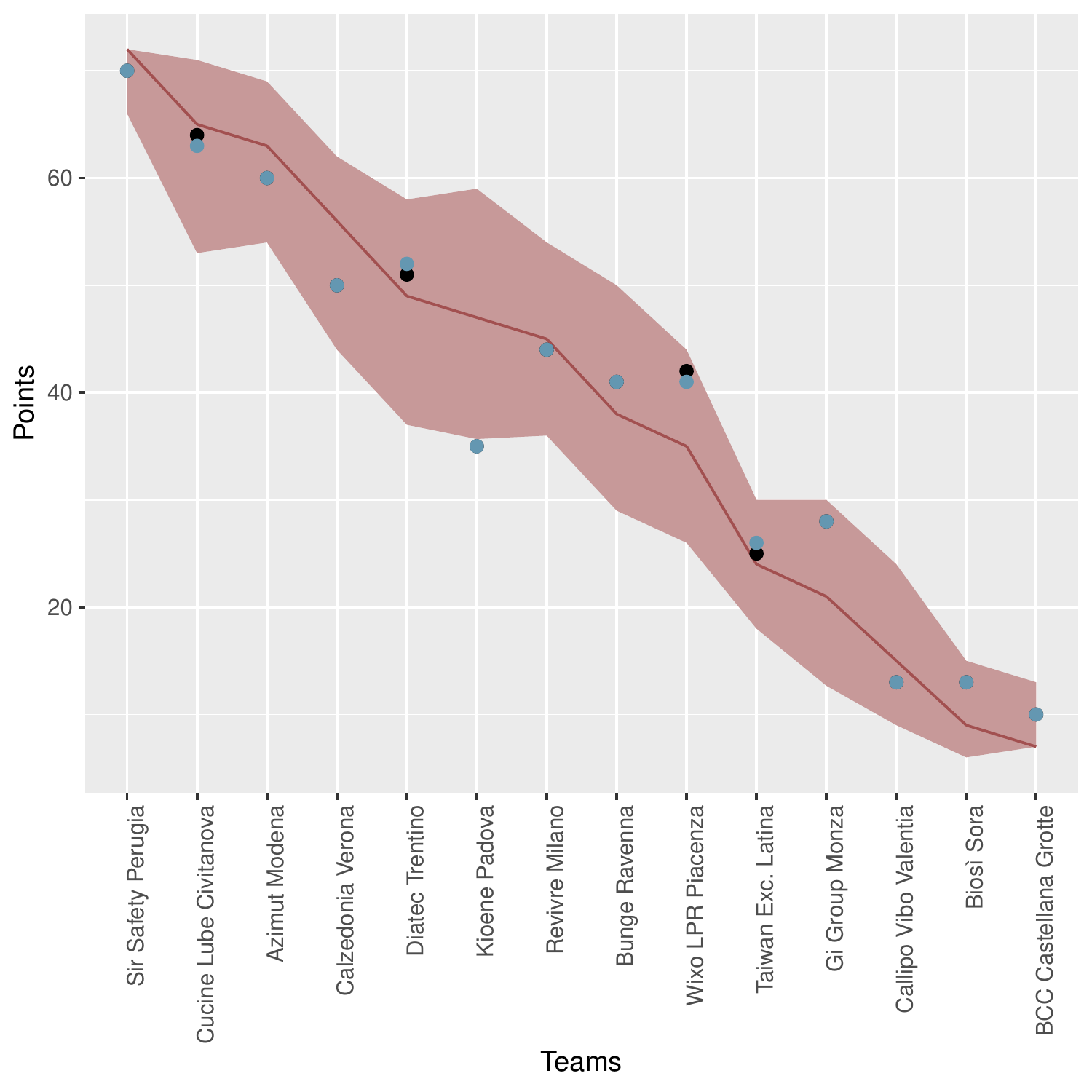}
\caption{Mid-season out-of-sample prediction: 95\% predictive intervals (red ribbon) from the posterior predictive distribution of the final points collected by each of the 14 teams of the Italian SuperLega 2017-2018 along with the observed final points (black dots) and the expected points from the in-sample league reconstruction (blue dots, see Table \ref{tab:05}). The red solid line represents the median.}
\label{fig10}
\end{figure}

\begin{figure}
\centering
\includegraphics[scale=0.5]{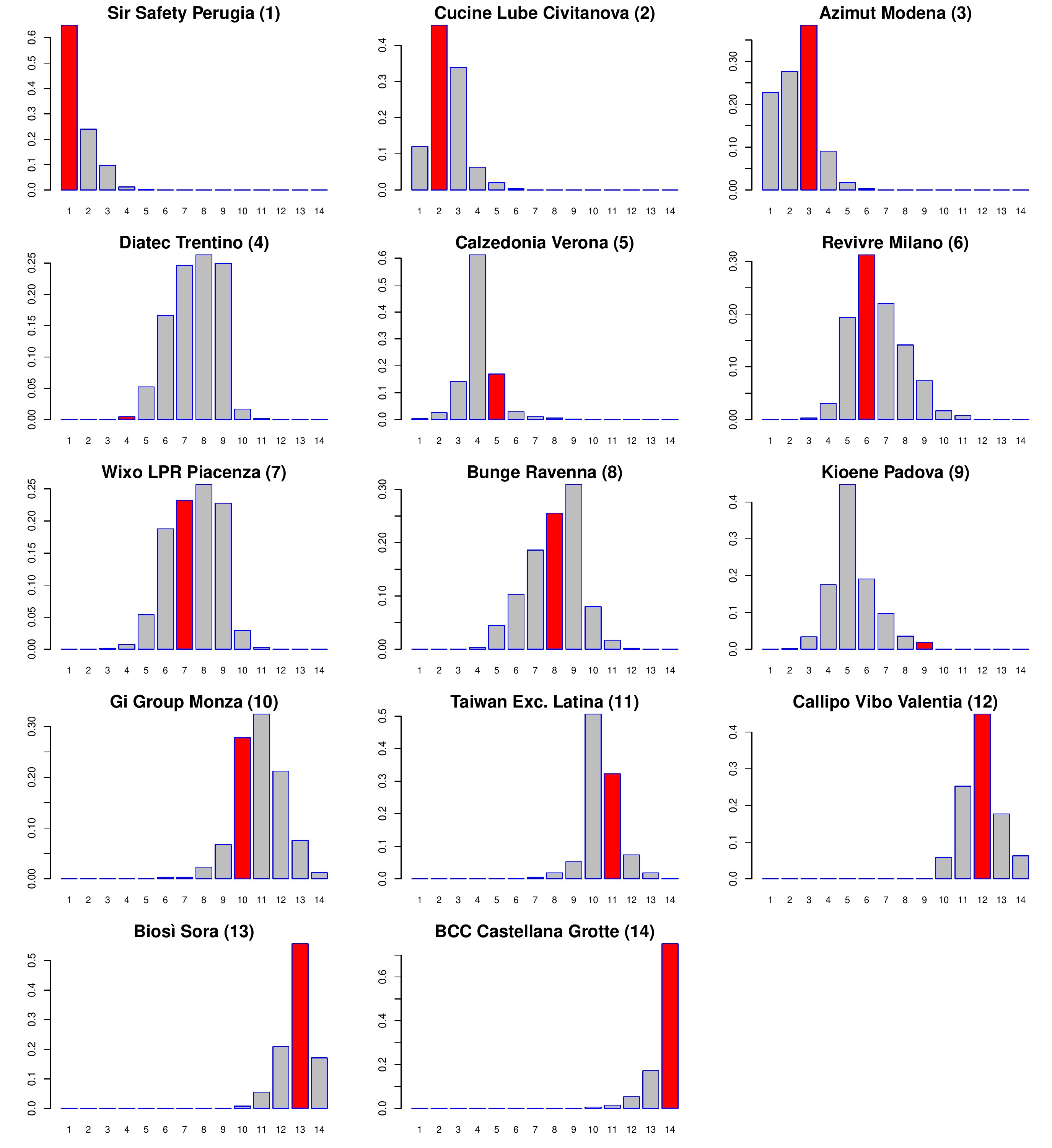}
\caption{Mid-season out-of-sample prediction: posterior predictive distribution of the final league ranking of each team in the Italian SuperLega 2017-2018. The red bar depicts the actual final rank position. The final league ranking is given within brackets in the title of each figure.}
\label{fig11}
\end{figure}

To preliminarily assess the model's predictive accuracy we use the percentage of agreement between predicted games/sets from the MCMC sample and the observed ones: the posterior distribution of the percentage of agreement of the correctly predicted games for the mid-season prediction scenario is presented in Figure~\ref{fig9} (left panel). The posterior mean of correct predictions concerning the final result of the game is found to be equal to 78.26\%($\pm$ 3\%). On the other hand, in the set level, the posterior agreement of correctly predicted sets (not displayed in the plot) is equal to 69.5\% ($\pm$ 1\%).
Figure~\ref{fig10} displays 95\% predictive intervals (red ribbon) for the predicted achieved points of the 14 teams competing in the Italian SuperLega 2017/2018, using the first half as training set, along with the observed final points (black dots), and the expected points from the in-sample league reconstruction (blue dots, see Table~\ref{tab:05}). At a first glance, the predicted rankings are in high agreement with the observed ones, especially for the top-three teams (Perugia, Civitanova and Modena) and the last ones (Vibo Valentia, Sora and Castellana Grotte): in these cases, the predicted points coincide with the median predictions. Padova is the only team whose observed points fall outside the 95\% predictive interval. Moreover, the majority of the observed final points (black dots) coincides with the in-sample simulated points (blue dots).
Figure~\ref{fig11} shows the posterior predictive distribution of the league ranking of each team for  the Italian SuperLega 2017-2018. The red bar, which is in correspondence of the actual rank, is the highest (i.e., is associated with the highest probability) both for the top-three teams and for the bottom three teams, suggesting again a good predictive ability for our model. 

\subsubsection{Play off prediction using regular season games}

Here we predict the games of the play off phase using the entire regular season as training set. Figure~\ref{fig9} (right panel) displays the posterior distribution of the percentage of agreement of the correctly predicted games: the posterior mean is 73.06\% ($\pm 6.05$\%). The posterior agreement of correctly predicted sets (not displayed in the plot) is 61.5\% ($\pm$ 2.54\%).

The play off phase consists of a small knockout tournament between  the best eight teams of the regular season: Sir Safety Perugia, Cucina Lube Civitanova, Azimut Modena, Diatec Trentino, Calzedonia Verona, Revivre Milano, Bunge Ravenna and Wixo LPR Piacenza. Table \ref{tab:06} shows for each team the probability to win in each play off stage 
and progress to the next round until winning the tournament: Perugia is associated with the highest probability (0.75) to win the play off phase (Perugia actually won this phase, defeating Civitanova in the final) and, generally, reports the highest probabilities to progress in each stage. Civitanova, the second best team during the regular season, is associated with a high probability to enter in the semifinals (0.87) and with the second highest probability to win the play off (0.14). Modena and Trentino, who reached the semifinals, report high probabilities to progress in the semifinals, 0.76 and 0.61 respectively, whereas Piacenza and Ravenna yield 0.13 and zero probabilities to reach the semifinals, respectively (they were actually eliminated in the quarter of finals). Globally, these probabilities seem to realistically mirror the actual strength of each team in the final stage of the season.

\begin{table}
	\caption{Play off out-of-sample prediction for the Italian SuperLega 2017/18: probability to progress in each stage of the play off phase along with the actual results.\label{tab:06}}
	\begin{tabular}{|rcccc|}
		\hline
		\emph{Teams} & \emph{Semi} & \emph{Final} & \emph{Winner} & Actual  \\ 
		\hline
		Sir Safety Perugia & 1.00 & 0.96 & 0.75 & Winner \\ 
		Cucine Lube Civitanova & 0.87 & 0.58 & 0.14 & Finalist \\ 
		Azimut Modena & 0.76 & 0.28 & 0.05 & Semi\\ 
		Diatec Trentino & 0.61 & 0.03 & 0.02 & Semi \\ 
		Calzedonia Verona & 0.39 & 0.01 & 0.00 & Quarter \\ 
		Revivre Milano & 0.24 & 0.07 & 0.02 & Quarter\\ 
		Bunge Ravenna   & 0.00 & 0.00 & 0.00 & Quarter \\ 
		Wixo LPR Piacenza   & 0.13 & 0.06 & 0.01& Quarter \\ 
		\hline
	\end{tabular}
\end{table}

\begin{figure}
\begin{tikzpicture}[
  level distance=5cm,every node/.style={minimum width=3cm,inner sep=0pt},
  edge from parent/.style={cyan!70!black,ultra thick,draw},
  level 1/.style={sibling distance=4cm},
  level 2/.style={sibling distance=2cm},
  legend/.style={draw=orange,fill=orange!30,inner sep=3pt}
]
\node (1) {\Pair{Perugia  \tiny{[0.78]} }{3}{\small{Lube Civ.} \tiny{[0.22]}  }{2}}
[edge from parent fork left,grow=left]
child {node (2) {\Pair{Perugia \tiny{[0.99]} }{3}{Trentino \tiny{[0.01] }}{2}}
child {node (3) {\Pair{Trentino \tiny{[0.61]}}{2}{Verona \tiny{[0.39]}}{1}}}
child {node {\Pair{Perugia \tiny{[1]}}{2}{Ravenna \tiny{[0]}}{1}}}
}
child {node {\Pair{\small{Lube Civ.} \tiny{ [0.65]}}{3}{Modena \tiny{ [0.35] }}{1}}
child  {node {\Pair{Modena \tiny{[0.76]}}{2}{Milano \tiny{[0.24]}}{0}}} 
child {node {\Pair{ \small{Lube Civ.} \tiny{[0.87]}}{2}{Piacenza \tiny{[0.13]}}{0}}}
};
\node[legend] at ([yshift=50pt]3) (QF) {Quarter Finals};
\node[legend] at (2|-QF) {Semi-Finals};
\node[legend] at (1|-QF) (QF) {Final};
\end{tikzpicture}
\caption{  Play off out-of-sample prediction using the full season data of the Italian SuperLeaga 2017/18: probabilities for each team to progress in each play off stage are reported within brackets.}
\label{fig12}
\end{figure}
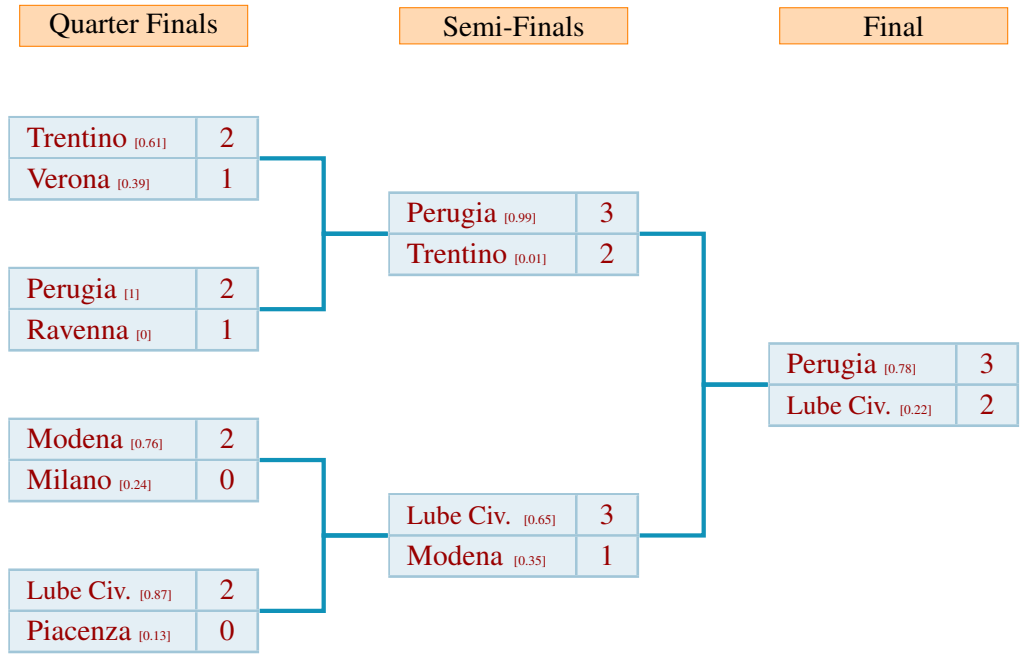

Figure \ref{fig12} displays the play off results of the matches actually played along with the posterior probabilities to progress in each play off stage. These probabilities have been obtained simply by considering the regular season results. As we can see, Perugia, the play off winner, is associated with the highest probabilities in each match, especially against Ravenna and Trentino: although it may seem a bit unrealistic that Perugia has probability one to beat Ravenna, this happens because the model's probability for Perugia is so high that during the MCMC simulation it is never defeated by Ravenna. To give an intuition for this issue, when playing at home and away Perugia has probabilities of 0.81 and 0.76 to win a set against Ravenna, respectively. In general, the highest probabilities are always associated with the teams that actually won the match and, consequently, progressed in the next stage.

Overall, our model yields good out-of-sample predictive performance, especially for the second mid-season.

\color{black}  

\section{Discussion}
\label{sec:disc}

With this work, we propose a unified hierarchical framework for modelling volleyball data using both outcomes (sets and points) of the game. 
The model follows a top-down approach (from sets to points) which initially seems counter-intuitive but it helps to capture the characteristics of the game itself.  
The core model structure is based on truncated versions of negative binomial for the points. 
Moreover, the two levels of the outcomes (set and points) are connected via common abilities with extra set abilities when needed. 
Finally, the main characteristics of the game are taken into consideration including that: 
(a) the winner of the set is the team that first scores a pre-specified number of points, and 
(b) the winner needs at least two points of difference to win the set (and the set continues until this is achieved). 
The latter is modelled via an extra latent component which is assumed to follow a zero inflated Poisson distribution. 
We have also tested for: the existence of correlations between sets (using random effects); the appropriateness of dynamic set and point abilities; finally, whether the probability of playing for extra points is influenced by the abilities of the teams (using a variety of functional forms). For the former check, there is some evidence that correlation might be present, as expected; however, the DIC provided similar predictive ability compared to the model without random effects, and therefore we proceeded with the simplified version of our model due to computational convenience. Dynamic abilities do not seem to improve the model (although point ability dynamics seem to be more useful than set ability dynamics). Finally, the difference in the abilities between the two competing teams does not seem to alter the probability of playing for extra time (as we might expect). 
We have concluded our modelling quest by selecting a ZIP truncated model with connected abilities and extra set abilities for Verona and Padova and constant probability to observe extra points of each set. Posterior predictive checks show a good agreement between our model and the observed  results, and an overall exceptional ability to replicate the final rankings of the league. Concerning future out-of-sample predictions, our proposed model is well behaved with acceptable predictive accuracy for future matches both for the mid-season and for the play off phase.

\subsection{Prior considerations}

\paragraph{Prior sensitivity analysis.} 
We have used priors of low information following an informal objective Bayes approach. For this reason, all prior distributions used are proper with relatively large variances. To ensure that the selected prior parameters had minor influence on the inference, we have also conducted sensitivity analysis. 
Detailed results in the form of 95\% posterior error bars can be found at the supplementary material of the article; see Figures A.1--A.3 at Appendix A. 

\paragraph{Prior elicitation.}  

Our modelling approach can be used to also incorporate prior information coming from historical data or from experts opinion. 
A standard method can be developed by using the power prior approach \citep{Chen2000} where historical data (even with incomplete or different covariate information) can be incorporated in our modelling approach. 
In this framework, the power parameter controls how reliable this prior information we believe that it is and, therefore, how much it will influence the posterior results. Empirical Bayes approaches or fully Bayesian hierarchical approaches can be used to estimate the power parameter from the data; see for example the works of \cite{Gravestock2017,Gravestock2019} in medicine. 
Simpler approaches can be used when proportions of wins for each game are available, which is very common in sports. A simple approach based on generalized linear models can be used to build prior estimates of the team abilities in the game level and further convert them in prior for the winning probabilities of sets and the associated team abilities. 
Finally, prior elicitation techniques for information coming from experts can be used to extract winning proportions and team abilities for each game.
This can be implemented by following the earlier work of \cite{Chen1999} or the more recent and general framework of \cite{Albert2012}. Nevertheless, experts, such as coaches, have rather limited skills and training on the quantification of their intuition or empirical knowledge. 
Therefore, the use of a down weight parameter (similar to the power parameter) is recommended as in \cite{Drikos_etal_2019}. 
By this way, most of the inference will be based on the actual data while a small portion of it will come from the experts knowledge. The latter information will correct for possible model misspecification or it may increase confidence for some specific parameters. Alternatively, we may incorporate predictions based on more reliable sources of prior information such as bookies as in \cite{Egidi2018} for football. 
Generally, prior elicitation in sports, and more specifically in volleyball, is an intriguing topic due to the general availability of historical data, the large amount of data published by betting companies and the easy access to sport experts (betting players, coaches and people working in sport industry). 
For any of the above cases, a more elaborate treatment is needed which is outside of the scope of this work. 

\subsection{Limitations of proposed methodology.}
Naturally, our approach embodies a number of assumptions which were tested using the data of Italian SuperLega
2017/2018 season. For example, for the final model we have assumed independence of sets and points conditional on the explanatory information. 
This was tested for the specific dataset using both random effects  and dynamic ability components. 
In both cases, no convincing evidence was found in order to incorporate either of these components in our final model. 
Moreover, we have assumed that the extra points follow a zero inflated distribution. 
This was found to be sufficient in our implementation but further investigation is needed in order to validate this result.

\paragraph{Limitation I: Team specific covariates.}
The main aim of this paper is to validate a general modelling formulation for volleyball data. Therefore we focus on modelling the main characteristics of the game and in developing a ``vanilla'' model using the two outcomes (sets and points) and the competing teams in each game without considering any extra information. 
Therefore, a limitation of our approach is that we do not consider any further covariates to improve both the interpretational and the predictive ability of the model. 
Towards this direction, \cite{Gabrio2020} has used a variety of team-specific covariates:  attack/defence types, service, service reception, blocks, passing abilities, roster quality. 
According the results of \cite{Gabrio2020},  the use of some of these covariates may be beneficial in terms of both game explanation and predictive power.  
The authors are currently working on more enriched volleyball dataset in order to include other characteristics of the game using two different approaches (a) descriptive, using the end of game statistics for interpretational reasons and (b) predictive, using statistics available at the beginning of the game to improve prediction. Both of these approaches will be applied in combination with Bayesian variable selection techniques.

\paragraph{Limitation II: Separate attacking and defensive team abilities.}
As a referee pointed out, we did not use separate attacking and defensive team abilities either on the set or the point level. Concerning the sets, it was not possible to separate attacking and defensive abilities due to identifiability reasons (also, all related models in bibliography do not consider separate attacking and defensive team abilities). For the points level, since the actual response is only represented by the points of the loosing team, conditionally on the winner of the set,
we believe that the data will not have enough information to accurately estimate both attacking and defensive abilities. This was confirmed by the empirical results appearing in Table \ref{tab:02}, see model 5. Although, in this work we illustrate this finding using a single season dataset, we intuitively believe that this result also holds generally for other leagues and datasets. 
  
\paragraph*{Limitation III: Not considering the fatigue in the model.}
Another  important athletic characteristic that we have not included in our modelling approach is fatigue. Although, this is of interest for every sport, it might be of prominent importance for volleyball since sets are terminal important time points of each volleyball game and these are achieved sequentially. This can be done by several ways: using fixed trend effects or random effects in the modelling of sets. Nevertheless, the exploration of the function that optimally influences the sets might be cumbersome and therefore we believe that should be treated separately, in another research work which is more focused on empirical results.

\subsection{Comparisons with other methods.} 
Early attempts to model volleyball were more focused on winning match and set probabilities through Markovian models \citep{ferrante2014winning}, whereas a Bayesian logistic regression to determine how
the performance of individual skills affects the probability of scoring a point is proposed by \cite{Miskin_etal_2010}. However, the majority of previous studies in volleyball is not oriented in modelling the entire game and validating the model strategy through league re-construction and prediction for future matches.


As far as we know from reviewing the literature, the only attempt to implement a generative model for volleyball results is proposed by \cite{Gabrio2020} for the women’s volleyball Italian SuperLega 2017/2018 season. Our work presents some similarities with this paper, such as the Bayesian framework and the posterior predictive validation of the model in terms of final points and ranking positions. However, distributional assumptions are deeply different: Gabrio's model is an adjustment of the double Poisson model adopted for football \citep{Maher_1982, Lee_1997}, whereas we propose a model which takes into consideration the special characteristics of the game itself which is different than the goal-scoring team sports for which the double Poisson and its extensions were introduced.


\subsection{Limitations regarding data specific results.}

\paragraph{Considering only one dataset.} 
Concerning the empirical implementation, a limitation of our results concerns the use of a single season dataset from a single league. In order to check the model's adequacy in a wider sense, we should apply our model on a variety of seasons and tournaments. One problem towards this direction, is that volleyball datasets are not so widely available as in other sports (e.g. basketball and football). The authors are currently in touch with volleyball experts in order to obtain richer datasets (including game specific covariates) and more seasons from the Greek league. 

\paragraph{No covariates in the model structure.}  
Moreover, additional covariates were not used here therefore we have not touched topics that other authors have dealt in the past (with simpler approaches),
such as 
fatigue \citep{Shaw2008},  
the effect of service \citep{Papadimitriou2004,Lopez_Martinez2009} and  
the effect of specific volleyball skills on final outcomes \citep{Miskin_etal_2010,Drikos_etal_2019,Gabrio2020}.

\paragraph{Not considering the service advantage.}
There is an increasing bibliography which focuses on which team serves first, 
not only for volleyball \citep{Shaw2008} but also for tennis \citep{Cohen_Zada2018}. 
This seems to be an important determinant at the point level (when we model the individual success of a point) but it might be less relevant at the accumulated point level for each set that we consider here. 
This is reasonable since the two teams are playing by having the serve advantage in turns, especially when the two competing teams are of high level. Nevertheless, it might be more relevant for the tie break where every small detail may count on determining the final winner. We believe that this effect will be minor when the game is unbalanced in terms of abilities (i.e. one team is much better than the other) but it might play a role (similar to home effect) if the two teams are close in terms of abilities. Unfortunately, for our dataset this information was not available but it might be of great interest to study this effect in the future.


\subsection{Final conclusion}

To conclude with, we have introduced an alternative model for volleyball data which uses a top-down approach modelling both sets and points and by considering the sport-specific characteristics such as the extra points played due to the required two points margin of win. Our work focuses on the validation of the simple ``vanilla'' model without considering extra covariates structure or characteristics such as fatigue, serve or specific sport skills. 
We expect and hope that this work will initiate further quests for finding new methods and models for predicting and understanding volleyball and other sports belonging to the group of net and ball games. 

\color{black}

\section*{Acknowledgements}
We would like to thank Dr. Sotirios Drikos for motivating us to work with Volleyball data and the two anonymous referees who improved the quality of the manuscript with their fruitful comments.
This research is financed by the Research Centre of Athens University of Economics and Business, 
in the framework of the project entitled ``Original Scientific Publications 2019''.

\section*{Supplementary material} Electronic Supplementary Material with further plots and sensitivity check is available at:\\ \url{https://github.com/LeoEgidi/Bayesian-Volleyball-paper}.

\bibliographystyle{chicago}
\bibliography{volley}

\begin{thebibliography}{}

\bibitem[\protect\citeauthoryear{Albert, Donnet, Guihenneuc-Jouyaux, Low-Choy,
  Mengersen, and Rousseau}{Albert et~al.}{2012}]{Albert2012}
Albert, I., S.~Donnet, C.~Guihenneuc-Jouyaux, S.~Low-Choy, K.~Mengersen, and
  J.~Rousseau (2012).
\newblock {Combining expert opinions in prior elicitation}.
\newblock {\em Bayesian Analysis\/}~{\em 7\/}(3), 503--532.

\bibitem[\protect\citeauthoryear{Albert}{Albert}{1992}]{albert1992bayesian}
Albert, J. (1992).
\newblock A {B}ayesian analysis of a poisson random effects model for home run
  hitters.
\newblock {\em The American Statistician\/}~{\em 46\/}(4), 246--253.

\bibitem[\protect\citeauthoryear{Barnett, Brown, and Jackson}{Barnett
  et~al.}{2008}]{Barnett_etal_2008}
Barnett, T.~J., A.~Brown, and K.~Jackson (2008).
\newblock {Modelling outcomes in volleyball}.
\newblock In {\em 9th Australasian Conference on Mathematics and Computers in
  Sport (9M\&CS) (Tweed Heads, Australia, 2008)}, pp.\  130--137. London:
  Chapman \& Hall.

\bibitem[\protect\citeauthoryear{Carpita, Ciavolino, and Pasca}{Carpita
  et~al.}{2019}]{carpita2019exploring}
Carpita, M., E.~Ciavolino, and P.~Pasca (2019).
\newblock Exploring and modelling team performances of the kaggle european
  soccer database.
\newblock {\em Statistical Modelling\/}~{\em 19\/}(1), 74--101.

\bibitem[\protect\citeauthoryear{Chen and Ibrahim}{Chen and
  Ibrahim}{2000}]{Chen2000}
Chen, M.-H. and J.~G. Ibrahim (2000).
\newblock {Power prior distributions for regression models}.
\newblock {\em Statistical Science\/}~{\em 15\/}(1), 46--60.

\bibitem[\protect\citeauthoryear{Chen, Ibrahim, and Yiannoutsos}{Chen
  et~al.}{1999}]{Chen1999}
Chen, M.~H., J.~G. Ibrahim, and C.~Yiannoutsos (1999).
\newblock {Prior elicitation, variable selection and Bayesian computation for
  logistic regression models}.
\newblock {\em Journal of the Royal Statistical Society. Series B: Statistical
  Methodology\/}~{\em 61\/}(1), 223--242.

\bibitem[\protect\citeauthoryear{Cohen-Zada, Krumer, and Shapir}{Cohen-Zada
  et~al.}{2018}]{Cohen_Zada2018}
Cohen-Zada, D., A.~Krumer, and O.~M. Shapir (2018).
\newblock {Testing the effect of serve order in tennis tiebreak}.
\newblock {\em Journal of Economic Behavior and Organization\/}~{\em 146},
  106--115.

\bibitem[\protect\citeauthoryear{Dixon and Coles}{Dixon and
  Coles}{1997}]{Dixon_Coles_1997}
Dixon, M. and S.~Coles (1997).
\newblock Modelling association football scores and inefficiencies in football
  betting market.
\newblock {\em Journal of the Royal Statistical Society C\/}~{\em 46},
  265--280.

\bibitem[\protect\citeauthoryear{Drikos, Ntzoufras, , and Apostolidis}{Drikos
  et~al.}{2019}]{Drikos_etal_2019}
Drikos, S., I.~Ntzoufras, , and N.~Apostolidis (2019).
\newblock {Bayesian Analysis of Skills Importance in World Champions Men’s
  Volleyball across Ages}.
\newblock {\em Journal of Computer Science in Sport\/}~{\em 18}, 24--44.

\bibitem[\protect\citeauthoryear{Egidi, Pauli, and Torelli}{Egidi
  et~al.}{2018}]{Egidi2018}
Egidi, L., F.~Pauli, and N.~Torelli (2018).
\newblock {Combining historical data and bookmakers' odds in modelling football
  scores}.
\newblock {\em Statistical Modelling\/}~{\em 18\/}(5-6), 436--459.

\bibitem[\protect\citeauthoryear{Facchinetti, Metulini, and
  Zuccolotto}{Facchinetti et~al.}{2019}]{Facchinetti_etal_2019}
Facchinetti, T., R.~Metulini, and P.~Zuccolotto (2019).
\newblock Detecting and classifying moments in basketball matches using sensor
  tracked data.
\newblock Technical report, arXiv: 1906.11720 [stat.AP].

\bibitem[\protect\citeauthoryear{Fellingham, Hinkle, and Hunter}{Fellingham
  et~al.}{2013}]{Fellingham_etal_2013}
Fellingham, G.~W., L.~J. Hinkle, and I.~Hunter (2013).
\newblock {Importance of attack speed in volleyball}.
\newblock {\em Journal of Quantitative Analysis in Sports\/}~{\em 9\/}(1),
  87--96.

\bibitem[\protect\citeauthoryear{Ferrante and Fonseca}{Ferrante and
  Fonseca}{2014}]{ferrante2014winning}
Ferrante, M. and G.~Fonseca (2014).
\newblock On the winning probabilities and mean durations of volleyball.
\newblock {\em Journal of Quantitative Analysis in Sports\/}~{\em 10\/}(2),
  91--98.

\bibitem[\protect\citeauthoryear{Gabrio}{Gabrio}{2020}]{Gabrio2020}
Gabrio, A. (2020).
\newblock {Bayesian Hierarchical Models for the Prediction of Volleyball
  Results}.
\newblock {\em Journal of Applied Statistics\/}~{\em (to appear)}, available at
  \url{https://doi.org/10.1080/02664763.2020.1723506}.

\bibitem[\protect\citeauthoryear{Gabry and Mahr}{Gabry and
  Mahr}{2019}]{bayesplot}
Gabry, J. and T.~Mahr (2019).
\newblock bayesplot: Plotting for {B}ayesian models.
\newblock R package version 1.7.0.

\bibitem[\protect\citeauthoryear{Gelman, Carlin, Stern, Dunson, Vehtari, and
  Rubin}{Gelman et~al.}{2013}]{gelman2013bayesian}
Gelman, A., J.~B. Carlin, H.~S. Stern, D.~B. Dunson, A.~Vehtari, and D.~B.
  Rubin (2013).
\newblock {\em Bayesian data analysis}.
\newblock Chapman and Hall/CRC.

\bibitem[\protect\citeauthoryear{Gelman, Jakulin, Pittau, Su, et~al.}{Gelman
  et~al.}{2008}]{gelman2008weakly}
Gelman, A., A.~Jakulin, M.~G. Pittau, Y.-S. Su, et~al. (2008).
\newblock A weakly informative default prior distribution for logistic and
  other regression models.
\newblock {\em The Annals of Applied Statistics\/}~{\em 2\/}(4), 1360--1383.

\bibitem[\protect\citeauthoryear{Gelman, Rubin, et~al.}{Gelman
  et~al.}{1992}]{gelman1992inference}
Gelman, A., D.~B. Rubin, et~al. (1992).
\newblock Inference from iterative simulation using multiple sequences.
\newblock {\em Statistical science\/}~{\em 7\/}(4), 457--472.

\bibitem[\protect\citeauthoryear{Gravestock and Held}{Gravestock and
  Held}{2017}]{Gravestock2017}
Gravestock, I. and L.~Held (2017).
\newblock {Adaptive power priors with empirical Bayes for clinical trials}.
\newblock {\em Pharmaceutical Statistics\/}~{\em 16\/}(5), 349--360.

\bibitem[\protect\citeauthoryear{Gravestock and Held}{Gravestock and
  Held}{2019}]{Gravestock2019}
Gravestock, I. and L.~Held (2019).
\newblock {Power priors based on multiple historical studies for binary
  outcomes}.
\newblock {\em Biometrical Journal\/}~{\em 61\/}(5), 1201--1218.

\bibitem[\protect\citeauthoryear{Harville}{Harville}{1977}]{Harville_1977}
Harville, D. (1977).
\newblock The use of linear-model methodology to rate high school or college
  football teams.
\newblock {\em Journal of the American Statistical Association\/}~{\em 72},
  278--289.

\bibitem[\protect\citeauthoryear{Hass and Craig}{Hass and
  Craig}{2018}]{Hass2018}
Hass, Z. and B.~A. Craig (2018).
\newblock {Exploring the potential of the plus/minus in NCAA women's volleyball
  via the recovery of court presence information}.
\newblock {\em Journal of Sports Analytics\/}~{\em 4\/}(4), 285--295.

\bibitem[\protect\citeauthoryear{Karlis and Ntzoufras}{Karlis and
  Ntzoufras}{2003}]{karlis2003analysis}
Karlis, D. and I.~Ntzoufras (2003).
\newblock Analysis of sports data by using bivariate {P}oisson models.
\newblock {\em Journal of the Royal Statistical Society D\/}~{\em 52\/}(3),
  381--393.

\bibitem[\protect\citeauthoryear{Karlis and Ntzoufras}{Karlis and
  Ntzoufras}{2009}]{Karlis_Ntzoufras_2009}
Karlis, D. and I.~Ntzoufras (2009).
\newblock Bayesian modelling of football outcomes: using the {S}kellam's
  distribution for the goal difference.
\newblock {\em IMA Journal of Management Mathematics\/}~{\em 20}, 133--146.

\bibitem[\protect\citeauthoryear{Koopman and Lit}{Koopman and
  Lit}{2015}]{Koopman_Lit_2015}
Koopman, S.~J. and R.~Lit (2015).
\newblock A dynamic bivariate {P}oisson model for analysing and forecasting
  match results in the {E}nglish {P}remier {L}eague.
\newblock {\em Journal of the Royal Statistical Society A\/}~{\em 178},
  167--186.

\bibitem[\protect\citeauthoryear{Lee}{Lee}{1997}]{Lee_1997}
Lee, A. (1997).
\newblock Modeling scores in the {P}remier {L}eague: Is {M}anchester {U}nited
  really the best?
\newblock {\em Chance\/}~{\em 10}, 15--19.

\bibitem[\protect\citeauthoryear{Lee and Chin}{Lee and Chin}{2004}]{Lee2004}
Lee, K.~T. and S.~T. Chin (2004).
\newblock {Strategies to serve or receive the service in volleyball}.
\newblock {\em Mathematical Methods of Operations Research\/}~{\em 59\/}(1),
  53--67.

\bibitem[\protect\citeauthoryear{Lopez-Martinez and Palao}{Lopez-Martinez and
  Palao}{2009}]{Lopez_Martinez2009}
Lopez-Martinez, A. and J.~Palao (2009).
\newblock {Effect of serve execution on serve efficacy in men's and women's
  beach volleyball}.
\newblock {\em International Journal of Applied Sports Sciences\/}~{\em
  21\/}(1), 1--16.

\bibitem[\protect\citeauthoryear{Maher}{Maher}{1982}]{Maher_1982}
Maher, M. (1982).
\newblock Modelling association football scores.
\newblock {\em Statistica Neerlandica\/}~{\em 36}, 109--118.

\bibitem[\protect\citeauthoryear{Marcelino, Mesquita, Palao, and
  Sampaio}{Marcelino et~al.}{2009}]{Marcelino2009}
Marcelino, R., I.~Mesquita, J.~M. Palao, and J.~Sampaio (2009).
\newblock {Home advantage in high-level volleyball varies according to set
  number}.
\newblock {\em Journal of Sports Science and Medicine\/}~{\em 8\/}(3),
  352--356.

\bibitem[\protect\citeauthoryear{Mendes, Nascimento, Souza, Collet, Milistetd,
  Côté, and Carvalho}{Mendes et~al.}{2018}]{Mendes_etal_2008}
Mendes, F.~G., J.~V. Nascimento, E.~R. Souza, C.~Collet, M.~Milistetd,
  J.~Côté, and H.~M. Carvalho (2018).
\newblock Retrospective analysis of accumulated structured practice: A bayesian
  multilevel analysis of elite brazilian volleyball players.
\newblock {\em High Ability Studies\/}, 1--15.

\bibitem[\protect\citeauthoryear{Metulini, Marisera, and Zuccolotto}{Metulini
  et~al.}{2017}]{Metulini_etal_2017}
Metulini, R., M.~Marisera, and P.~Zuccolotto (2017).
\newblock Space-time analysis of movements in basketball using sensor data.
\newblock In {\em Statistics and Data Science: New Challenges, New Generations
  -- SIS 2017 Proceedings}. Firenze University Press.

\bibitem[\protect\citeauthoryear{Miskin, Fellingham, and Florence}{Miskin
  et~al.}{2010}]{Miskin_etal_2010}
Miskin, M.~A., G.~W. Fellingham, and L.~W. Florence (2010).
\newblock Skill importance in women’s volleyball.
\newblock {\em Journal of Quantitative Analysis in Sports\/}~{\em 6\/}(2),
  Article 5.

\bibitem[\protect\citeauthoryear{Mosteller}{Mosteller}{1952}]{Mosteller_1952}
Mosteller, F. (1952).
\newblock The world series competition.
\newblock {\em Journal of the American Statistical Association\/}~{\em 47},
  355--380.

\bibitem[\protect\citeauthoryear{Mosteller}{Mosteller}{1970}]{Mosteller_1970}
Mosteller, F. (1970).
\newblock Collegiate football scores, {U.S.A.}
\newblock {\em Journal of the American Statistical Association\/}~{\em 65},
  35--48.

\bibitem[\protect\citeauthoryear{Owen}{Owen}{2011}]{Owen_2011}
Owen, A. (2011).
\newblock Dynamic {B}ayesian forecasting models of football match outcomes with
  estimation of the evolution variance parameter.
\newblock {\em IMA Journal of Management Mathematics\/}~{\em 22}, 99--113.

\bibitem[\protect\citeauthoryear{Papadimitriou, Pashali, Sermaki, Mellas, and
  Papas}{Papadimitriou et~al.}{2004}]{Papadimitriou2004}
Papadimitriou, K., E.~Pashali, I.~Sermaki, S.~Mellas, and M.~Papas (2004).
\newblock {The effect of the opponents' serve on the offensive actions of Greek
  setters in volleyball games}.
\newblock {\em International Journal of Performance Analysis in Sport\/}~{\em
  4}, 23--33.

\bibitem[\protect\citeauthoryear{Plummer}{Plummer}{2018}]{rjags}
Plummer, M. (2018).
\newblock {\em rjags: {B}ayesian Graphical Models using MCMC}.
\newblock R package version 4-8.

\bibitem[\protect\citeauthoryear{Reep and Benjamin}{Reep and
  Benjamin}{1967}]{Reep_Benjamin_1967}
Reep, C. and B.~Benjamin (1967).
\newblock Skill and chance in association football.
\newblock {\em Journal of the Royal Statistical Society A\/}~{\em 131},
  581--585.

\bibitem[\protect\citeauthoryear{Rue and Salvesen}{Rue and
  Salvesen}{2000}]{Rue_Salvesen_2000}
Rue, H. and O.~Salvesen (2000).
\newblock Prediction and retrospective analysis of soccer matches in a league.
\newblock {\em Journal of the Royal Statistical Society D\/}~{\em 49},
  399--418.

\bibitem[\protect\citeauthoryear{Schwertman, McCready, and Howard}{Schwertman
  et~al.}{1991}]{Schwertman_etal_1991}
Schwertman, N.~C., T.~A. McCready, and L.~Howard (1991).
\newblock Probability models for the ncaa regional basketball tournaments.
\newblock {\em The American Statistician\/}~{\em 45}, 35--38.

\bibitem[\protect\citeauthoryear{Sepulveda, Lleida, and Alcaraz}{Sepulveda
  et~al.}{2017}]{Sepulveda2017}
Sepulveda, D., U.~D. Lleida, and A.~G.~D. Alcaraz (2017).
\newblock {Analysis of Volleyball Attack from the Markov Chain Model}.
\newblock In {\em Complex Systems in Sport, International Congress. Linking
  Theory and Practice.}, pp.\  166--168. Frontiers Media SA.

\bibitem[\protect\citeauthoryear{Shaw, Gribble, Atc, Frye, and Atc}{Shaw
  et~al.}{2008}]{Shaw2008}
Shaw, M.~Y., P.~A. Gribble, {\`{A}}.~Atc, J.~L. Frye, and {\`{A}}.~Atc (2008).
\newblock {Ankle Bracing, Fatigue, and Time to Stabilization in Collegiate
  Volleyball Athletes}.
\newblock {\em Journal of Athletic Training\/}~{\em 43}, 164--171.

\bibitem[\protect\citeauthoryear{Shonkwiler}{Shonkwiler}{2016}]{shonkwiler2016variance}
Shonkwiler, J. (2016).
\newblock Variance of the truncated negative binomial distribution.
\newblock {\em Journal of Econometrics\/}~{\em 195\/}(2), 209--210.

\bibitem[\protect\citeauthoryear{Sonnabend}{Sonnabend}{2020}]{Sonnabend_2020}
Sonnabend, H. (2020).
\newblock On discouraging environments in team contests: Evidence from
  top-level beach volleyball.
\newblock {\em Managerial and Decision Economics\/}, 1--12.

\bibitem[\protect\citeauthoryear{Spiegelhalter, Best, Carlin, and Van
  Der~Linde}{Spiegelhalter et~al.}{2002}]{spiegelhalter2002bayesian}
Spiegelhalter, D.~J., N.~G. Best, B.~P. Carlin, and A.~Van Der~Linde (2002).
\newblock Bayesian measures of model complexity and fit.
\newblock {\em Journal of the Royal Statistical Society B\/}~{\em 64\/}(4),
  583--639.

\bibitem[\protect\citeauthoryear{Stefani}{Stefani}{1980}]{Stefani_1980}
Stefani, R. (1980).
\newblock Improved least squares football, basketball, and soccer predictions.
\newblock {\em IEEE Transactions on Systems, Man, and Cybernetics\/}~{\em
  SMC--10}, 116--23.

\bibitem[\protect\citeauthoryear{Tsokos, Narayanan, Kosmidis, Baio, Cucuringu,
  Whitaker, and KirÃ¡ly}{Tsokos et~al.}{2019}]{Tsokos_etal_2019}
Tsokos, A., S.~Narayanan, I.~Kosmidis, G.~Baio, M.~Cucuringu, G.~Whitaker, and
  F.~KirÃ¡ly (2019).
\newblock Modeling outcomes of soccer matches.
\newblock {\em Machine Learning\/}~{\em 108}, 77--95.

\end{thebibliography}

\color{black}

\label{sect:bib}

\numberwithin{figure}{section}
\numberwithin{table}{section}

\newpage
\appendix

\section{Sensitivity analysis}

\begin{figure}
\centering
\includegraphics[scale=0.25]{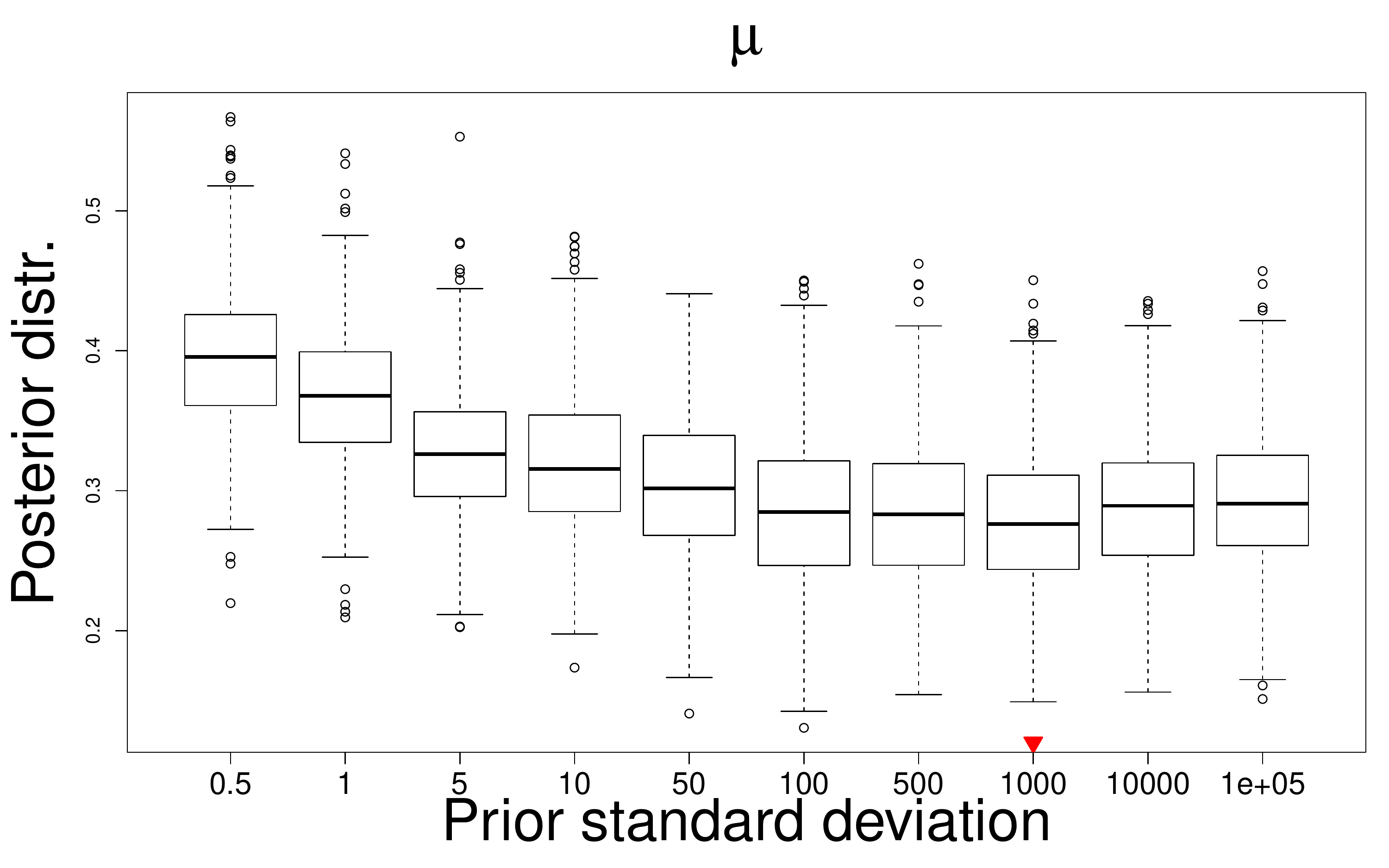}~
\includegraphics[scale=0.25]{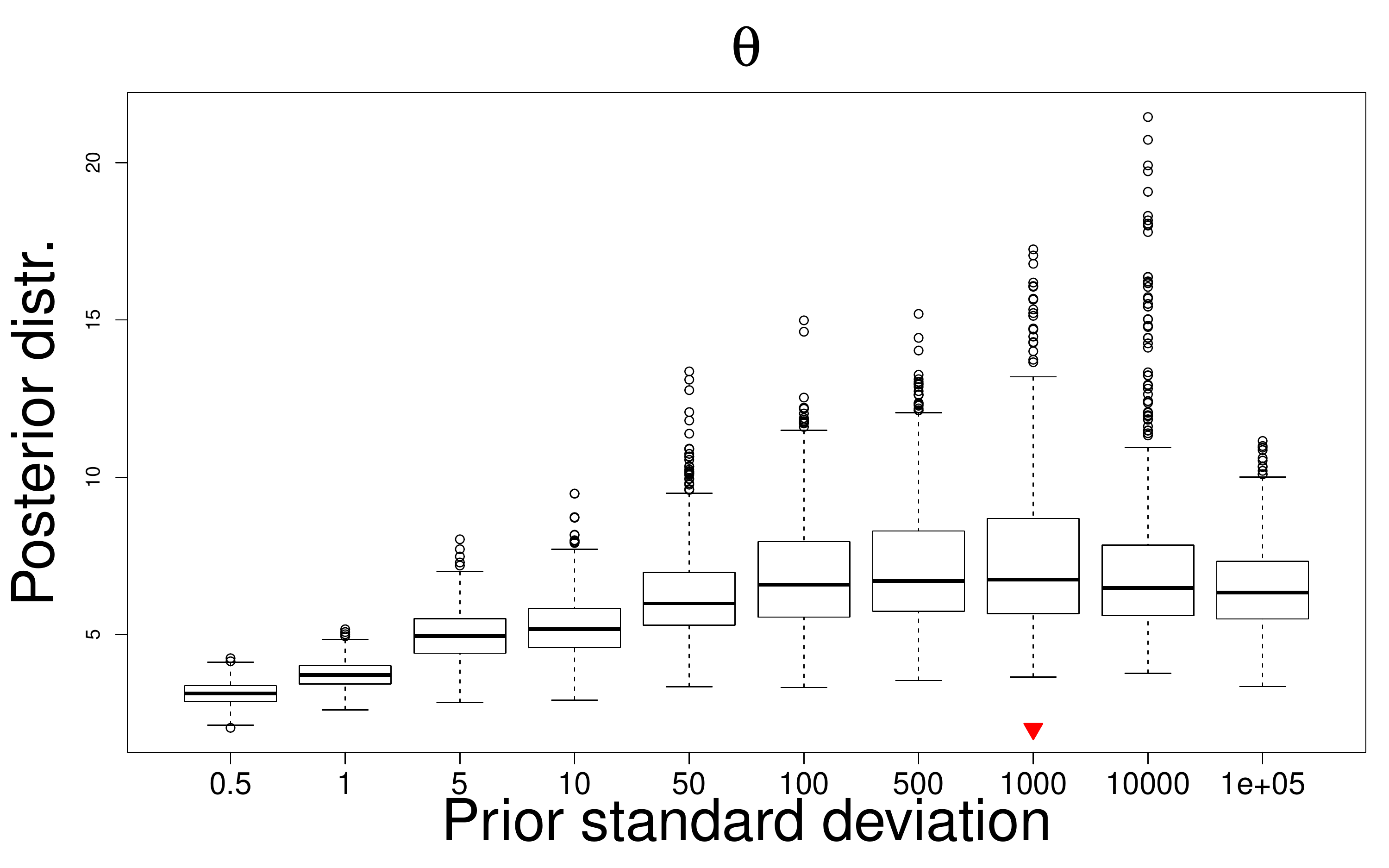}\\
\includegraphics[scale=0.25]{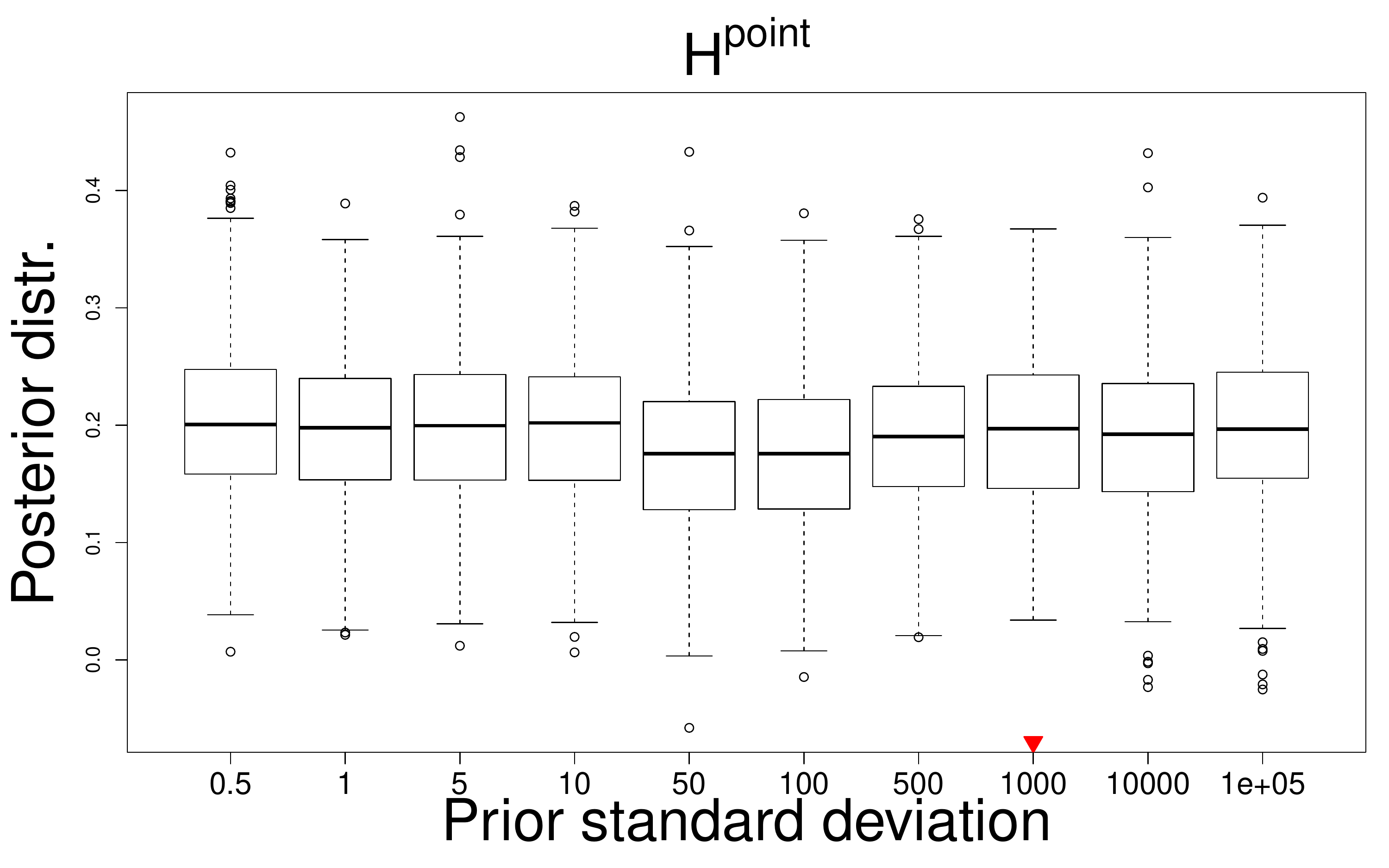}~
\includegraphics[scale=0.25]{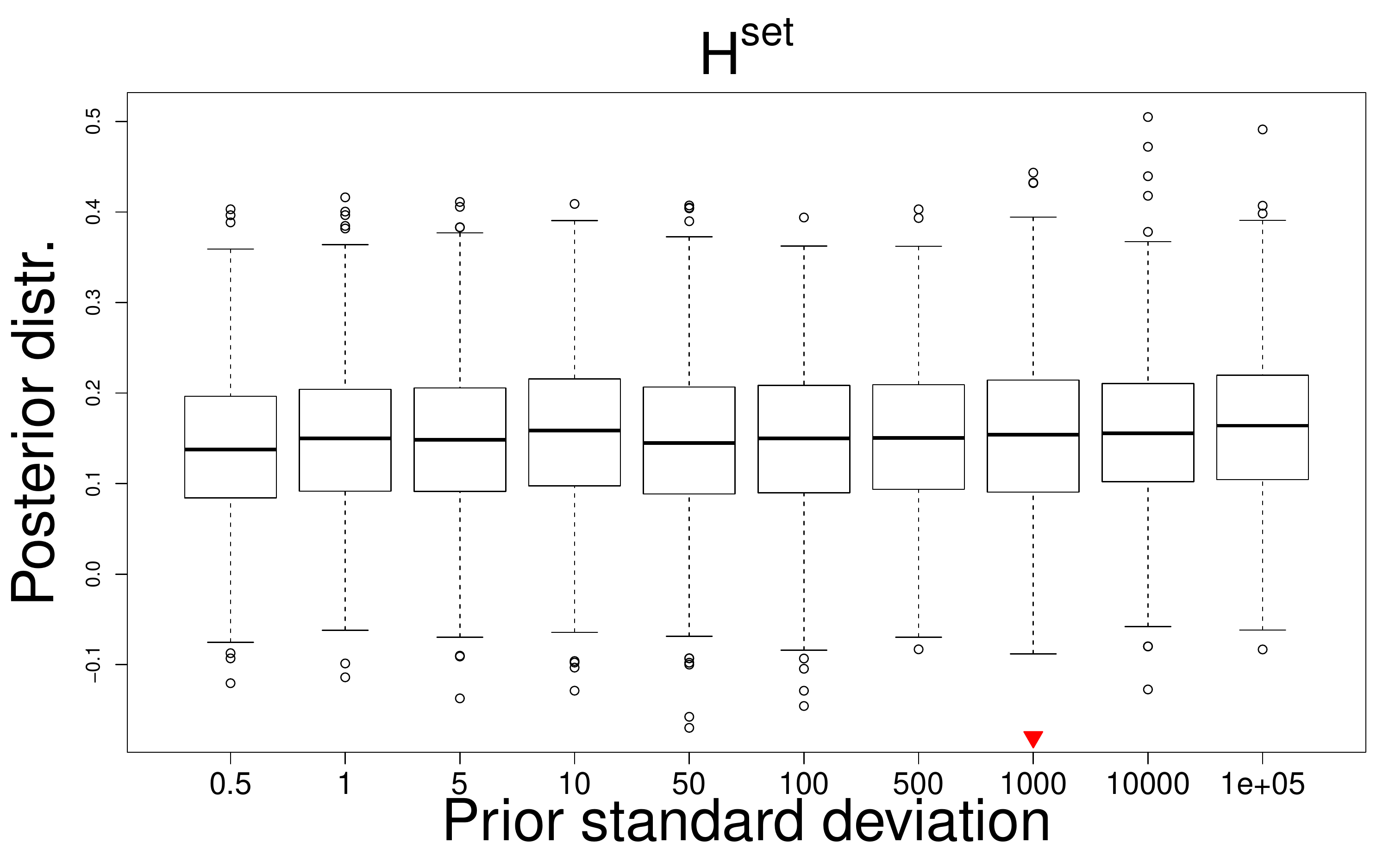}\\
\includegraphics[scale=0.25]{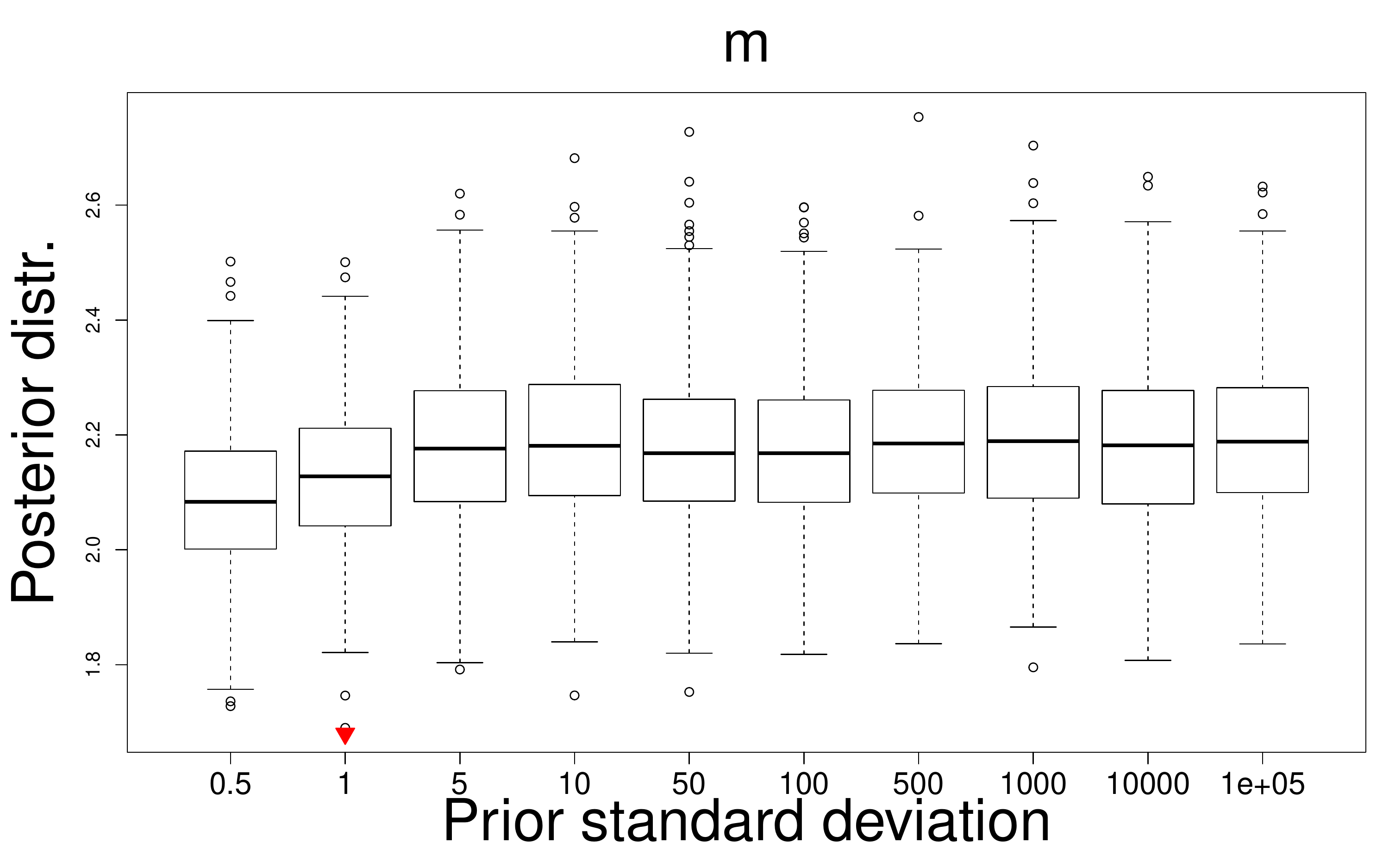}~
\includegraphics[scale=0.25]{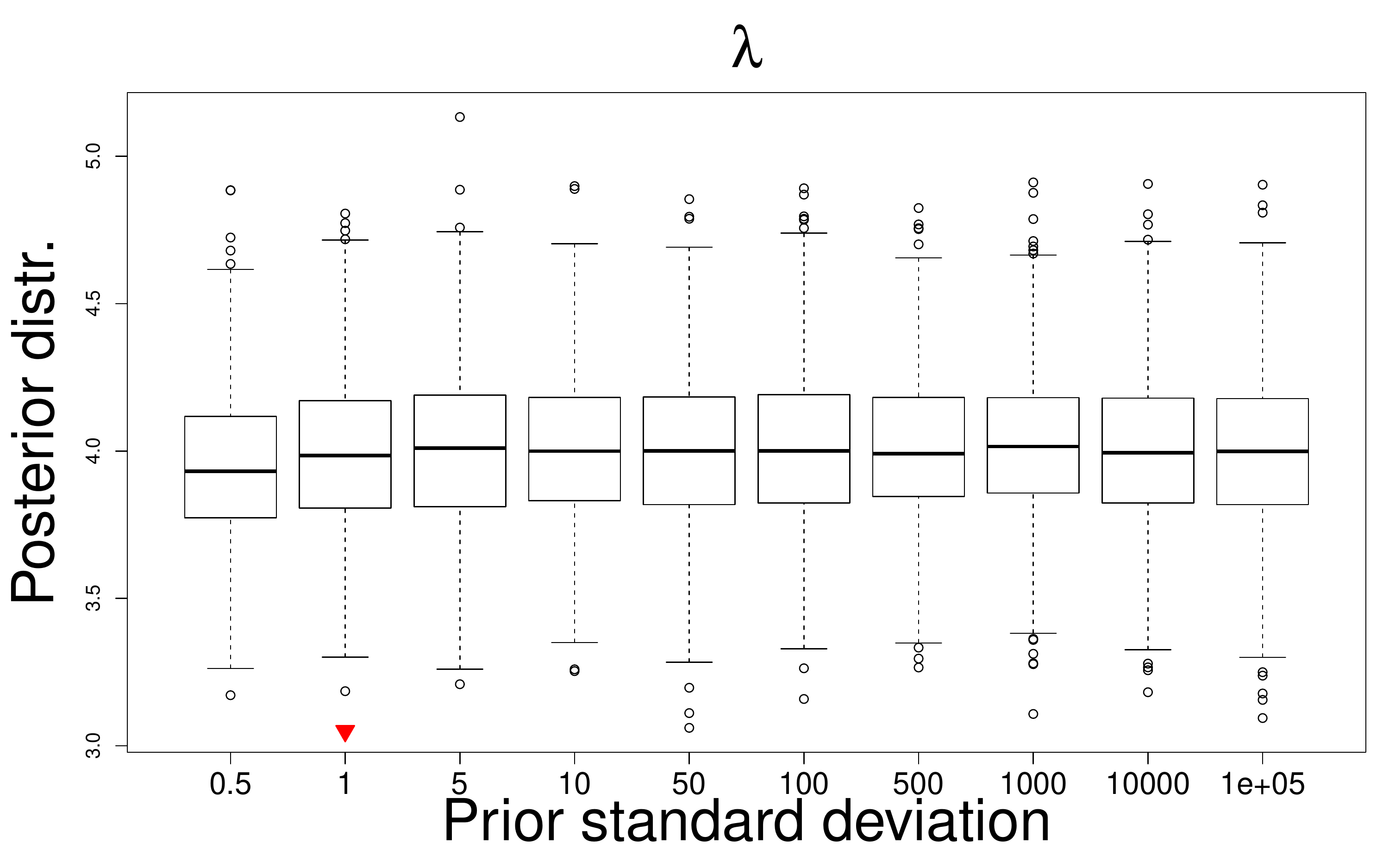}\\
\caption{Sensitivity analysis for parameters $\mu, \theta, H^{point}, H^{set}, m, \lambda$: Posterior box-plots for a variety of the prior standard deviations. 3000 MCMC iterations obtained by 3 chains with burn-in period of 100 iterations using {\tt rjags}.}
\label{figS1}
\end{figure}

\begin{figure}
\centering
\includegraphics[scale=0.25]{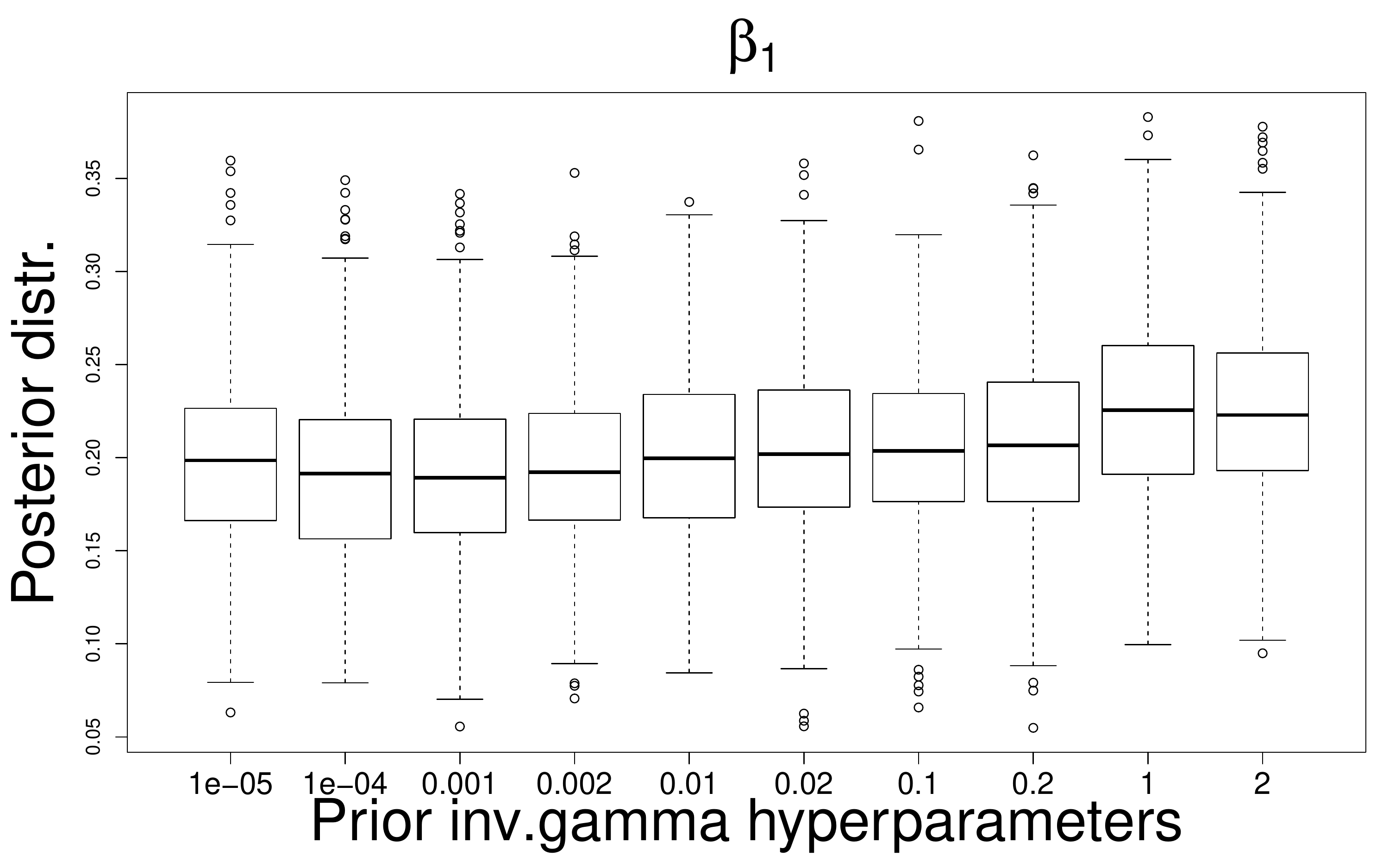}~
\includegraphics[scale=0.25]{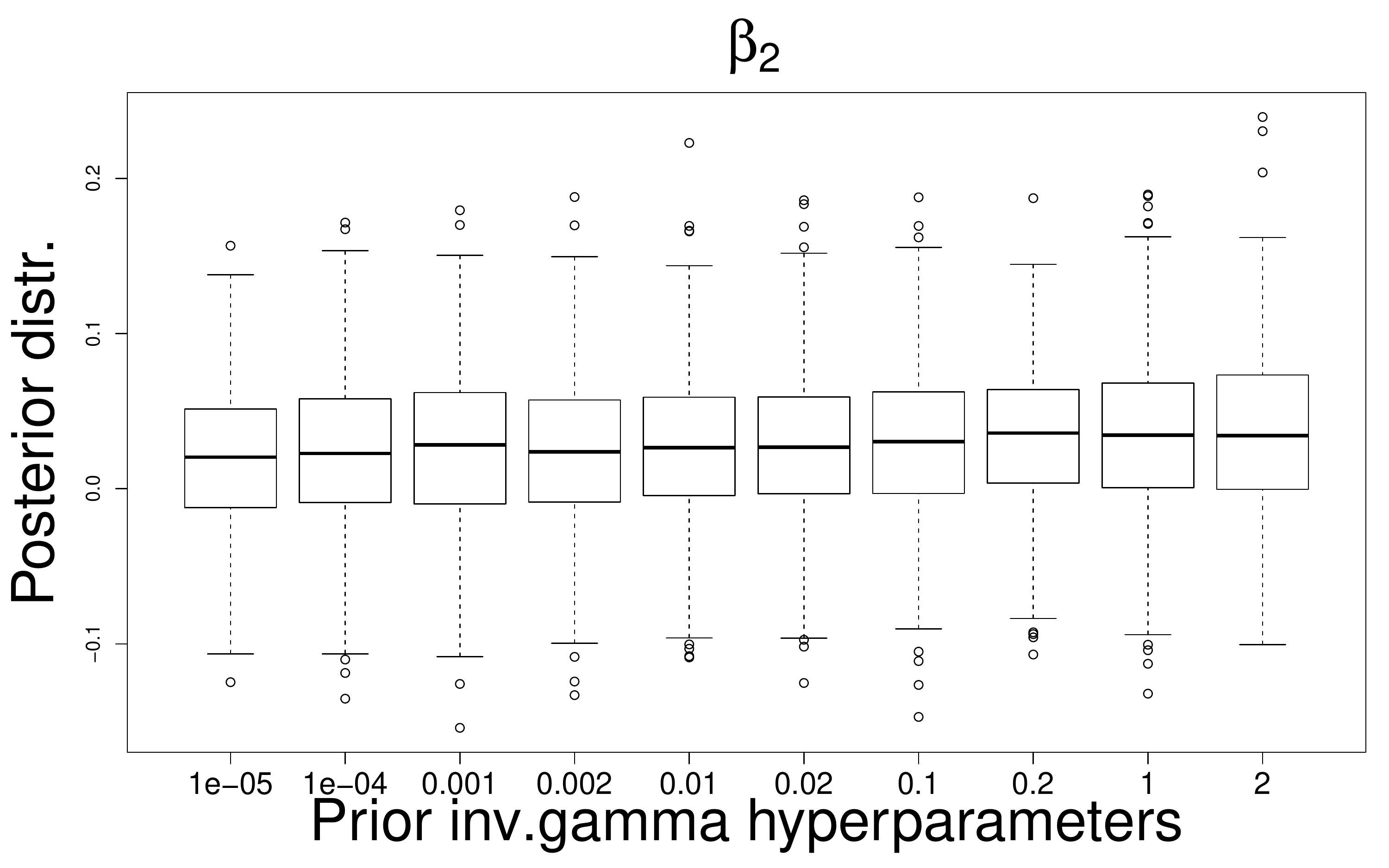}\\
\includegraphics[scale=0.25]{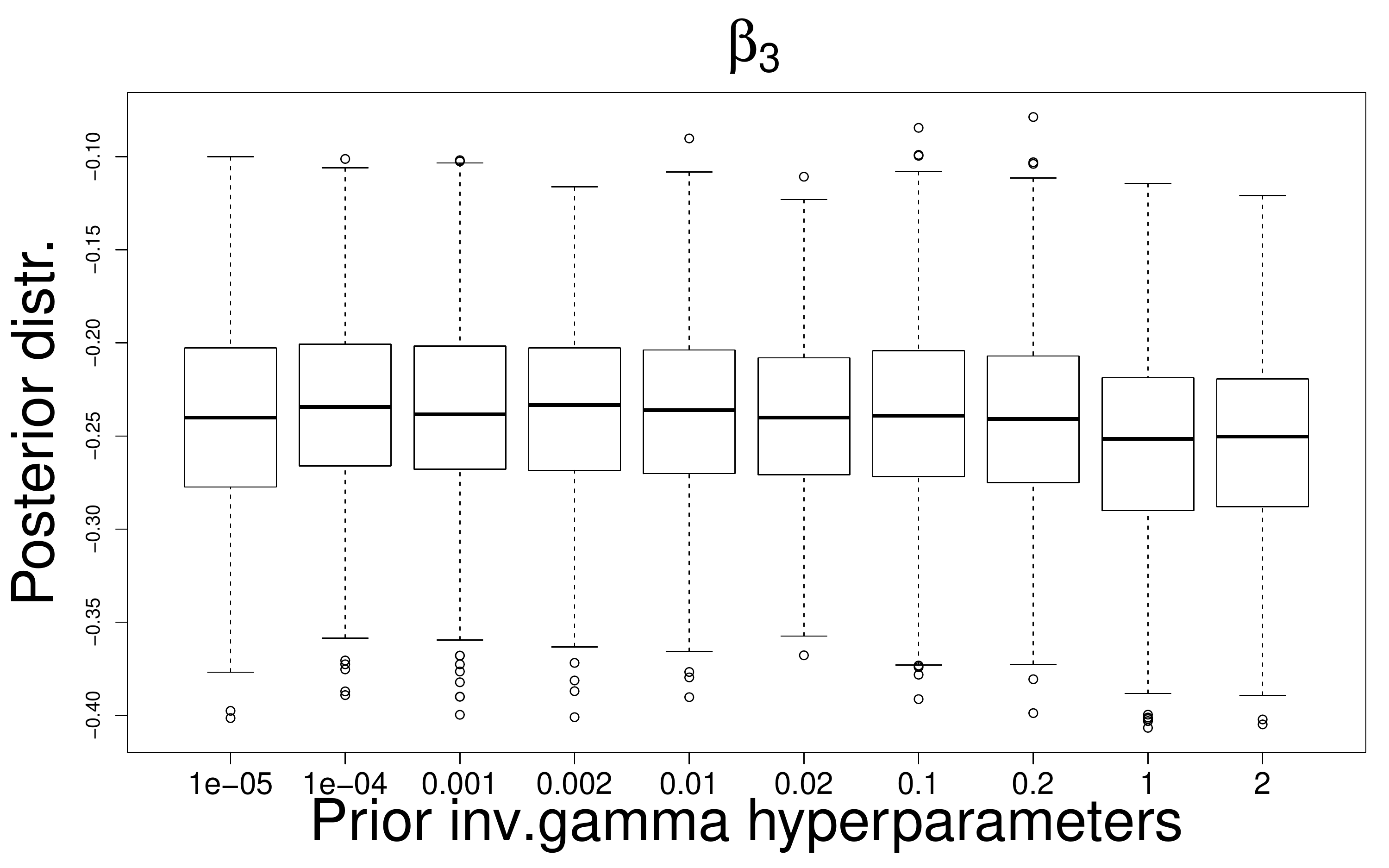}~
\includegraphics[scale=0.25]{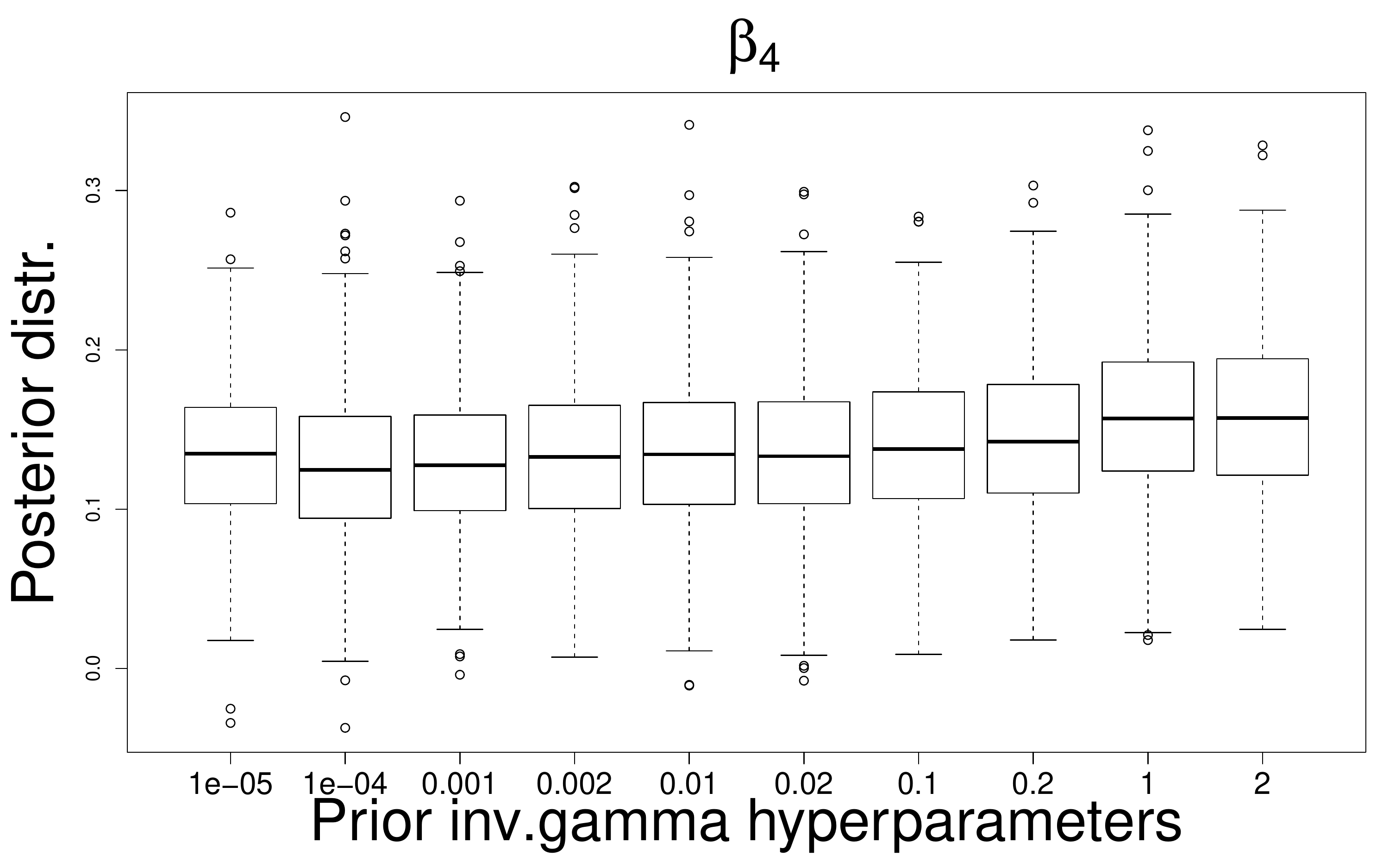}\\
\includegraphics[scale=0.25]{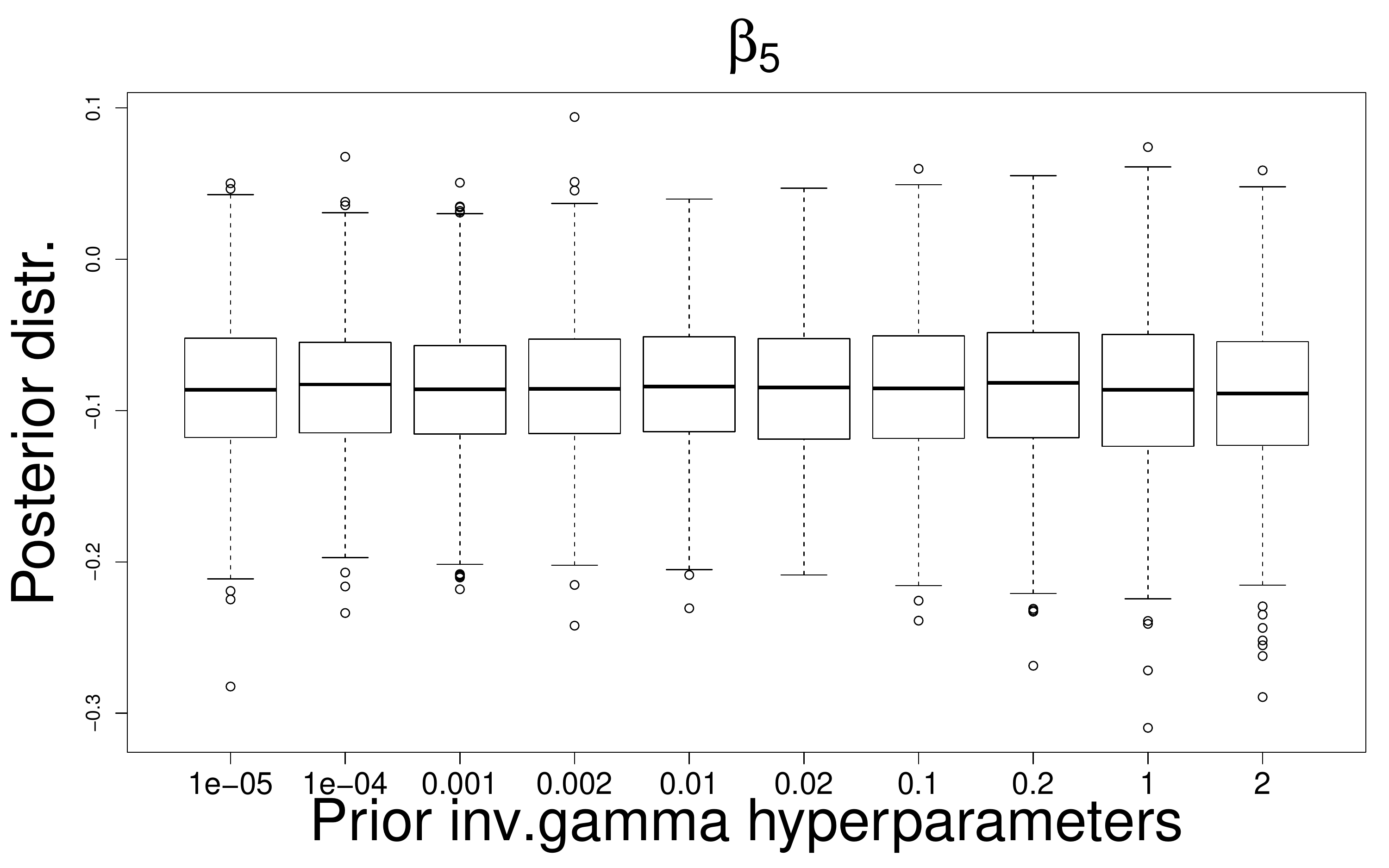}~
\includegraphics[scale=0.25]{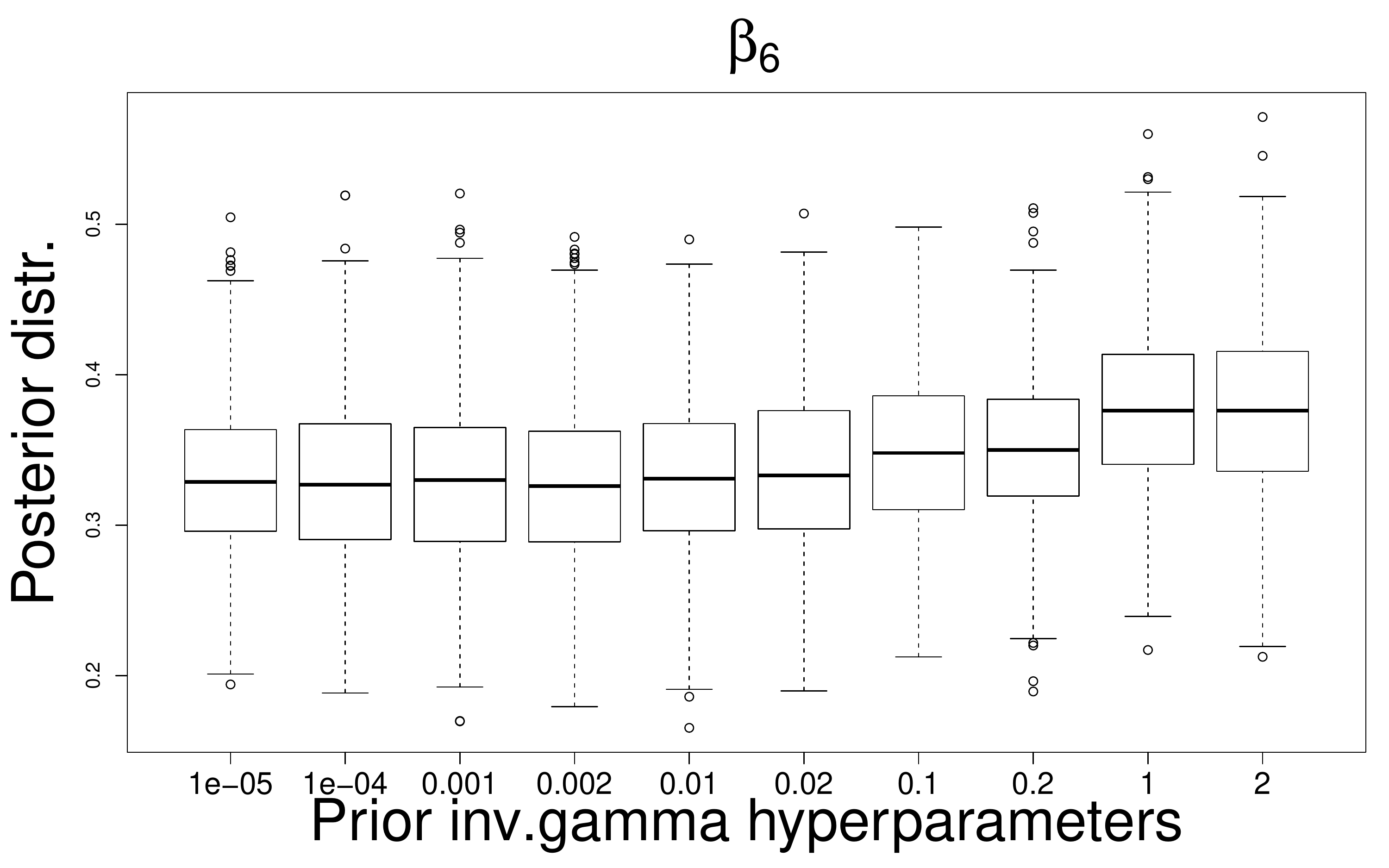}\\
\includegraphics[scale=0.25]{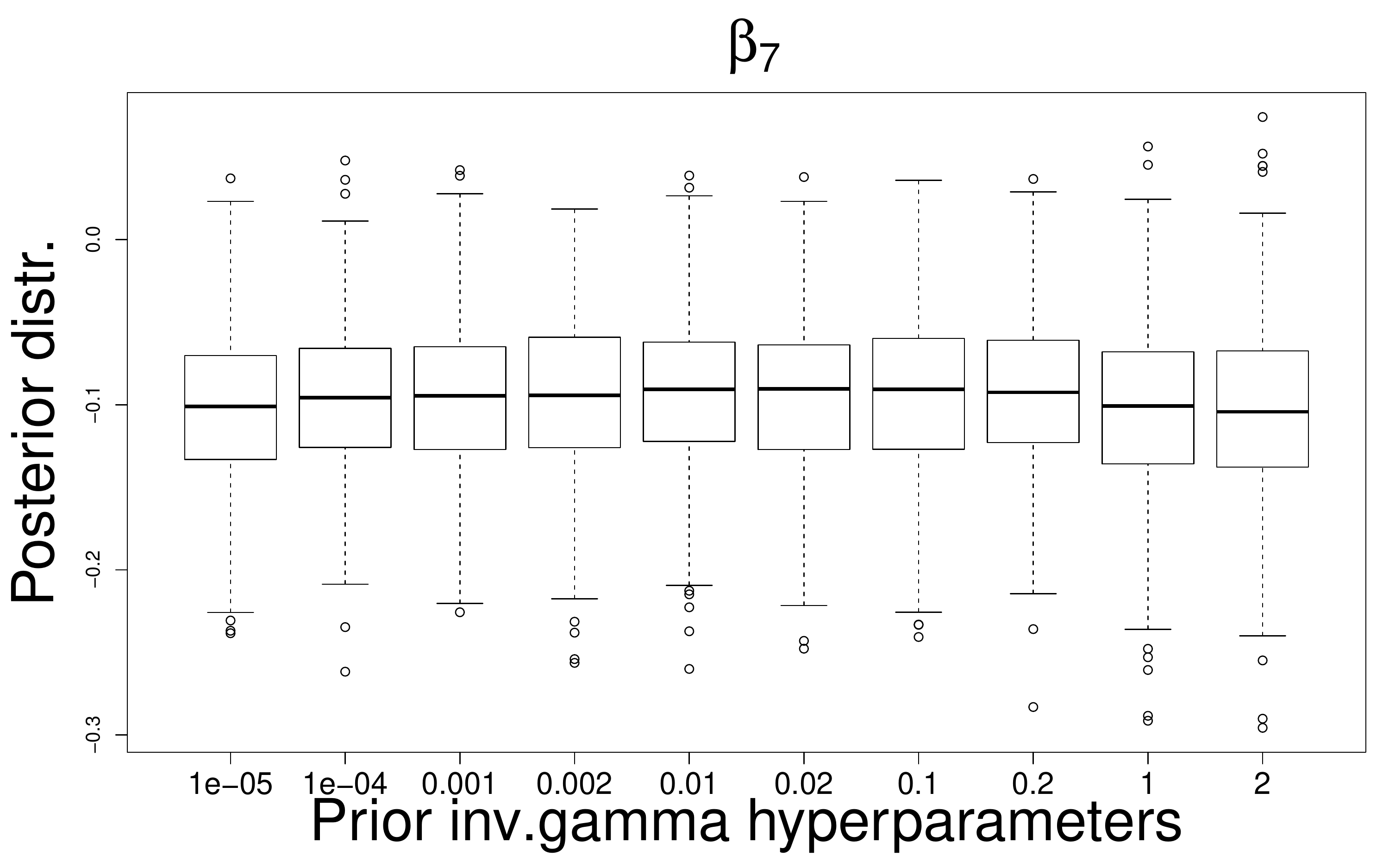}~
\includegraphics[scale=0.25]{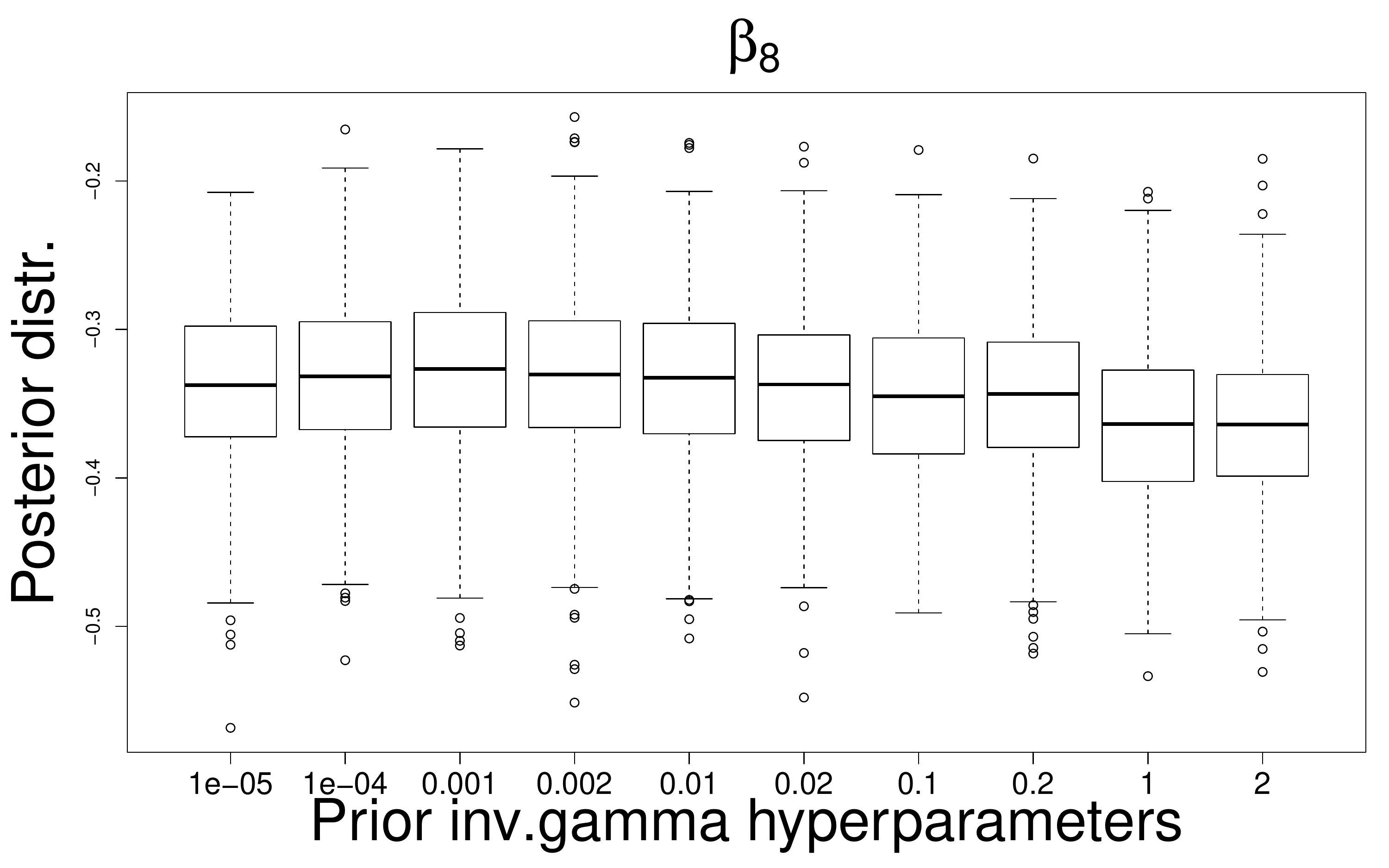}\\
\caption{Sensitivity analysis  for point abilities parameters $\beta_1, \ldots,\beta_8$: Posterior box-plots for a variety of hyperparameter values of the inverse-gamma prior assigned to $\tau^2_{\beta}$. 3000 MCMC iterations obtained from 3 parallel chains with burn-in period of 100 iterations using {\tt rjags}.}
\label{figS3}
\end{figure}

\begin{figure}
\centering

\includegraphics[scale=0.25]{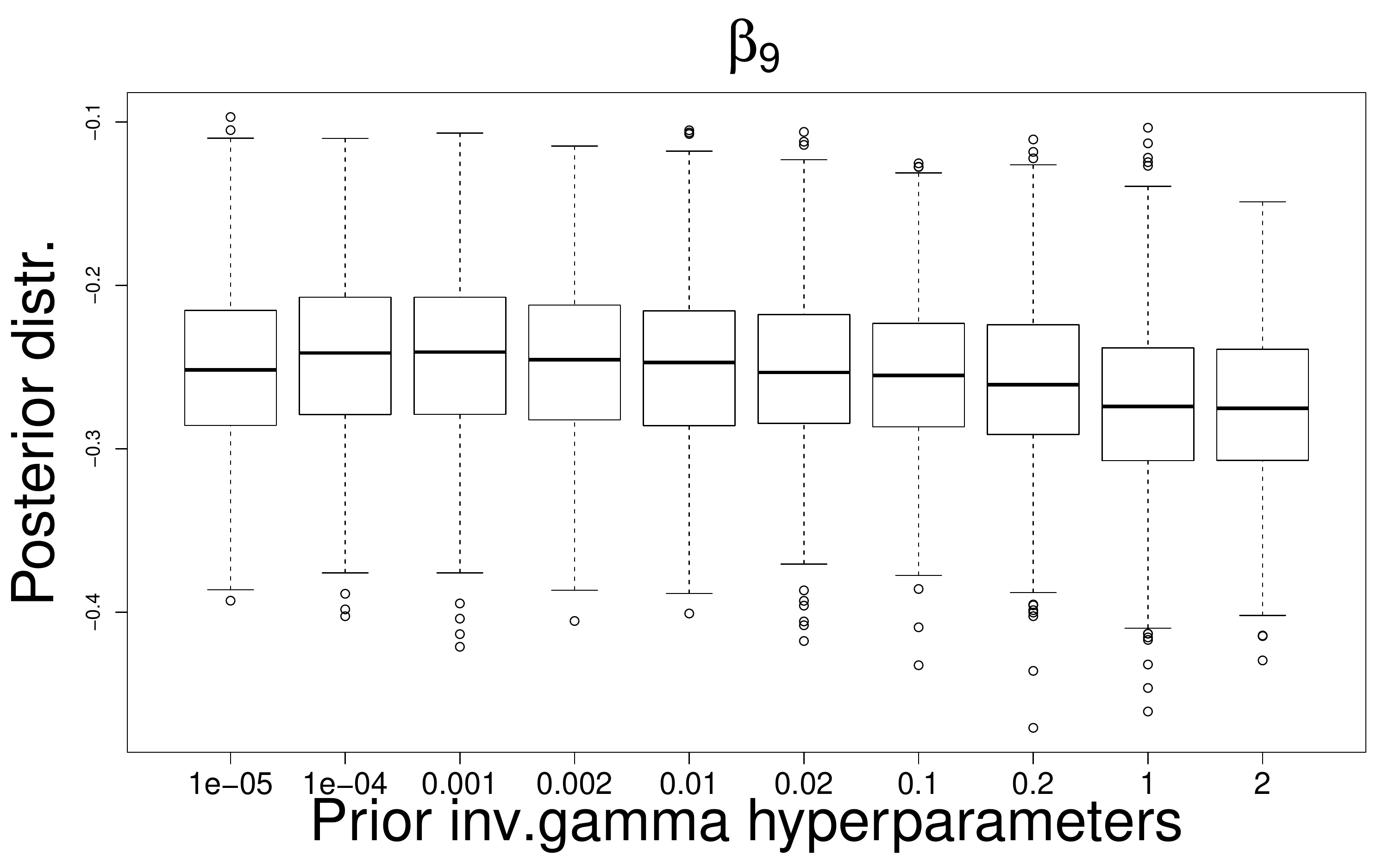}~
\includegraphics[scale=0.25]{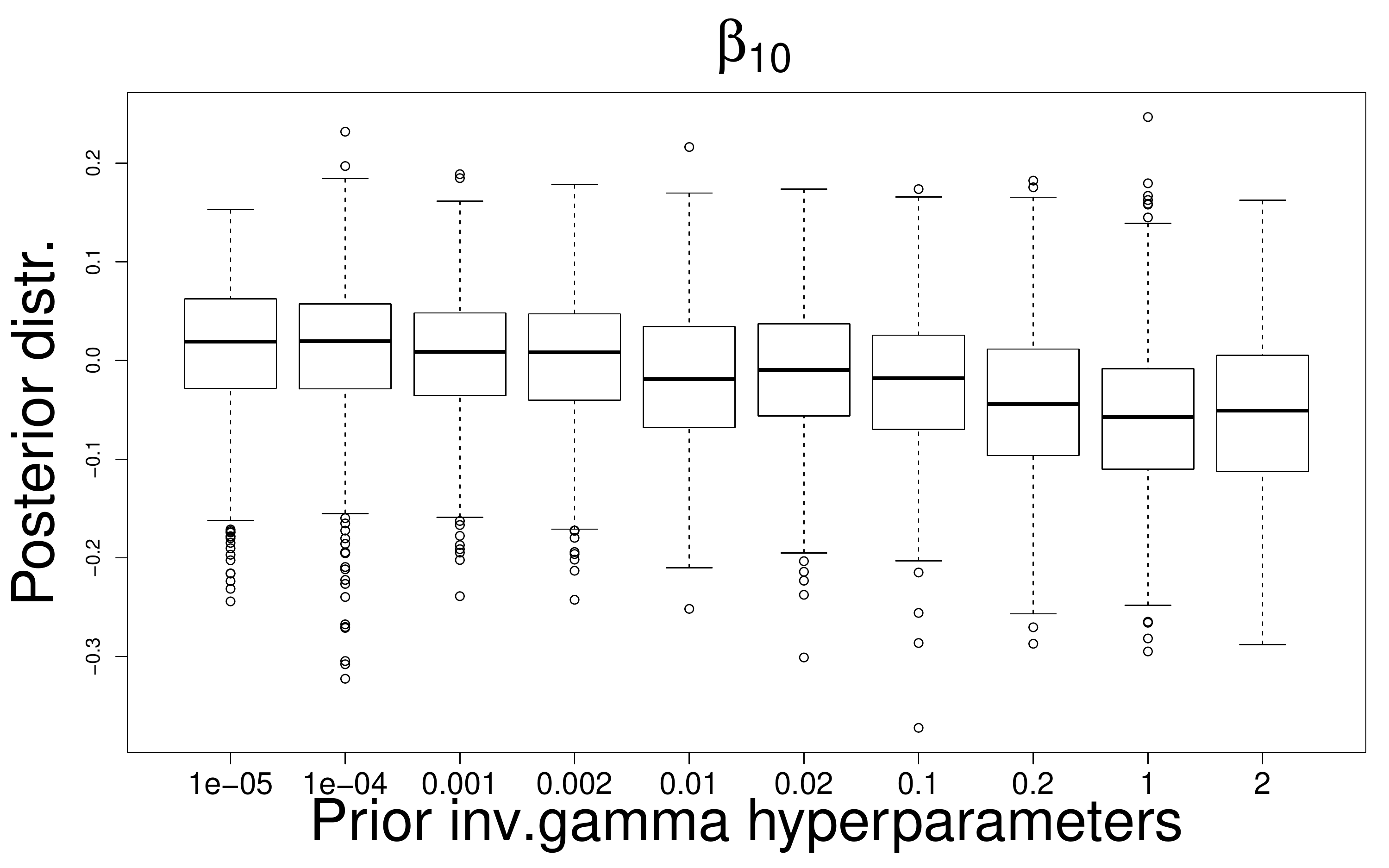}\\
\includegraphics[scale=0.25]{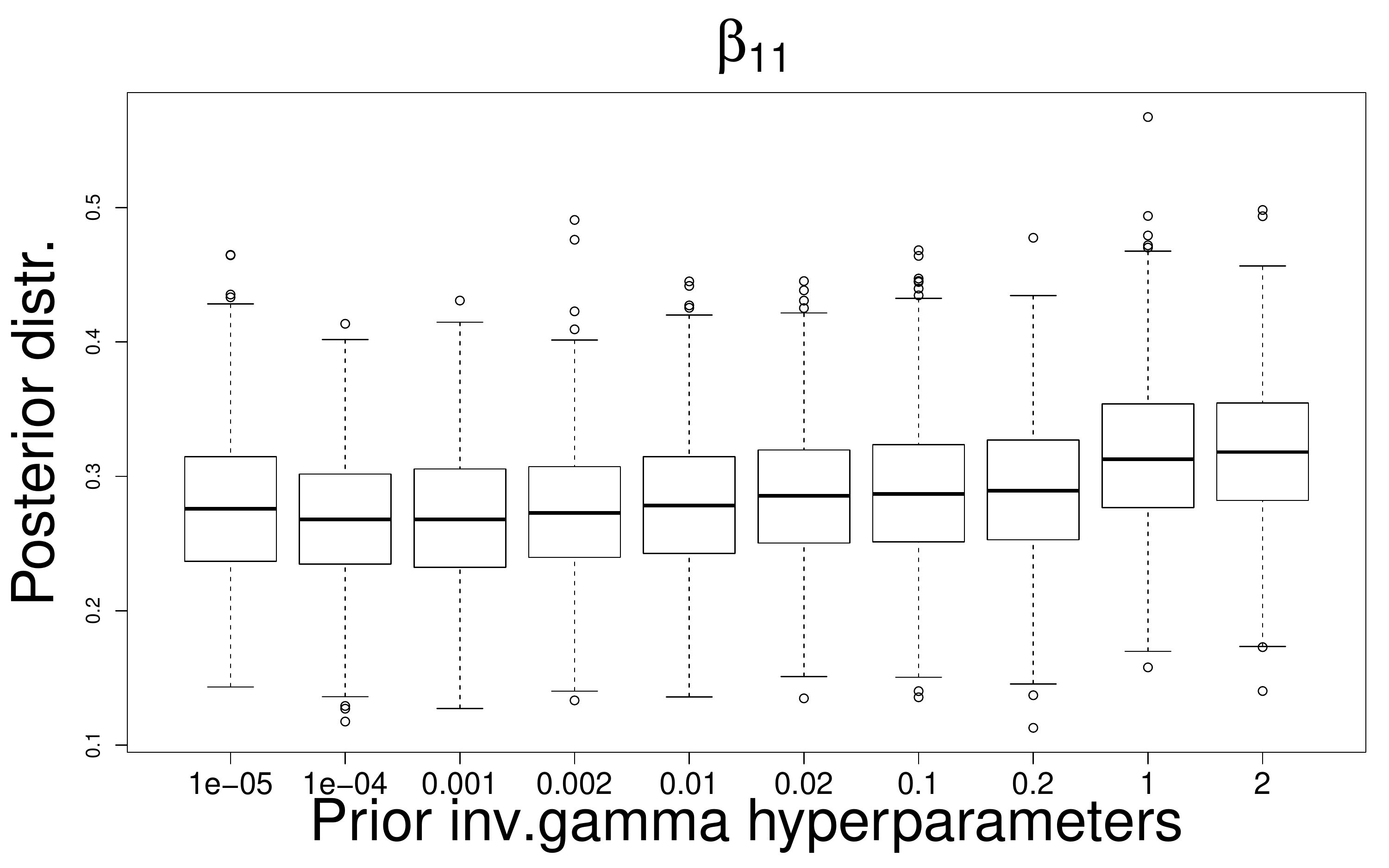}~
\includegraphics[scale=0.25]{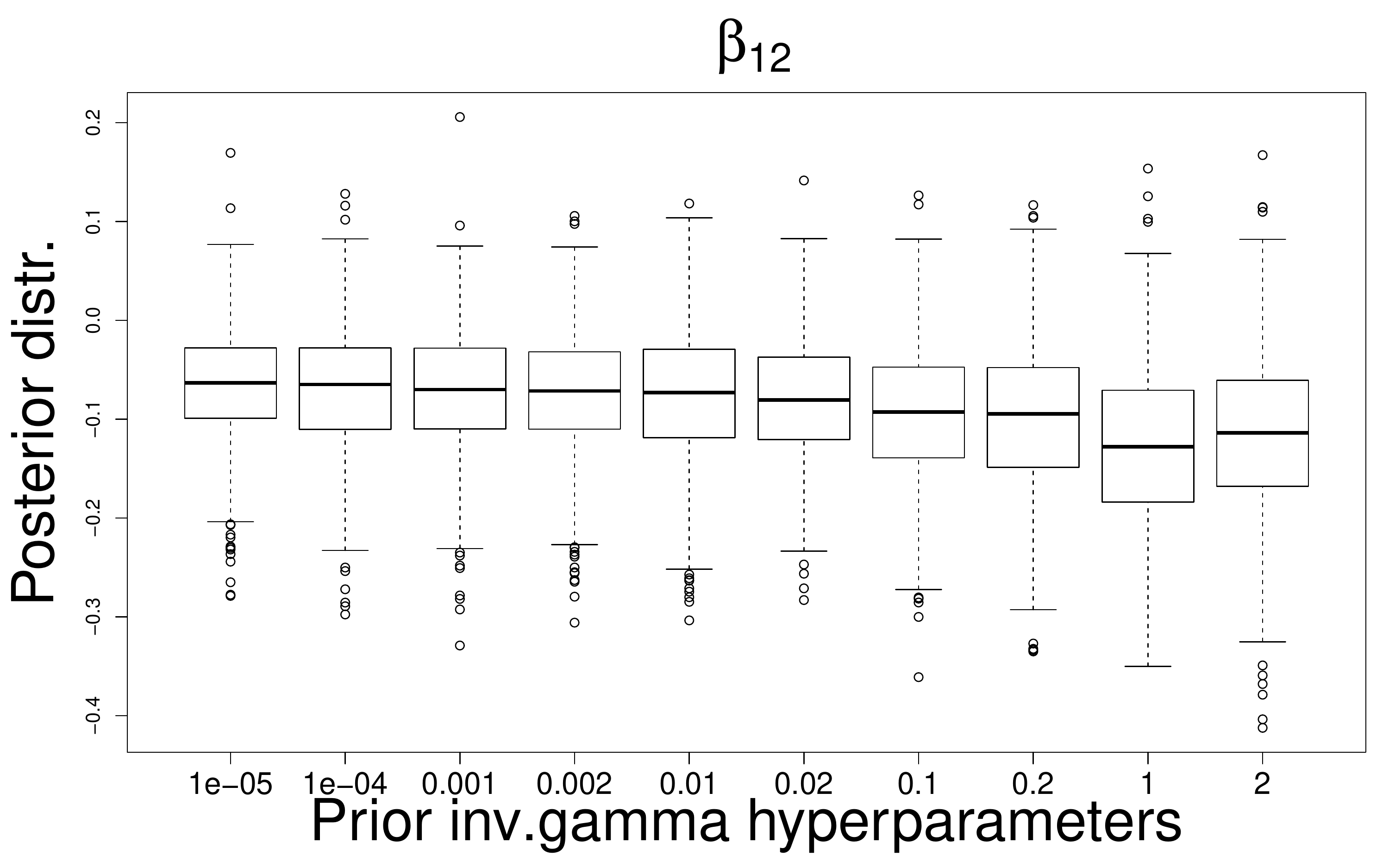}\\
\includegraphics[scale=0.25]{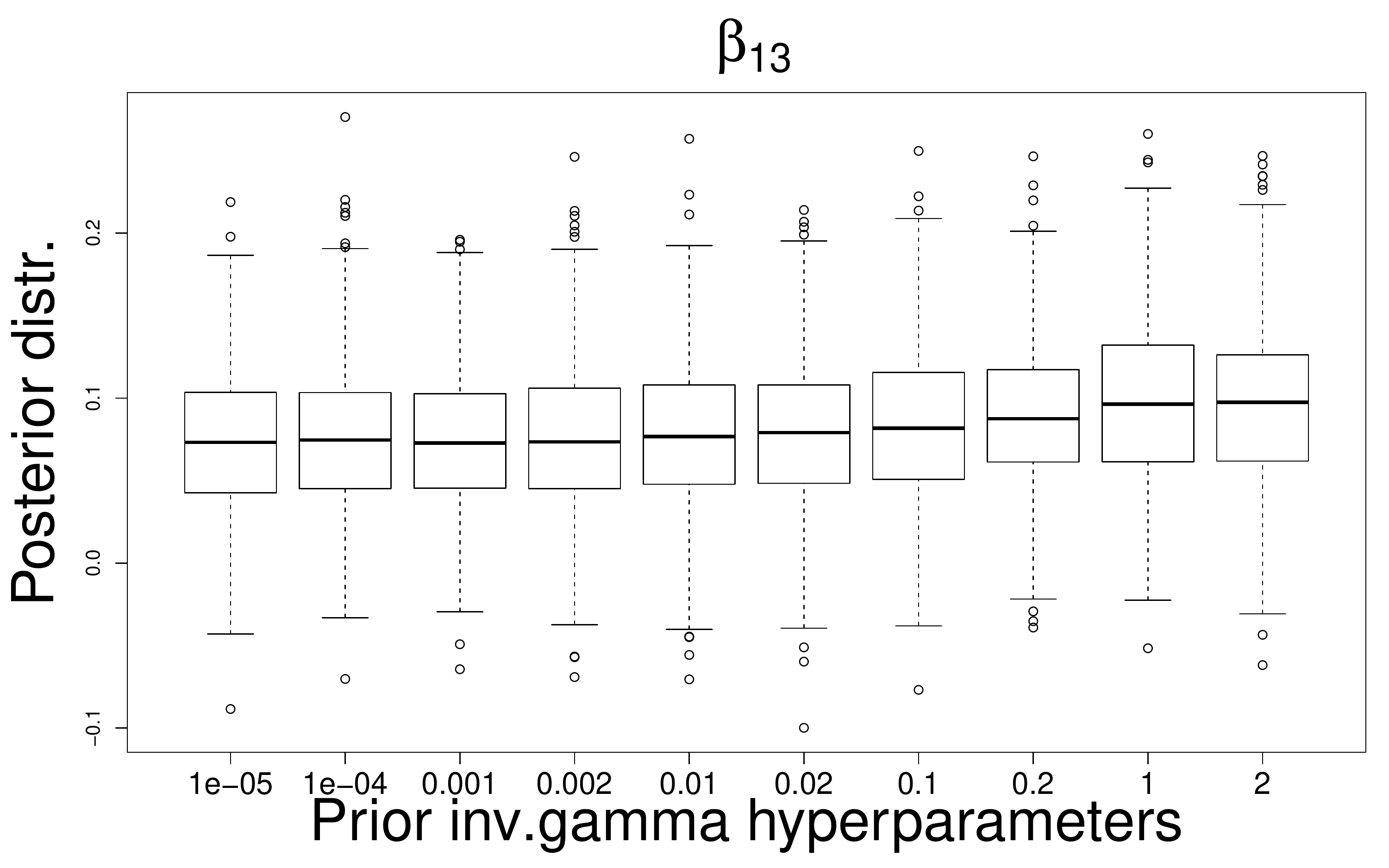}~
\includegraphics[scale=0.25]{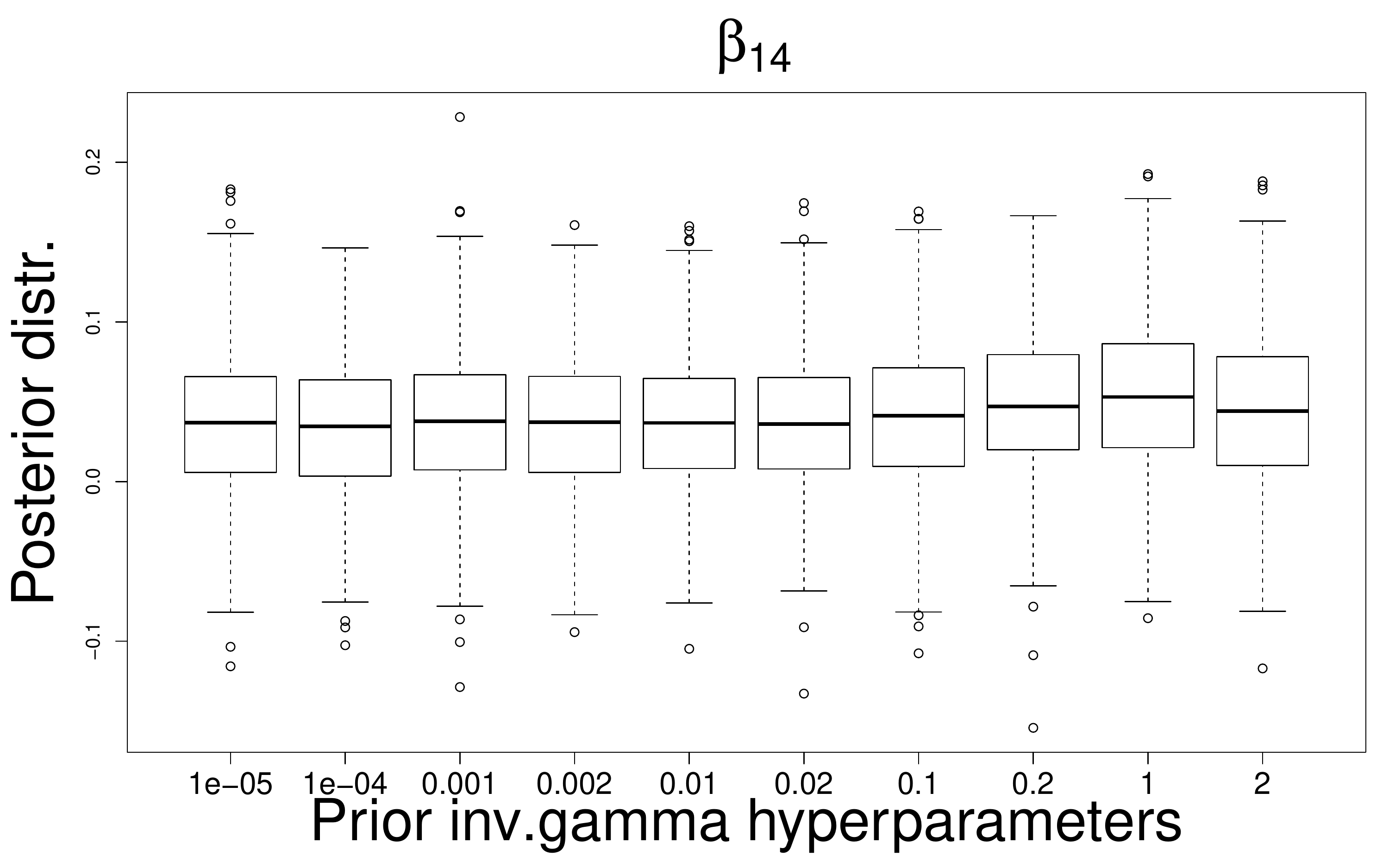}\\
\includegraphics[scale=0.25]{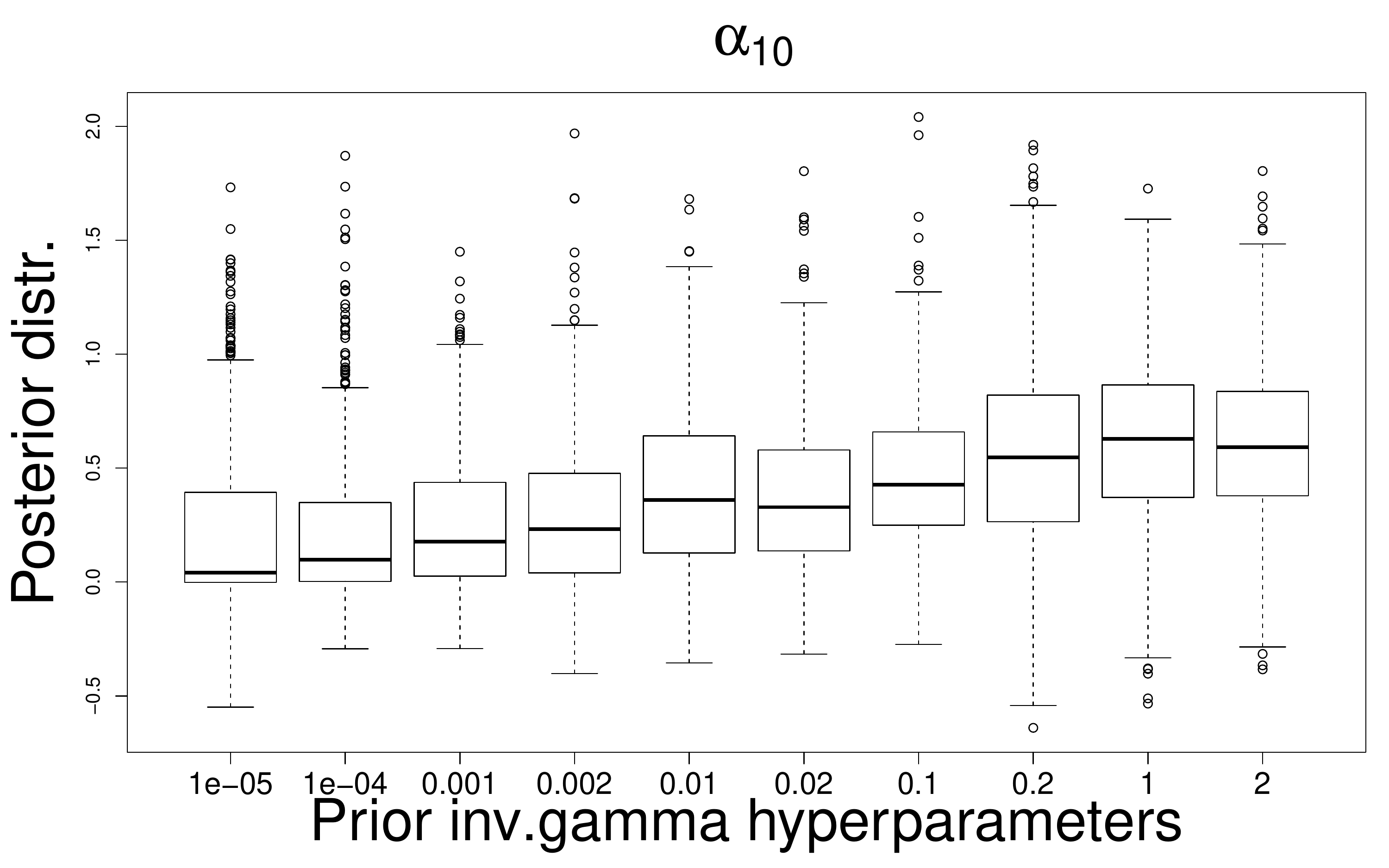}~
\includegraphics[scale=0.25]{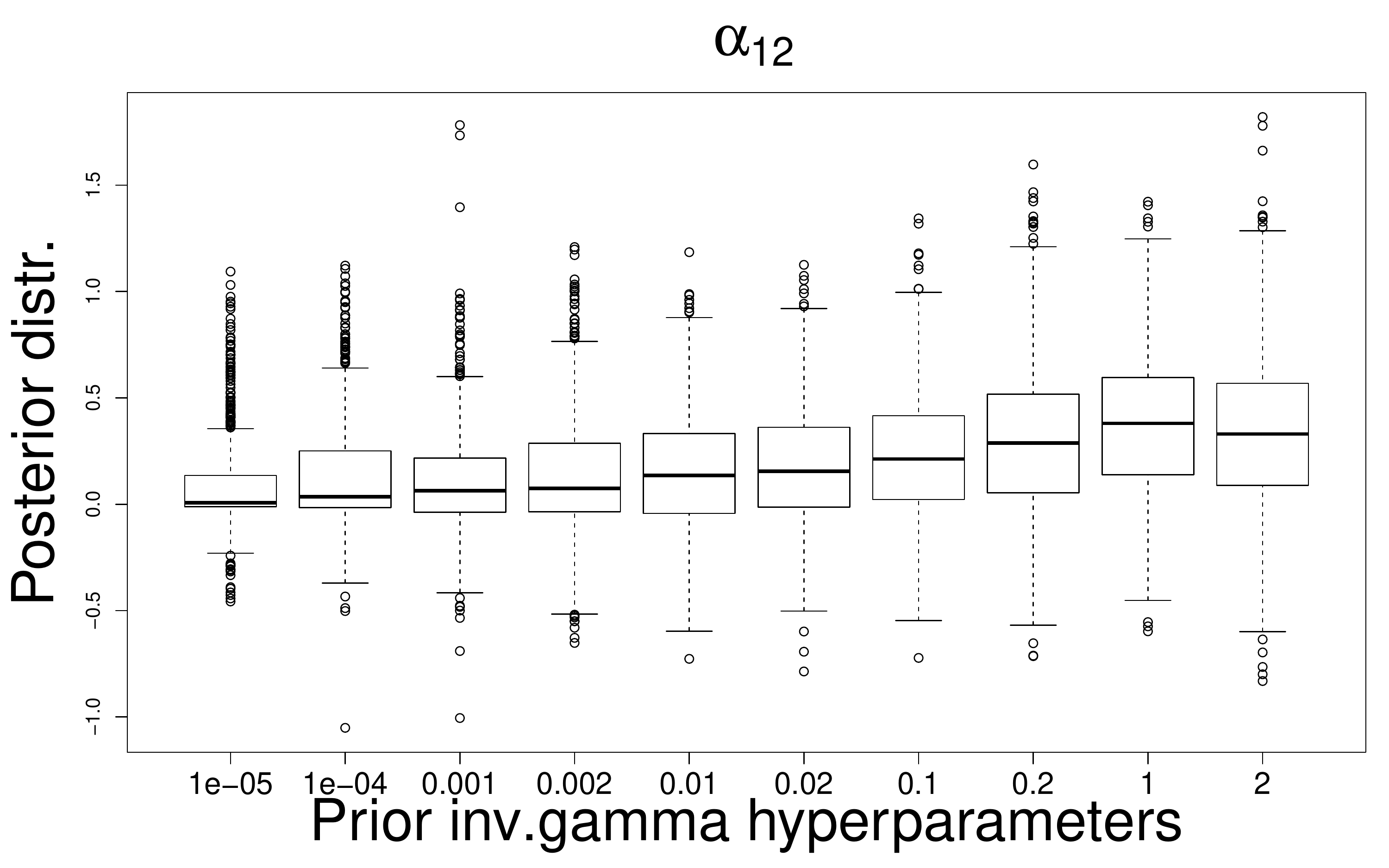}\\
\caption{Sensitivity analysis for the point abilities parameters $\beta_9, \ldots,\beta_{14}$ and the extra set abilities $\alpha_{10}, \alpha_{12}$ Posterior box-plots for a variety of hyperparameter values of the inverse-gamma prior assigned to $\tau^2_{\alpha}$ and $\tau^2_{\beta}$. 3000 MCMC iterations obtained from 3 parallel chains with burn-in period of 100 iterations using {\tt rjags}} 
\label{figS4}
\end{figure}

In order to ensure the robustness of our posterior estimates with regard to the prior specification, we performed sensitivity analysis for the model's parameters (see Table 4 in the paper for a comprehensive summary of the selection procedure of the final model). Sensitivity plots may be retrieved in the Supplementary Material folder available at \url{https://github.com/LeoEgidi/Bayesian-Volleyball-paper}. Specifically, we let vary:

\begin{itemize}
\item the standard deviations of the normal priors for the parameters $\mu, \theta, H^{point}, H^{set}, m$
(first five panels in Figure~\ref{figS1});
\item  the scale parameter of the log-normal prior for $\lambda$ (sixth panel in Figure~\ref{figS1});
\item the inverse-gamma parameters $a_1, a_2, b_1, b_2$ for the hyper-priors $\tau^2_{\alpha} \sim \mathsf{InvGamma}(a_1,a_2), \ \tau^2_{\beta} \sim \mathsf{InvGamma}(b_1,b_2),$ assuming that the set and point abilities are assigned two normal priors, $\bm{\alpha} \sim \mathcal{N}(0, \tau^2_{\alpha}),\ \bm{\beta} \sim \mathcal{N}(0, \tau^2_{\beta}) $, respectively. For simplicity, we assume $a_1=a_2,\ b_1=b_2$ (Figures~\ref{figS3}--\ref{figS4}).
\end{itemize}
Red triangles on the $x$-axis in Figures~\ref{figS1} denote the values used in the final model. As a general comment, the posterior estimates seem to be not much sensitive with respect to the hyperparameters choices. All selected values seem to lie on the ``non-informative'' region where the priors are stabilized to a specific posterior solely based on data information. 

Regarding the point and set abilities in Figures~\ref{figS3}--\ref{figS4}, we note that there are not sensitive changes of the posterior estimates in correspondence of distinct hyperparameters' values of the inverse-gamma distributions for $\tau^2_{\alpha}, \tau^2_{\beta}$.

This sensitivity analysis suggests that our posterior results are robust to changes of the prior parameters when they are changing towards distributions of greater uncertainty (and hence, of lower information).  Therefore, our selected prior distributions can be regarded as ``non-informative'' priors resulting to an informal ``objective'' Bayes analysis. 


\clearpage 
\newpage

\section{Set dynamic abilities}

\begin{figure}[h!]
\centering
\includegraphics[scale=0.7]{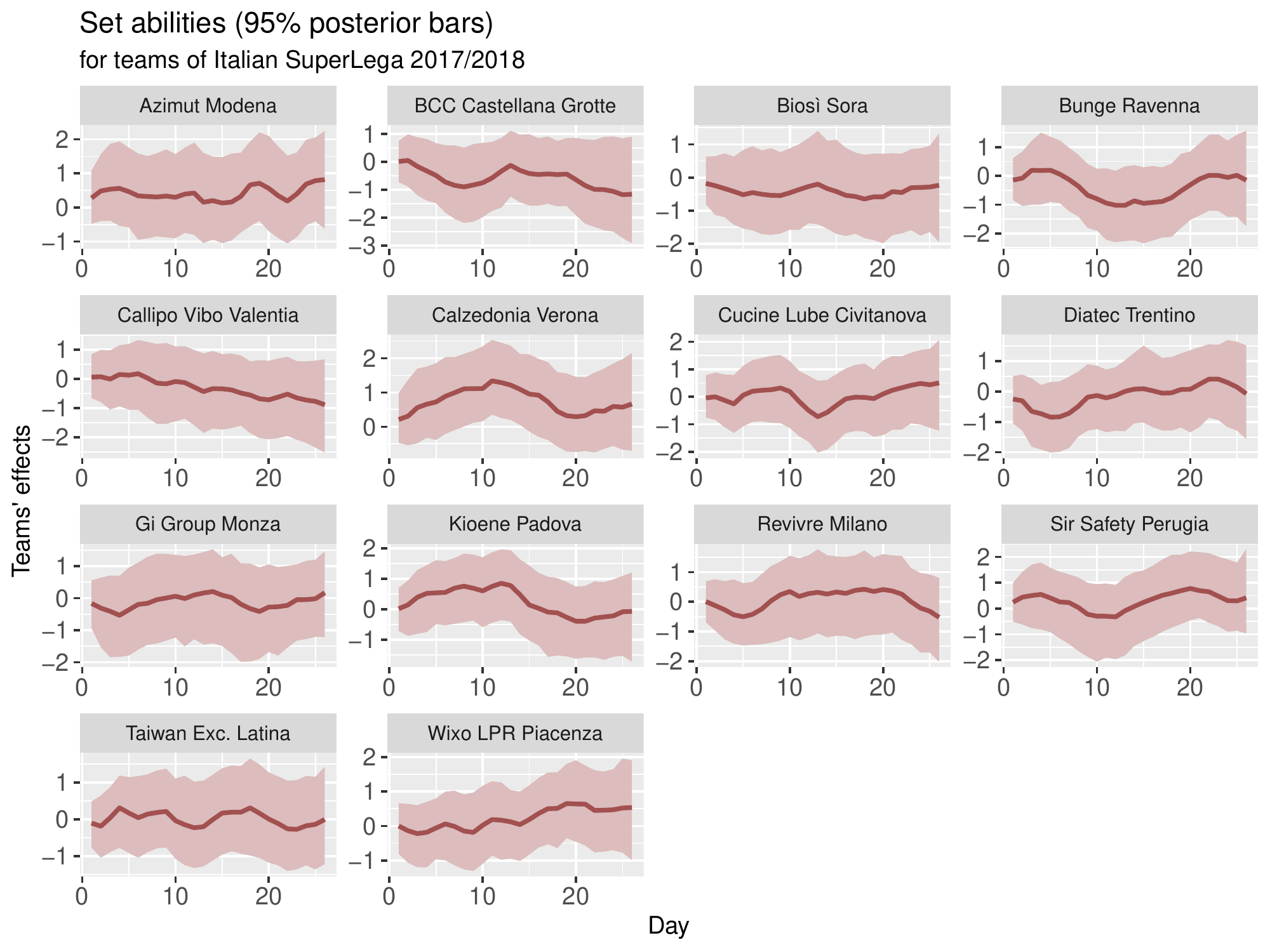}
\caption{Posterior mean and 95\% density interval for the dynamic set ability parameters $\bm{\alpha}$ for the Italian SuperLega 2017/2018 data (model 11 in Table 2 of the paper).}
\label{fig16}
\end{figure}

Here,  in Figure~\ref{fig16}, we provide the predictive intervals for the dynamic set abilities (model 11 of Table 2 in the paper), with $\sigma^2_{\alpha} \sim \mathsf{InvGamma}(2,2)$ which was not included in the paper.

\clearpage 
\newpage 
\section{Graphical convergence diagnostics}

\begin{figure}[h!]
\centering
\includegraphics[scale=0.6]{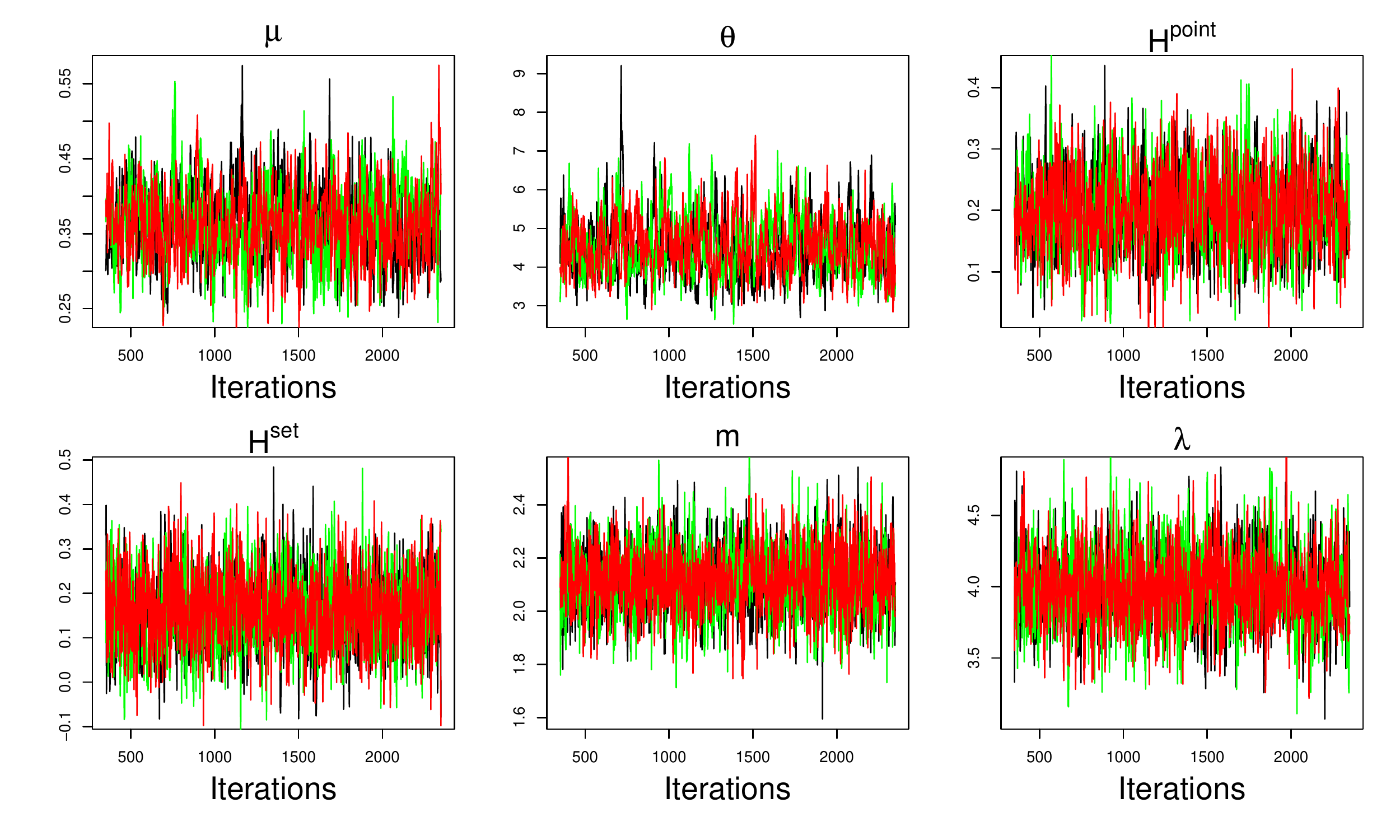}\\
\caption{Trace plots for the parameters: $\mu, \theta, H^{point}, H^{set},\lambda, m$ (model 9 in the paper).}
\label{figS6}
\end{figure}

\begin{figure}
\centering
\includegraphics[scale=0.6]{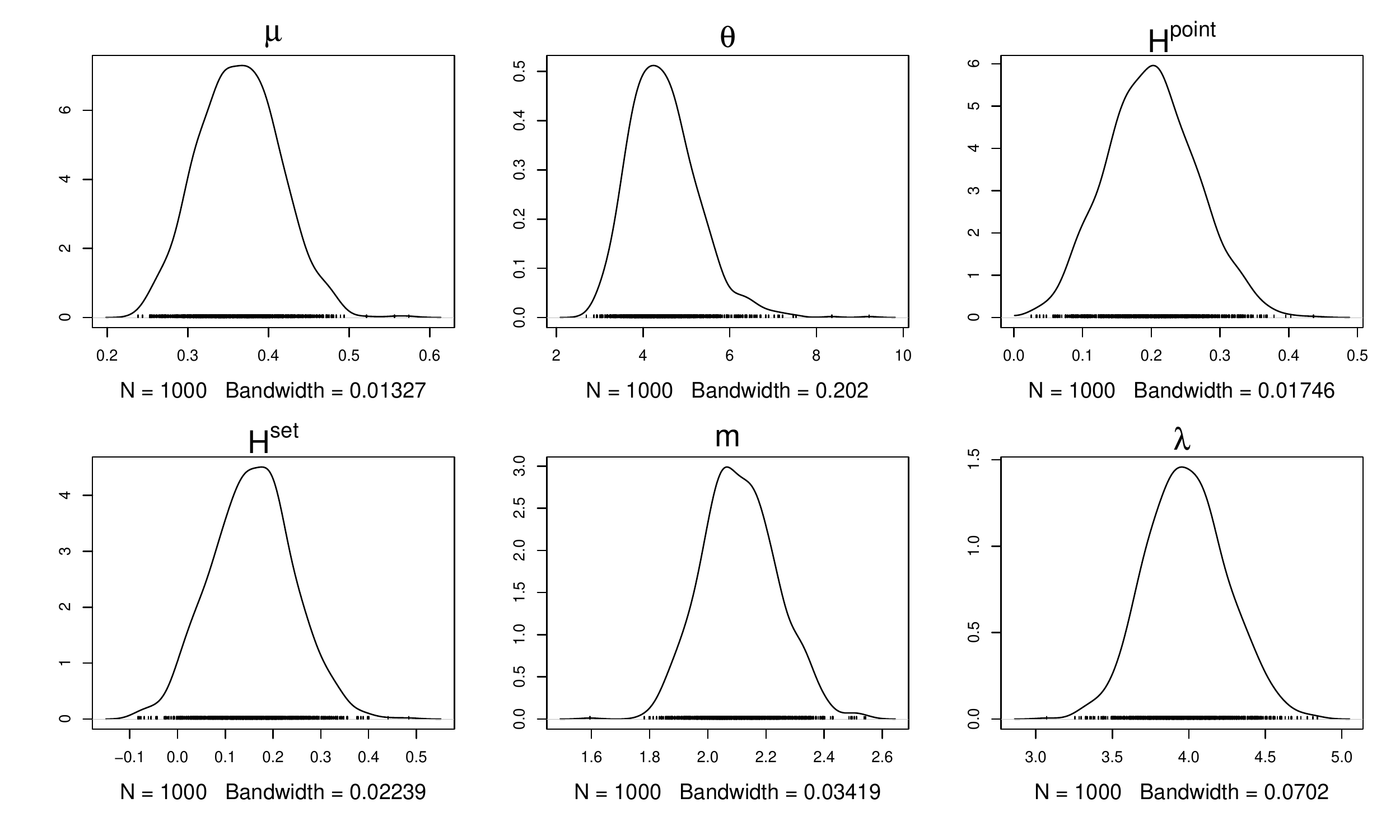}\\
\caption{Posterior density plots for the parameters: $\mu, \theta, H^{point}, H^{set}, m, \lambda$  (model 9 in the paper).}
\label{figS7}
\end{figure}

Figures~\ref{figS6} and \ref{figS7} depict the trace and the density plots from the MCMC sampling for the parameters: $\mu, \theta, H^{point}, H^{set},\lambda, m$ (3000 total iterations obtained using 3 chains of 1000 iterations and additional 100 iterations as a burn-in period in \texttt{rjags}). 

\end{document}